\documentclass[aps,prd,showpacs,superscriptaddress,groupedaddress,nofootinbib]{revtex4-2} 
\usepackage[T1]{fontenc}
\usepackage{lmodern} 
\usepackage{graphicx}  
\usepackage{dcolumn}   
\usepackage{bm}        
\usepackage{amssymb}   
\usepackage{amsmath}
\usepackage{float}
\usepackage{bbold}
\usepackage{array}
\usepackage{makecell}
\usepackage{braket}
\usepackage{longtable}
\usepackage{supertabular,booktabs}
\usepackage{amsmath}
\usepackage{amsmath}

\usepackage[english]{babel}
\usepackage{xspace}

\usepackage[perpage]{footmisc}

\renewcommand{\thefootnote}{\fnsymbol{footnote}}
\usepackage{graphicx,lipsum}
\newcommand{\PRLsep}{\noindent\makebox[\linewidth]{\resizebox{0.6999\linewidth}{1.5pt}{$\bullet$}}\bigskip}

\usepackage{titlesec} 
\usepackage[colorlinks,citecolor=blue,urlcolor=blue,hypertexnames=true]{hyperref}
\usepackage{subfigure}

\usepackage[usenames,dvipsnames,svgnames,table]{xcolor} 

\newcommand{\be}{\begin{equation}}
\newcommand{\ee}{\end{equation}}
\newcommand{\bea}{\begin{eqnarray}}
\newcommand{\eea}{\end{eqnarray}}
\newcommand{\bml}{\begin{subequations}}
\newcommand{\eml}{\end{subequations}}
\newcommand{\bfig}{\begin{figure}}
\newcommand{\efig}{\end{figure}}

\newcommand{\bmat}{\begin{pmatrix}}
\newcommand{\emat}{\end{pmatrix}}

\begin{document}


\renewcommand{\thefootnote}{\fnsymbol{footnote}}

\pagenumbering{roman}
	\baselineskip=3.5pt \thispagestyle{empty}
	
	
	\vspace{0.2cm}
\title{\textcolor{Sepia}{\textbf \Huge \Large {Circuit Complexity in $Z_{2}$ EEFT}}}


	 \vspace{0.25cm}
\author{{\large  Kiran Adhikari${}^{1}$}}
\email{kiran.adhikari@tum.de}
\author{{\large  Sayantan Choudhury${}^{2,3,4}$}}
\email{sayantan_ccsp@sgtuniversity.org\\  sayanphysicsisi@gmail.com}

\author{ \large Sourabh Kumar${}^{5,6}$}
\email{sourabh.phys@gmail.com}
\author{\large Saptarshi Mandal${}^{7,8}$}
\email{saptarshijhikra@gmail.com }
\author{ \large Nilesh Pandey${}^{9}$}
\email{nilesh911999@gmail.com  }
\author{ \large Abhishek Roy${}^{10}$}
\email{roy.1@iitj.ac.in }
\author{\large Soumya Sarkar${}^{11}$}
\email{sarkarsoumya65@gmail.com }
\author{\large Partha Sarker${}^{12}$}
\email{sarker239@gmail.com }
\author{\large Saadat Salman Shariff${}^{13,14}$}
\email{saadatsalman342@gmail.com}

\affiliation{ ${}^{1}$RWTH Aachen University, Aachen, Germany.}
\affiliation{${}^{2}$Centre For Cosmology and Science Popularization (CCSP),\\
        SGT University, Gurugram, Delhi- NCR, Haryana- 122505, India.}
\affiliation{${}^{3}$National Institute of Science Education and Research, Bhubaneswar, Odisha - 752050, India.}
\affiliation{${}^{4}$Homi Bhabha National Institute, Training School Complex, Anushakti Nagar, Mumbai-400085, India.}
\affiliation{${}^{5}$Department of Physics and Astronomy, University of Calgary, Calgary, AB T2N 1N4, Canada.}
\affiliation{${}^{6}$Institute for Quantum Science and Technology, University of Calgary, Calgary, AB T2N 1N4, Canada.}
\affiliation{${}^{7}$  Department of Physics, Jadavpur University, Kolkata-700032.}
\affiliation{${}^{8}$ Department of Physics, Indian Institute of Technology Kharagpur, Kharagpur-721302, India.}
\affiliation{${}^{9}$Department of Applied Physics, Delhi Technological University, Delhi-110042, India.}
\affiliation{${}^{10}$ Department of Physics, Indian Institute of Technology Jodhpur,Karwar, Jodhpur - 342037, India.}
\affiliation{${}^{11}$National Institute of Technology Karnataka, Mangalore, Karnataka-575025.}
\affiliation{${}^{12}$Department of Physics, University of Dhaka, Curzon Hall, Dhaka 1000, Bangladesh.}	
\affiliation{${}^{13}$Department of Theoretical Physics, University of Madras, Guindy Campus, Chennai - 600025.}
\affiliation{${}^{14}$Department of Physics,  Indian Institute of Science and Educational Research, Behrampur-760010.}
    



	{
	

\begin{abstract}
Motivated by recent studies of circuit complexity in weakly interacting scalar field theory, we explore the computation of circuit complexity in $\mathcal{Z}_2$ Even Effective Field Theories ($\mathcal{Z}_2$ EEFTs). We consider a massive free field theory with higher-order Wilsonian operators such as $\phi^{4}$, $\phi^{6}$, and $\phi^8.$ To facilitate our computation, we regularize the theory by putting it on a lattice. First, we consider a simple case of two oscillators and later generalize the results to $N$ oscillators. The study has been carried out for nearly Gaussian states. In our computation, the reference state is an approximately Gaussian unentangled state, and the corresponding target state, calculated from our theory, is an approximately Gaussian entangled state. We compute the complexity using the geometric approach developed by Nielsen, parameterizing the path ordered unitary transformation and minimizing the geodesic in the space of unitaries. The contribution of higher-order operators, to the circuit complexity, in our theory has been discussed. We also explore the dependency of complexity with other parameters in our theory for various cases. 

\end{abstract}




\setcounter{page}{2}

\pagenumbering{arabic}


\maketitle
\textcolor{Sepia}{\section{\sffamily Prologue}\label{sec:introduction}}
In recent years, tools and techniques from Quantum Information have played a vital role in developing new perspectives in areas such as Quantum Field Theory and Holography, in particular the AdS/CFT duality. A particular line of study in the context of the AdS/CFT correspondence is to decipher the emergence of bulk physics using information from the boundary CFT \cite{Harlow:2018fse}. It has been shown in \cite{Ryu:2006bv,Hubeny:2007xt,Rangamani:2016dms} that the codimension-2 extremal surfaces in the AdS are associated with the Entanglement Entropy (EE) of the boundary CFT. However, in recent years, studies from black hole physics suggest that EE is not sufficient to capture the complete information, which led Susskind et al. to introduce a new measure, known as Quantum Computational Complexity (QCC) \cite{Susskind:2014rva,Stanford:2014jda,Susskind:2014jwa,Susskind:2014moa,Brown:2015lvg,Brown:2015bva,Brown:2016wib,Couch:2016exn,Susskind:2018pmk}. In the context of AdS/CFT, QCC of the dual CFT is proposed to be associated with the properties of codimension-0 and codimension-1 extremal surfaces. This stirred the study of QCC in QFTs.\par
Not only in the context of holography, but the complexity of quantum states has emerged as a significant quantity of interest across different subfields of physics, from quantum computing and information to many-body physics, as it appears to be a better measure of information. In \cite{Jefferson:2017sdb, Chapman:2017rqy}, the notion of circuit complexity has been defined and studied for free bosonic field theory and in \cite{Khan:2018rzm,Hackl:2018ptj} for free fermionic field theory. For a weakly interacting field theory, \cite{Bhattacharyya:2018bbv} extends the study to the $\phi^4$ theory, where in addition to the study of QCC, its relationship with Renormalization Group Flows has also been explored. The growth of complexity in the quantum circuit model has been studied in \cite{Haferkamp:2021uxo}. Circuit complexity has also been discussed in the context of chaos, quantum mechanics and quantum computing in \cite{Bhattacharyya:2020art,Ali:2019zcj,Eisert:2021mjg,Roberts:2016hpo}. It has been probed in relation to conformal and topological field theories, and Chern-Simmons theory \cite{Camilo:2019bbl,Couch:2021wsm,Chagnet:2021uvi,Flory:2020dja}.  An active study of this quantity in the context of many-body quantum systems has been also gaining interest in recent years \cite{Jaiswal:2020snm}. QCC has been studied in many other contexts. It has been explored extensively in holography \cite{Barbon:2015ria,Alishahiha:2015rta,Yang:2016awy,Chapman:2016hwi,Carmi:2016wjl,Reynolds:2016rvl,Zhao:2017iul,Flory:2017ftd,Reynolds:2017lwq,Carmi:2017jqz,Couch:2017yil,Yang:2017czx,Abt:2017pmf,Swingle:2017zcd,Reynolds:2017jfs,Fu:2018kcp,An:2018xhv,Bolognesi:2018ion,Chen:2018mcc,Abt:2018ywl,Hashimoto:2018bmb,Flory:2018akz,Couch:2018phr,HosseiniMansoori:2018gdu,Chapman:2018dem,Chapman:2018lsv,Caceres:2019pgf,Ben-Ami:2016qex,Abad:2017cgl}. The thermodynamic properties of QCC have been studied in \cite{Brown:2017jil,Bernamonti:2019zyy,Bernamonti:2020bcf}. Also, various applications and properties of QCC have been investigated in \cite{Cai:2016xho,Lehner:2016vdi,Moosa:2017yvt,Moosa:2017yiz,Hashimoto:2017fga,Chapman:2018hou,Guo:2018kzl,Camargo:2018eof,Doroudiani:2019llj,Chapman:2019clq,Bhattacharyya:2019kvj,Bhargava:2020fhl,Lehners:2020pem,Bhattacharyya:2020kgu,Choudhury:2020lja,Choudhury:2020hil,Adhikari:2021pvv,Adhikari:2021ked,Choudhury:2021qod,Bai:2021ldj,Caputa:2017yrh,Caputa:2018kdj,Boruch:2020wax,Boruch:2021hqs}.

In this paper, we extend the work of \cite{Bhattacharyya:2018bbv} by including even higher-order Wilsonian operators, which we denote by $\mathcal{Z}_{2}$ EEFT (Even Effective Field Theory). Our theory contains the interaction terms $\phi^{4}$, $\phi^{6}$ and $\phi^{8}$. These are weakly coupled to the free scalar field theory via the coupling constants $\lambda_{4}$, $\lambda_{6}$ and $\lambda_{8}$ respectively. The primary motivation for studying QCC in this context is to compute and understand QCC by including higher-order terms. The organization of the paper is as follows. In section \ref{sec:complexity}, we summarize Nielsen's method for computing the circuit complexity. In section \ref{sec:Effective Field Theory}, we briefly discuss the pertinent details of EFT related to our work. In section \ref{sec:Interactioneven}, we illustrate the computation of QCC for our theory, first by taking an example of two coupled oscillators. In the following section \ref{sec:5}, we generalize the calculation to the $N$-oscillator case. Since we could not observe any analytical expression for the relevant eigenvalues for $N$-oscillators, in section \ref{sec:continuous}, we resorted to numerical computation of the QCC. We plot the corresponding graphs of QCC with the relevant parameters in our theory. We finish up by summarizing and providing possible future prospects of our work.     


\textcolor{Sepia}{\section{\sffamily Circuit Complexity and its purposes}\label{sec:complexity}}

~~Computationally, Circuit Complexity is defined as a measure of the minimum number of elementary operations required by a computer to solve a certain computational problem \cite{Nielsen1,Nielsen2,Nielsen3,Nielsen4,Watrous2009,Aaronson:2016vto}. In quantum computation, a quantum operation is described by a unitary transformation. So, Quantum Circuit Complexity is the length of the optimized circuit that performs this unitary operation. As the size of the input increases, if the complexity grows polynomially, the problem is called ``easy", but if it grows exponentially, the problem is called ``hard". 

 Quantum information-theoretic concepts, such as entanglement, have proven to be helpful in areas other than quantum computing, such as \cite{Orus:2018dya,Nishioka:2009un,Almheiri:2014lwa,Swingle:2009bg}. Quantum Circuit Complexity (QCC) is emerging to be one such quantum information-theoretic concept that has the potential to explain phenomena in several areas of quantum physics.~However, lower bounding quantum circuit complexity is an extremely challenging open problem.
 
For our purpose, we will consider the geometric approach to compute quantum circuit complexity developed by  Nielsen et al.
\cite{Nielsen1,Nielsen3}. The prime reason to consider a geometric approach is that it is much easier to minimize a smooth function on a smooth space than to minimize an arbitrary function on a discrete space. Since the unitaries are continuous, this method of optimization suits well. Interestingly, this approach allows us to formulate the optimal circuit finding problem in the language of the Hamiltonian control problem, for which a mathematical method called the calculus of variations can be employed to find the minima. Another reason is that this method is similar to the general Lagrangian formalism, where the motion of the test particle is obtained from minimizing a global functional. For example, in general relativity, test particles move along geodesics of spacetime described by the geodesic equation,
\begin{equation*}
    \frac{d^2x^j}{dt^2} + \Gamma ^j_{kl}\frac{dx^k}{dt}\frac{dx^l}{dt} = 0
\end{equation*}
where $x^j$ are the coordinates for the position on the manifold, and $\Gamma ^j_{kl}$ are Christoffel symbols given by the geometry of the space-time. Then, the problem of finding an optimal quantum circuit is related to ``freely falling" along the minimal geodesic curve connecting identity to the desired operation, and the path is given by the ``local shape" of the manifold. If we have information about the local velocity and the geometry, it is possible to predict the rest of the path. In this regard, geometric analysis of quantum computation is quite powerful as it allows to design the rest of the shortest quantum circuit with information about only a part of it.
\vspace{0.5cm}
\subsection{Main Mathematical Ideas}
 Our goal is to understand how difficult it is to implement an arbitrary unitary operation $\mathbb{U}$ generated by a time-dependent Hamiltonian $H(t)$:
\begin{equation}
\mathbb{U}(s)=\overleftarrow{\mathcal{P}}\exp\Big[\displaystyle -i\int_{0}^{s}ds' ~H(s')\Big]
\end{equation}
Where $\overleftarrow{\mathcal{P}}$ is the path ordering operator, and the space of circuits is parameterized by `$s$'. The path ordering operator $\overleftarrow{\mathcal{P}}$ is the same as the time ordering operator, which indicates that the circuit is from right to left. We can expand the Hamiltonian $H(s)$ as,
\begin{equation}
H(s)= \sum_{I}Y^{I}(s)M_{I}
\end{equation}
where $M_I$ represents the generalized Pauli matrices, and the coefficients $Y^I(s)$ are the control functions that tell us the gate to be applied at particular values of `$s$'.

Schrödinger equation $d\mathbb{U}/dt = -iH\mathbb{U}$ describes the evolution of the unitary
\begin{equation}
\label{eq:schequnitary}
\frac{d\mathbb{U}(s)}{ds}=-iY(s)^I M_I \mathbb{U}(s)
\end{equation}
where at the final time $t_f$, $\mathbb{U}(t_f) = \mathbb{U}$.

We can impose a cost function $F(\mathbb{U},\Dot{\mathbb{U}})$ on the Hamiltonian control $H(t)$, which will tell us how difficult it is to apply a specific unitary operation $\mathbb{U}$. One can then define a Riemannian geometry on the space of unitaries with this cost function. Then, the problem of finding an optimal control function is translated to the problem of finding the minimal geodesic on this geometry, and we can define a notion of distance in $SU(2^n)$.
For this, we have to define a curve $\mathbb{U}$ between the identity operation $I$ and the desired unitary $\mathbb{U}$, which is a smooth 
function $\mathbb{U}:[0,t_f]\rightarrow SU(2^n)$ such that $\mathbb{U}(0) = I$ and $\mathbb{U}(t_f) = \mathbb{U}$. The length of this curve is defined as:
\begin{equation}
d([\mathbb{U}]) = \int_0^{t_f} dt F(\mathbb{U},\Dot{\mathbb{U}})
\end{equation}
This length $d([\mathbb{U}])$ gives the total cost of synthesizing the Hamiltonian that describes the motion along the curve. In particular, distance $d(I,\mathbb{U})$ is also a lower bound on the number of one- and two-qubit quantum gates necessary to exactly simulate U. The proof is available in the original papers of Nielsen \cite{nielsen2005geometric}. Therefore, one can also consider the distance $d([\mathbb{U}])$ as an alternative description of complexity. 

The cost function $F$ has to satisfy certain properties, such as continuity, positivity, positive homogeneity, and triangle inequality \cite{Adhikari:2021pvv}. If we also demand $F$ to be smooth, i.e. $ F \in C^\infty$, then the manifold is referred to as the Finsler manifold.
Since the field of differential geometry is relatively mature, we hope that borrowing tools from differential geometry can provide a unique perspective on quantum complexity. 
\par
In literature, there are several alternative definitions of the cost function $F(\mathbb{U},v)$. Some of them are:
\begin{align} \label{cost_function}
\begin{split}
F_1(\mathbb{U},Y) &= \sum_I |Y^I| \\ F_p(\mathbb{U},Y) &= \sum_I p_I|Y^I| \\
F_2(\mathbb{U},Y) &= \sqrt{\sum_I |Y^I|^2 }   \\ F_q(\mathbb{U},Y) &= \sqrt{\sum_I q_I|Y^I|^2 } 
\end{split}
\end{align}
 $F_1$, the linear cost functional measure, is the concept closest to the classical concept of counting gates. $F_2$, the quadratic cost functional, can be understood as the proper distance in the manifold. $F_{p}$ is similar to $F_1$ but with penalty parameters, $p_I$, used to favor certain directions over others.

\begin{figure}
\centering
\includegraphics[width=5.5in]{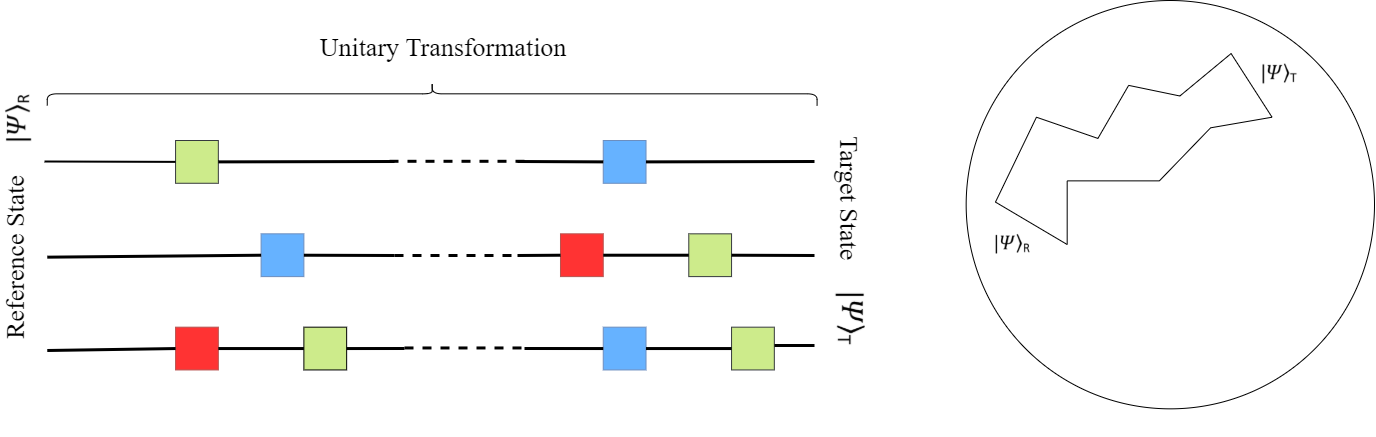}
\caption{The left figure represents a unitary transformation from a reference state to a target state using quantum gates (Square Blocks) and the right figure represents geometrizing the problem of calculating the minimum number of gates representing the transformation}
\end{figure}

\subsection{Geometric algorithm to compute Circuit Complexity}
\label{subsectionStepsAlgo}
We will now describe the algorithm for computing the circuit complexity. These algorithms are not rigorously proven, but from an operational point, these general steps are implemented to calculate the circuit complexity.

\begin{enumerate}
\item Give the Hamiltonian corresponding to a particular physical system.
\item Specify the reference state $\ket{\psi}_R$, the target state $\ket{\psi}_T$ and the unitary $\mathbb{U}$ that takes the former to the latter, $\ket{\psi}_T = \mathbb{U}\ket{\psi}_R$ .
\item Now, we need to choose some set of elementary gates $Q_{ab} = \exp[\epsilon M_{ab}]$, where $M_I$ are the generators of the group corresponding to the choice of gates and $\epsilon$ is a controllable parameter. For simplicity, we often choose generators satisfying $Tr[M_I M^T_J] = \delta_{IJ}$.

\item With the basis of generators $M_I$, parametrize the unitary $\mathbb{U}$ as $\mathbb{U}(s)$
\item Velocity component $Y^{I}(s)$ can be explicitly computed using:
\begin{align}
  Y^{I}(s)M_{I} = i(\partial_{s} \mathbb{U}(s))\mathbb{U}^{-1}(s)) \rightarrow  Y^{I}(s)=\frac{1}{\text{Tr}\left[M^{I}\left(M^{I}\right)^{T}\right]} \text{Tr}\left[\partial_{s} \mathbb{U}(s) \mathbb{U}^{-1}\left(M^{I}\right)^{T}\right]
\end{align}
For generators obeying $Tr[M_I M^T_J] = \delta_{IJ}$, $Y^{I}(s)$ reduces to:
\begin{equation}
  Y^I(s) = \text{Tr}[i(\partial_s \mathbb{U}(s))\mathbb{U}^{-1}(s)M^T_I]
\end{equation}
 The right invariant metric on the space is given by:
 \begin{equation}
 ds^{2}= G_{I J} Y^{I}Y^{J} 
 \end{equation}
where $G_{IJ}$ gives the penalty parameters.
If $G_{IJ} = \delta_{IJ}$, i.e. assigning an equal cost to every choice of gate, and having an extra condition $Tr[M_I M^T_J] = \delta_{IJ}$, we obtain a metric of the reduced simple form:
\begin{equation}
\label{eq:rightInvariantStepMetric}
    ds^2 = \delta_{IJ}\text{Tr}[i(\partial_s \mathbb{U}(s))\mathbb{U}^{-1}(s)M^T_I]\text{Tr}[i(\partial_s \mathbb{U}(s))\mathbb{U}^{-1}(s)M^T_J]
\end{equation}
\item
The general form of circuit complexity would be:
\begin{equation}
    \mathcal{C}[\mathbb{U}] = \int_0^1 ds \sqrt{G_{IJ}Y^I(s)Y^J(s)}
\end{equation}
 The circuit complexity for $F_2$ metric i.e. $G_{IJ} = \delta_{IJ} $ is then:
\begin{equation}
     \mathcal{C}[\mathbb{U}]= \int_0^1 ds \sqrt{g_{ij}\Dot{x}^i\Dot{x}^j}
\end{equation}
\item From the boundary conditions of the evolution of unitaries, we can compute the geodesic path and geodesic length. This length then gives a measure of circuit complexity.
\end{enumerate}

In the literature, circuit complexity using this geometric approach is computed mostly for Gaussian wave functions because of its simpler structure as compared to non-Gaussian wave functions. A Gaussian wave function can be represented as:
\begin{equation}
    \psi  \approx \exp\left[ -\frac{1}{2}v_{a} A(s)_{ab} \ v_{b} \right], \text{  where } v = \{ x_a, x_b\}
\end{equation}

where $x_a$ and $x_b$ are the bases of vector $v$.
If we can simultaneously diagonalize the reference and the target states, then a common pattern observed in the complexity is that it will be given by some function of the ratio of the eigenvalues of $A(s = 0)$ and $A(s = 1)$. Here, $A(s = 0)$ represents the reference state, and $A(s = 1)$ represents the target state.

We would like to mention that our approach to computing complexity is based on  Nielsen's geometric approach which suffers from ambiguity  in choosing the elementary quantum gates and states. However, these choices of our gates  significantly
   simplify the calculation. Furthermore, the previous works on complexity in QFT and interacting QFT \cite{Bhattacharyya:2018bbv, Jefferson:2017sdb}, using similar quantum gates like ours, have connected to Holographic proposal which is the original motivation to study Quantum Circuit complexity in QFT. Recently, Krylov complexity has been proposed as a tool for studying operator growth and  associated quantum chaos \cite{Caputa:2021sib,Parker:2018yvk,Roberts:2018mnp,Rabinovici:2020ryf,Barbon:2019wsy,Jian:2020qpp,Dymarsky:2019elm,Dymarsky:2021bjq,Balasubramanian:2022tpr}. In contrary to Nielsen's geometric approach, Krylov complexity is independent of such arbitrary choices making it a good candidate for complexity in QFT and holography. However, Krylov complexity doesn't have a good operational meaning like Nielsen's geometric measure. Nielsen's measure not only gives the state complexity but also gives us a method of constructing an optimal quantum circuit. This feature makes it more appealing than Krylov complexity. In the future, we would like to study Krylov complexity for our case too. 
   
\textcolor{Sepia}{\section{\sffamily Effective Field Theory in a nutshell}\label{sec:Effective Field Theory}}
    
An effective field theory (EFT) is a theory corresponding to the dynamics of a physical system at energies that are smaller than cutoff energy. EFTs have made a significant impact on several areas of theoretical physics, including condensed matter physics \cite{RShankar1997EffectiveFT}, cosmology \cite{Cheung:2007st,Weinberg:2008hq,Agarwal:2012mq,Burgess:2017ytm,Choudhury:2016wlj,Choudhury:2014sua,Naskar:2017ekm}, particle physics \cite{Pich:1998xt,Burgess:2007pt}, gravity \cite{Donoghue:1995cz,Donoghue:2012zc} and hydrodynamics \cite{PhysRevD.85.085029,Crossley:2015evo}. The idea behind EFT is that we can compute results without knowing the full theory. In the context of quantum field theory, this implies that using the method of EFT, one can study the low energy aspect of the theory without having a full theory in the high energy limit. If the high-energy theory is known, one can obtain EFT using the ``top-down" approach \cite{CMBSAYANTAN}, where one has to eliminate high energy effects. Using the ``bottom-up" approach, one can obtain an EFT if the theory for high energy is not available. Here, one has to impose constraints given by symmetry and ``naturalness" on suitable Lagrangians.

The Hamiltonian of our theory is, 
 \begin{equation}
H=\frac{1}{2} \int d^{d-1} x\Bigg[\pi(x)^{2}+(\nabla \phi(x))^{2}+m^{2} \phi^{2}(x)+2\sum_{n=2}^4C_{2n}\phi^{2n}(x)\Bigg]
\label{Eq_3.1}
\end{equation}
where the coefficients $C_{2n}=2\hat{\lambda}_{2n}/(2n)!$ are called the `Wilson Coefficients' for the $\mathcal{Z}_2$ EEFTs in arbitrary dimensions. These coefficients depend on the scaling of the theory. These coefficients are expected to be functions of the $\lambda$'s, the cut-off of our theory, and this functional dependence can be found by solving the Renormalization Group equations or Callan-Symanzik equations. $\phi^{2n}$'s are called the `Wilson Operators' in $\mathcal{Z}_2$ EEFTs. $\phi^2(x)$ and $\phi^4(x)$ are called `Relevant Operators of EEFTs' and this theory is renormalizable up to $\phi^4(x)$. Beyond that, all the higher-order even terms, in our case, $\phi^6(x)$ and $\phi^{8}(x)$, are called `Non-renormalizable Irrelevant Operators of $\mathcal{Z}_2$ EEFTs'. But it should be noted that even though this theory goes up in the `Wilson Operator' order, the contributions from those terms decrease gradually. So, it is an infinite convergent series. Building upon this, we go on to compute the circuit complexity in $\mathcal{Z}_{2}$ EEFT.

\textcolor{Sepia}{\section{\sffamily Circuit Complexity with $(\hat{\lambda}_{4}\phi^4+\hat{\lambda}_{6}\phi^6+\hat{\lambda}_{8}\phi^8)$ interaction for the case of two harmonic oscillators}\label{sec:Interactioneven}}

We work with massive scalar field theory with the even interaction terms $\phi^4$, $\phi^6$ and $\phi^8$, which are weakly coupled to the free field theory via the coupling constants $\hat{\lambda}_{4}$, $\hat{\lambda}_{6}$ and $\hat{\lambda}_{8}$ respectively. The inequality between the coupling constants are $\frac{\hat{\lambda}_{4}}{4!}>\frac{\hat{\lambda}_{6}}{6!}>\frac{\hat{\lambda}_{8}}{8!}$. The Hamiltonian for this scalar field in $d$ spacetime dimensions is 
\vspace{0.5cm}
\begin{equation}
H=\frac{1}{2} \int d^{d-1} x\Bigg[\pi(x)^{2}+(\nabla \phi(x))^{2}+m^{2} \phi(x)^{2}+2\sum_{n=2}^4C_{2n}\phi^{2n}(x)\Bigg]
\label{Eq_3.1}
\end{equation}
where the mass of the scalar field $\phi$ is $m$. We work in the weak-coupling regime \big($\hat{\lambda} \ll 1$\big) so that perturbative methods can be used to investigate the theory. The system can be reduced to a chain of harmonic oscillators if we regulate the theory by placing it on a \big($d-1$\big) dimensional square lattice with lattice spacing $\delta$. We are taking the infinite system in Eq. (\ref{Eq_3.1}) and discretizing it to a finite $N$-oscillator system because if we have an infinite convergent theory and an infinite number of terms in the Hamiltonian then we don't have the finite symmetries that we are interested in. So the discretized Hamiltonian becomes,

\begin{equation}\footnotesize
H=\frac{1}{2} \sum_{\vec{n}}\Bigg\{\frac{\pi(\vec{n})^{2}}{\delta^{d-1}}+\delta^{d-1}\Bigg[\frac{1}{\delta^{2}} \sum_{i}\left(\phi(\vec{n})-\phi\left(\vec{n}-\hat{x}_{i}\right)\right)^{2}+m^{2} \phi(\vec{n})^{2}+\frac{2\hat{\lambda}_4}{4!} \phi(\vec{n})^{4}+\frac{2\hat{\lambda}_6}{6!} \phi(\vec{n})^{6}+\frac{2\hat{\lambda}_8}{8!} \phi(\vec{n})^{8}\Bigg]\Bigg\}
\label{Eq_3.2}
\end{equation}

where the $\vec{n}$ denotes the spatial position vectors of the points on the lattice in $d$-dimension and $\hat{x}_{i}$ are the unit vectors along the lattice. We make the following substitutions to simplify the form of the Hamiltonian. 

\begin{align*}
X(\vec{n})&=\delta^{d / 2} \phi(\vec{n}) & P(\vec{n})&=\pi(\vec{n}) / \delta^{d / 2} &  M&=\frac{1}{\delta}, \omega=m , \Omega=\frac{1}{\delta}\\
\lambda_4&=\frac{\hat{\lambda}_4}{4!} \delta^{-d}  & \lambda_6&=\frac{\hat{\lambda}_6}{6!} \delta^{-2d} & \lambda_8&=\frac{\hat{\lambda}_8}{8!} \delta^{-3d}
\end{align*}

After the substitutions, we get,
\begin{equation}\footnotesize
H= \sum_{\vec{n}}\Big\{\frac{P(\vec{n})^{2}}{2 M}+\frac{1}{2} M\Big[\omega^{2} X(\vec{n})^{2}+\Omega^{2} \sum_{i}\left(X(\vec{n})-X\left(\vec{n}-\hat{x}_{i}\right)\right)^{2}+2\big\{ \lambda_4 X(\vec{n})^{4}\\
+ \lambda_6 X(\vec{n})^{6}+\lambda_8 X(\vec{n})^{8}\big\}\Big]\Big\}
\label{Eq_3.3}
\end{equation}

We observe that the Hamiltonian obtained is identical to that of an infinite family of coupled anharmonic oscillators. The nearest term interaction is coming from the kinetic part, and the self-interactions are coming from the remaining portion of the Hamiltonian. We start with the simple case of two coupled oscillators and generalize it to the case of $N$-oscillators later in the paper. Setting $M=1$, the Hamiltonian takes the form,

\begin{equation}
H=\textstyle \frac{1}{2}\Big[p_{1}^{2}+p_{2}^{2}+\omega^{2}\left(x_{1}^{2}+x_{2}^{2}\right)+\Omega^{2}\left(x_{1}-x_{2}\right)^{2}+2\left\{ \lambda_4\left(x_{1}^{4}+x_{2}^{4}\right)+ \lambda_6\left(x_{1}^{6}+x_{2}^{6}\right)+ \lambda_8\left(x_{1}^{8}+x_{2}^{8}\right)\right\}\Big] 
\label{Eq_3.4}
\end{equation}

Now, let's consider the normal mode basis:
\begin{align}
&\bar{x}_{0}=\frac{1}{\sqrt{2}}\left(x_{1}+x_{2}\right), \quad \bar{x}_{1}=\frac{1}{\sqrt{2}}\left(x_{1}-x_{2}\right), \\ \nonumber
&\bar{p}_{0}=\frac{1}{\sqrt{2}}\left(p_{1}+p_{2}\right), \quad \bar{p}_{1}=\frac{1}{\sqrt{2}}\left(p_{1}-p_{2}\right) \\ \nonumber
&\tilde{\omega}_{0}^{2}=\omega^{2}, \quad \tilde{\omega}_{1}^{2}=\omega^{2}+2 \Omega^{2}
\label{Eq_3.5}
\end{align}

In the normal mode basis, the unperturbed Hamiltonian becomes decoupled. Then, the eigenfunctions and eigenvalues for the unperturbed Hamiltonian can be easily solved, which is just the product of the ground-state eigenfunctions of the oscillators in the normal basis

\begin{equation}
\psi_{n_{1}, n_{2}}^{0}\left(\bar{x}_{0}, \bar{x}_{1}\right)=\frac{1}{\sqrt{2^{n_{1}+n_{2}} n_{1} ! n_{2} !}} \frac{\left(\tilde{\omega}_{0} \tilde{\omega}_{1}\right)^{1 / 4}}{\sqrt{\pi}} e^{-\frac{1}{2} \tilde{\omega}_{0} \bar{x}_{0}^{2}-\frac{1}{2} \tilde{\omega}_{1} \bar{x}_{1}^{2}} H_{n_{1}}\left(\sqrt{\tilde{\omega}_{0}} \bar{x}_{0}\right) H_{n_{2}}\left(\sqrt{\tilde{\omega}_{1}} \bar{x}_{1}\right)
\label{Eq_3.6}
\end{equation}
Here, $H_{n}(x)$'s denote Hermite polynomials of order $n$. The ground state wavefunction with first order perturbative correction in $\lambda_{4}$, $\lambda_{6}$, $\lambda_{8}$ has the following expression:
\begin{equation}
\psi_{0, 0}\left(\bar{x}_{0}, \bar{x}_{1}\right)=\psi_{0,0}^{0}\left(\bar{x}_{0}, \bar{x}_{1}\right)+\lambda_4 \psi_{0,0}^{1}\left(\bar{x}_{0}, \bar{x}_{1}\right)_4+\lambda_6 \psi_{0,0}^{1}\left(\bar{x}_{0}, \bar{x}_{1}\right)_6+\lambda_8 \psi_{0,0}^{1}\left(\bar{x}_{0}, \bar{x}_{1}\right)_8
\label{Eq_3.7}
\end{equation}

The $\psi_{0,0}^{1}\left(\bar{x}_{0}, \bar{x}_{1}\right)_4$, $ \psi_{0,0}^{1}\left(\bar{x}_{0}, \bar{x}_{1}\right)_6$, $\psi_{0,0}^{1}\left(\bar{x}_{0}, \bar{x}_{1}\right)_8$ are the terms representing the first order perturbative corrections to the ground state wavefunction due to the $\phi^{4},\phi^{6},\phi^{8}$ interactions respectively, which are as follows:

\begin{align*} 
\psi_{0,0}^{1}\left(\bar{x}_{0}, \bar{x}_{1}\right)_4=&-\frac{3(\tilde{\omega}_0+\tilde{\omega}_1)}{4 \sqrt{2} \tilde{\omega}_0 \tilde{\omega}_1^{3}}\psi_{0,2}^0-\frac{\sqrt{3}}{8\sqrt{2} \tilde{\omega}_1^{3}}\psi_{0,4}^0-\frac{3(\tilde{\omega}_0+\tilde{\omega}_1)}{4 \sqrt{2} \tilde{\omega}_0^{3} \tilde{\omega}_1}\psi_{2,0}^0-\frac{3}{4 \tilde{\omega}_0( \tilde{\omega}_0+ \tilde{\omega}_1) \tilde{\omega}_1}\psi_{2,2}^0\\
&-\frac{\sqrt{3}}{8\sqrt{2} \tilde{\omega}_0^{3}}\psi_{4,0}^0
 \end{align*}
 
 \begin{align*} 
\psi_{0,0}^{1}\left(\bar{x}_{0}, \bar{x}_{1}\right)_6=&-\frac{45(\tilde{\omega}_0+\tilde{\omega}_1)^{2}}{32 \sqrt{2} \tilde{\omega}_0^{2} \tilde{\omega}_1^{4}}\psi_{0,2}^0-\frac{15 \sqrt{3}(\tilde{\omega}_0+\tilde{\omega}_1)}{32\sqrt{2}\tilde{\omega}_0 \tilde{\omega}_1^{4}}\psi_{0,4}^2-\frac{\sqrt{5}}{16 \tilde{\omega}_1^{4}}\psi_{0,6}^0-\frac{45(\tilde{\omega}_0+\tilde{\omega}_1)^{2}}{32 \sqrt{2} \tilde{\omega}_0^{4} \tilde{\omega}_1^{2}}\psi_{2,0}\\ \nonumber
 &-\frac{45(\tilde{\omega}_0+\tilde{\omega}_1)}{16 \tilde{\omega}_{0}^{2}( \tilde{\omega}_0+ \tilde{\omega}_1) \tilde{\omega}_1^{2}}\psi_{2,2}^0-\frac{15 \sqrt{3}}{16 \tilde{\omega}_0( \tilde{\omega}_{0}+2 \tilde{\omega}_1) \tilde{\omega}_1^{2}}\psi_{2,4}^0-\frac{15 \sqrt{3/2}(\tilde{\omega}_0+\tilde{\omega}_1)}{32 \tilde{\omega}_0^{4} \tilde{\omega}_1}\psi_{4,0}^0\\ 
 &-\Big.
 \Big.\frac{15 \sqrt{3}}{16 \tilde{\omega}_0^{2}(2 \tilde{\omega}_0+ \tilde{\omega}_1) \tilde{\omega}_1}\psi_{4,2}^0-\frac{\sqrt{5}}{16 \tilde{\omega}_0^{4}}\psi_{6,0}^0
 \end{align*}
 
 \begin{align*}
 \psi_{0,0}^{1}\left(\bar{x}_{0}, \bar{x}_{1}\right)_8=&\Big(\frac{105 \sqrt{2}}{8 \tilde{\omega}_{0}^{5}}+\frac{315 \sqrt{2}}{8 \tilde{\omega}_{0}^{4} \tilde{\omega}_{1}}+\frac{315 \sqrt{2}}{8 \tilde{\omega}_{0}^{3} \tilde{\omega}_{1}^{2}}+\frac{105 \sqrt{2}}{8 \tilde{\omega}_{0}^{2} \tilde{\omega}_{1}^{3}}\Big) \psi_{2,0}^0+\Big(\frac{105 \sqrt{2}}{8 \tilde{\omega}_{1}^{5}}+\frac{105 \sqrt{2}}{8 \tilde{\omega}_{0}^{3} \tilde{\omega}_{1}^{2}}+\frac{315 \sqrt{2}}{8 \tilde{\omega}_{0}^{3} \tilde{\omega}_{1}^{2}}\Big.\Big. \Big.\\ \nonumber
&+\frac{315 \sqrt{2}}{8 \tilde{\omega}_{1}^{4} \tilde{\omega}_{0}}\Big) \psi_{0,2}^0+\Big(\frac{315}{4 \tilde{\omega}_{0}^{3} \tilde{\omega}_{1}(\tilde{\omega}_{0}+\tilde{\omega}_{1})}+\frac{315}{2 \tilde{\omega}_{0}^{2} \tilde{\omega}_{1}^{2}(\tilde{\omega}_{0}+\tilde{\omega}_{1})}\Big.+\frac{315}{4 \tilde{\omega}_{1}^{3} \tilde{\omega}_{0}(\tilde{\omega}_{0}+\tilde{\omega}_{1})}\Big)\\ \nonumber 
&*\psi_{2,2}^0+\Big(\frac{105 \sqrt{6}}{16 \tilde{\omega}_{0}^{5}}+\frac{105 \sqrt{6}}{8 \tilde{\omega}_{0}^{4} \tilde{\omega}_{1}}+\frac{105 \sqrt{6}}{16 \tilde{\omega}_{0}^{3} \tilde{\omega}_{1}^{2}}\Big)\psi_{4,0}^0 +\Big(\frac{105 \sqrt{6}}{16 \tilde{\omega}_{1}^{5}}+\frac{105 \sqrt{6}}{8 \tilde{\omega}_{1}^{4} \tilde{\omega}_{0}}+\frac{105 \sqrt{6}}{16 \tilde{\omega}_{0}^{2} \tilde{\omega}_{1}^{3}}\Big)\\ \nonumber
& *\psi_{0,4}^0+\Big(\frac{105 \sqrt{3}}{2 \tilde{\omega}_{0}^{3} \tilde{\omega}_{1}(2 \tilde{\omega}_{0}+\tilde{\omega}_{1})}\Big. \Big.+\frac{105 \sqrt{3}}{2 \tilde{\omega}_{0}^{2} \tilde{\omega}_{1}^{2}(2 \tilde{\omega}_{0}+\tilde{\omega}_{1})}\Big) \psi_{4,2}^0+\Big(\frac{105 \sqrt{3}}{2 \tilde{\omega}_{1}^{3} \tilde{\omega}_{0}(2 \tilde{\omega}_{1}+\tilde{\omega}_{0})}\Big. \\ \nonumber
 &\Big.+\frac{105 \sqrt{3}}{2 \tilde{\omega}_{0}^{2} \tilde{\omega}_{1}^{2}(\tilde{\omega}_{0}+2 \tilde{\omega}_{1})}\Big) \psi_{2,4}^0+\frac{105}{4 \tilde{\omega}_{0}^{2} \tilde{\omega}_{1}^{2}(\tilde{\omega}_{0}+\tilde{\omega}_{1})}\psi_{4,4}^0+\Big(\frac{7 \sqrt{5}}{2 \tilde{\omega}_{0}^{5}}\Big. +\frac{7 \sqrt{5}}{2 \tilde{\omega}_{0}^{4} \tilde{\omega}_{1}}\Big) \psi_{6,0}^0+\\ \nonumber
 &\Big(\frac{7 \sqrt{5}}{2 \tilde{\omega}_{1}^{5}}+\frac{7 \sqrt{5}}{2 \tilde{\omega}_{1}^{4} \tilde{\omega}_{0}}\Big) \psi_{0,6}^0+\frac{21 \sqrt{10}}{{2 \tilde{\omega}_{1}^{3} \tilde{\omega}_{0}(3 \tilde{\omega}_{1}+\tilde{\omega}_{0})}}\psi_{2,6}^0 +\frac{21 \sqrt{10}}{{2 \tilde{\omega}_{1}^{3} \tilde{\omega}_{0}(3 \tilde{\omega}_{1}+\tilde{\omega}_{0})}}\psi_{2,6}^0\\ 
 &+\frac{3\sqrt{70}}{\tilde{\omega}_0^5}\psi_{8,0}^0+\frac{3\sqrt{70}}{\tilde{\omega}_1^5}\psi_{0,8}^0
\end{align*}

We can approximate the total ground state wave function in Eq. (\ref{Eq_3.7}) in an exponential form as the values of $\lambda_{4}$, $\lambda_{6}$, $\lambda_{8}<<1$.

\begin{multline}
\psi_{0,0}\left(\bar{x}_{0}, \bar{x}_{1}\right)  \approx  \frac{\left(\tilde{\omega}_{0} \tilde{\omega}_{1}\right)^{1 / 4}}{\sqrt{\pi}} \exp \left[\alpha_{0}\right] \exp \Big[-\frac{1}{2}\Big(\alpha_{1}  \bar{x}_{0}^{2}+\alpha_{2}  \bar{x}_{1}^{2}+\alpha_{3}  \bar{x}_{0}^{2}  \bar{x}_{1}^{2}+\alpha_{4}  \bar{x}_{0}^{4}+\alpha_{5}  \bar{x}_{1}^{4} +\alpha_{6}  \bar{x}_{0}^{4}  \bar{x}_{1}^{2}+\alpha_{7}  \bar{x}_{0}^{2}  \bar{x}_{1}^{4}\\+\alpha_{8}  \bar{x}_{0}^{6}+ \alpha_{9}  \bar{x}_{1}^{6}+\alpha_{10}  \bar{x}_{0}^{2}  \bar{x}_{1}^{6}+\alpha_{11}  \bar{x}_{0}^{6}  \bar{x}_{1}^{2}+\alpha_{12}  \bar{x}_{0}^{4}  \bar{x}_{1}^{4}+
\alpha_{13}  \bar{x}_{0}^{8}+\alpha_{14}  \bar{x}_{1}^{8}\Big)\Big]
\label{Eq_3.9}
\\
\end{multline}

 We shall take $ \psi_{0,0}\left(\bar{x}_{0}, \bar{x}_{1}\right)$ as the general target state wavefunction for calculating complexity in the following sections. The Coefficients $\alpha_{0},\alpha_{1},\alpha_{2}\dots\alpha_{14}$ involved in the approximate wavefunction Eq. (\ref{Eq_3.9}) are given in the table below
 \begin{longtable}{|c|c|}

		\hline
		\rowcolor{Gray}
		  $\alpha_{i}$ & Coefficient of $\alpha_{i}$\\
\hline
$\alpha_{0}$ & 

{\Large \scalebox{.70}{
\parbox[t]{20cm}
{ $\\-2\bigg[\frac{9 \lambda_{4}}{32 \tilde{\omega}_{0}^{3}}+\frac{9 \lambda_{4}}{32 \tilde{\omega}_{1}^{3}}+\frac{3 \lambda_{4}}{8 \tilde{\omega}_{0} \tilde{\omega}_{1}{ }^{2}}+\frac{3 \lambda_{4}}{8 \tilde{\omega}_{0}{ }^{2} \tilde{\omega}_{1}}+\frac{3 \lambda_{4}}{4 \tilde{\omega}_{0}\left(-2 \tilde{\omega}_{0}-2 \tilde{\omega}_{1}\right) \tilde{\omega}_{1}}+\frac{55 \lambda_{6}}{128 \tilde{\omega}_{0}{ }^{4}}+\frac{55 \lambda_{6}}{128{\tilde{\omega}}_{1}{ }^{4}}+\frac{135 \lambda_{6}}{128 \tilde{\omega}_{0} \tilde{\omega}_{1}{ }^{3}}+\frac{45 \lambda_{6}}{32 \tilde{\omega}_{0}{ }^{2} \tilde{\omega}_{1}{ }^{2}}\\ - \frac{45 \lambda_{6}}{32 \tilde{\omega}_{0}\left(-2 \tilde{\omega}_{0}-4 \tilde{\omega}_{1}\right) \tilde{\omega}_{1}{ }^{2}}+\frac{45 \lambda_{6}}{16 \tilde{\omega}_{0}\left(-2 \tilde{\omega}_{0}-2 \tilde{\omega}_{1}\right) \tilde{\omega}_{1}{ }^{2}}+\frac{135 \lambda_{6}}{128 \tilde{\omega}_{0}{ }^{3} \tilde{\omega}_{1}}-\frac{45 \lambda_{6}}{32 \tilde{\omega}_{0}{ }^{2}\left(-4 \tilde{\omega}_{0}-2 \tilde{\omega}_{1}\right) \tilde{\omega}_{1}}+\frac{45 \lambda_{6}}{16 \tilde{\omega}_{0}{ }^{2}\left(-2 \tilde{\omega}_{0}-2 \tilde{\omega}_{1}\right) \tilde{\omega}_{1}}\\+ \frac{875 \lambda_{8}}{1024 \tilde{\omega}_{0}{ }^{5}}+\frac{875 \lambda_{8}}{1024 \tilde{\omega}_{1}{ }^{5}}+\frac{385 \lambda_{8}}{128 \tilde{\omega}_{0} \tilde{\omega}_{1}{ }^{4}}+\frac{105 \lambda_{8}}{256 \tilde{\omega}_{0}{ }^{2} \tilde{\omega}_{1}{ }^{3}}+\frac{2625 \lambda_{8}}{256 \tilde{\omega}_{0}{ }^{3} \tilde{\omega}_{1}{ }^{2}}+\frac{385 \lambda_{8}}{128 \tilde{\omega}_{0}{ }^{4} \tilde{\omega}_{1}}-\frac{315 \lambda_{8}}{64 \tilde{\omega}_{0} \tilde{\omega}_{1}{ }^{3}\left(\tilde{\omega}_{0}+\tilde{\omega}_{1}\right)}\\-\frac{2835 \lambda_{8}}{256 \tilde{\omega}_{0}{ }^{2} \tilde{\omega}_{1}{ }^{2}\left(\tilde{\omega}_{0}+\tilde{\omega}_{1}\right)}-\frac{315 \lambda_{8}}{64 \tilde{\omega}_{0}{ }^{3} \tilde{\omega}_{1}\left(\tilde{\omega}_{0}+\tilde{\omega}_{1}\right)}+\frac{315 \lambda_{8}}{64 \tilde{\omega}_{0}^{2} \tilde{\omega}_{1}^{2}\left(2 \tilde{\omega}_{0}+\tilde{\omega}_{1}\right)}+\frac{315 \lambda_{8}}{64 \tilde{\omega}_{0}^{3} \tilde{\omega}_{1}\left(2 \tilde{\omega}_{0}+\tilde{\omega}_{1}\right)}-\frac{105 \lambda_{8}}{64 \tilde{\omega}_{0}{ }^{3} \tilde{\omega}_{1}\left(3 \tilde{\omega}_{0}+\tilde{\omega}_{1}\right)}\\+\frac{315 \lambda_{8}}{64 \tilde{\omega}_{0} \tilde{\omega}_{1}{ }^{3}\left(\tilde{\omega}_{0}+2 \tilde{\omega}_{1}\right)}+ \frac{315 \lambda_{8}}{64 \tilde{\omega}_{0}{ }^{2} \tilde{\omega}_{1}{ }^{2}\left(\tilde{\omega}_{0}+2 \tilde{\omega}_{1}\right)}-\frac{105 \lambda_{8}}{64 \tilde{\omega}_{0} \tilde{\omega}_{1}{ }^{3}\left(\tilde{\omega}_{0}+3 \tilde{\omega}_{1}\right)}\bigg]\\$}}
	  }\\ \hline
$\alpha_{1}$ & 

{\Large \scalebox{0.70}{
\parbox[t]{20cm}
 {$\\\omega_0 -2\bigg[\frac{-3 \lambda_{4}}{8 \tilde{\omega}_{0}{ }^{2}}-\frac{3 \lambda_{4}}{4 \tilde{\omega}_{0} \tilde{\omega}_{1}}-\frac{3 \lambda_{4}}{2 \left(-2 \tilde{\omega}_{0}-2 \tilde{\omega}_{1}\right) \tilde{\omega}_{1}}-\frac{15 \lambda_{6}}{32 \tilde{\omega}_{0}{ }^{3}}-\frac{45 \lambda_{6}}{32 \tilde{\omega}_{0} \tilde{\omega}_{1}{ }^{2}}+\frac{45 \lambda_{6}}{16 \left(-2 \tilde{\omega}_{0}-4 \tilde{\omega}_{1}\right) \tilde{\omega}_{1}{ }^{2}}-\frac{45 \lambda_{6}}{8 \left(-2 \tilde{\omega}_{0}-2 \tilde{\omega}_{1}\right) \tilde{\omega}_{1}{ }^{2}}\\-\frac{45 \lambda_{6}}{32 \tilde{\omega}_{0}{ }^{2} \tilde{\omega}_{1}}+\frac{45 \lambda_{6}}{8 \tilde{\omega}_{0}\left(-4 \tilde{\omega}_{0}-2 \tilde{\omega}_{1}\right) \tilde{\omega}_{1}}-\frac{45 \lambda_{6}}{8 \tilde{\omega}_{0}\left(-2 \tilde{\omega}_{0}-2 \tilde{\omega}_{1}\right) \tilde{\omega}_{1}}-\frac{105 \lambda_{8}}{128 \tilde{\omega}_{0}{ }^{4}}-\frac{105 \lambda_{8}}{32 \tilde{\omega}_{0} \tilde{\omega}_{1}{ }^{3}}-\frac{315 \lambda_{8}}{64 \tilde{\omega}_{0}{ }^{2} \tilde{\omega}_{1}{ }^{2}}-\frac{105 \lambda_{8}}{32 \tilde{\omega}_{0}{ }^{3} \tilde{\omega}_{1}}\\+ \frac{315 \lambda_{8}}{32 \tilde{\omega}_{1}{ }^{3}\left(\tilde{\omega}_{0}+\tilde{\omega}_{1}\right)}+\frac{1575 \lambda_{8}}{64 \tilde{\omega}_{0} \tilde{\omega}_{1}{ }^{2}\left(\tilde{\omega}_{0}+\tilde{\omega}_{1}\right)}+\frac{315 \lambda_{8}}{32 \tilde{\omega}_{0}{ }^{2} \tilde{\omega}_{1}\left(\tilde{\omega}_{0}+\tilde{\omega}_{1}\right)}-\frac{315 \lambda_{8}}{16 \tilde{\omega}_{0} \tilde{\omega}_{1}{ }^{2}\left(2 \tilde{\omega}_{0}+\tilde{\omega}_{1}\right)}-\frac{315 \lambda_{8}}{16 \tilde{\omega}_{0}{ }^{2} \tilde{\omega}_{1}\left(2 \tilde{\omega}_{0}+\tilde{\omega}_{1}\right)}\\+\frac{315 \lambda_{8}}{32 \tilde{\omega}_{0}^{2} \tilde{\omega}_{1}\left(3 \tilde{\omega}_{0}+\tilde{\omega}_{1}\right)}-\frac{315 \lambda_{8}}{32 \tilde{\omega}_{1}{ }^{3}\left(\tilde{\omega}_{0}+2 \tilde{\omega}_{1}\right)}-\frac{315 \lambda_{8}}{32 \tilde{\omega}_{0} \tilde{\omega}_{1}{ }^{2}\left(\tilde{\omega}_{0}+2 \tilde{\omega}_{1}\right)}+\frac{105 \lambda_{8}}{32 \tilde{\omega}_{1}{ }^{3}\left(\tilde{\omega}_{0}+3 \tilde{\omega}_{1}\right)}\bigg]
 \\$}}
	  }\\ \hline
$\alpha_{2}$ & 
\scalebox{0.70}{
\parbox[t]{20cm}
 {\Large $\\\omega_1 -2\bigg[
\frac{-3 \lambda_{4}}{8 \omega_1^{2}}-\frac{3 \lambda_{4}}{4 \omega_0 \omega_1}-\frac{3 \lambda_{4}}{2 \omega_0(-2 \omega_0-2 \omega_1)}-\frac{15 \lambda_{6}}{32 \omega_1^{3}}-\frac{45 \lambda_{6}}{32 \omega_0^{2} \omega_1}+\frac{45 \lambda_{6}}{16 \omega_0^{2}(-4 \omega_0-2 \omega_1)}
-\frac{45 \lambda_{6}}{8 \omega_0^{2}(-2 \omega_0-2 \omega_1)}\\-\frac{45 \lambda_{6}}{32 \omega_0 \omega_1^{2}}+\frac{45 \lambda_{6}}{8 \omega_0(-2 \omega_0-4 \omega_1) \omega_1} 
-\frac{45 \lambda_{6}}{8 \omega 0(-2 \omega_0-2 \omega_1) \omega_1}-\frac{105 \lambda_{8}}{128 \omega_1^{4}}-\frac{105 \lambda_{8}}{8 \omega_0^{3} \omega_1}+\frac{315 \lambda_{8}}{64 \omega_0^{2} \omega_1^{2}}-\frac{105 \lambda_{8}}{32 \omega_0 \omega_1^{3}}\\ 
+\frac{315 \lambda_{8}}{32 \omega_0^{3}(\omega_0+\omega 1)}+\frac{1575 \lambda_{8}}{64 \omega_0^{2} \omega 1(\omega_0+\omega_1)}+\frac{315 \lambda_{8}}{32 \omega_0 \omega 1^{2}(\omega_0+\omega 1)} 
-\frac{315 \lambda_{8}}{32 \omega_0^{3}(2 \omega_0+\omega_1)}-\frac{315 \lambda_{8}}{32 \omega_0^{2} \omega_1(2 \omega_0+\omega_1)}+\frac{105 \lambda_{8}}{32 \omega_0^{3}(3 \omega_0+\omega_1)} \\
-\frac{315 \lambda_{8}}{16 \omega_0 \omega_1^{2}(\omega_0+2 \omega_1)}-\frac{315 \lambda_{8}}{16 \omega_0^{2} \omega_1(\omega_0+2 \omega_1)}+\frac{315 \lambda_{8}}{32 \omega 0 \omega_1^{2}(\omega_0+3 \omega_1)}
\bigg]
 \\$}
	  }\\ \hline
$\alpha_{3}$ & 
\scalebox{0.70}{
\parbox[t]{20cm}
{\Large $\\-2\bigg[
\frac{3 \lambda_{4}}{-2 \tilde{\omega}_{0}-2 \tilde{\omega}_{1}}-\frac{45 \lambda_{6}}{4 \tilde{\omega}_{0}\left(-4 \tilde{\omega}_{0}-2 \tilde{\omega}_{1}\right)}+\frac{45 \lambda_{6}}{4 \tilde{\omega}_{0}\left(-2 \tilde{\omega}_{0}-2 \tilde{\omega}_{1}\right)}-\frac{45 \lambda_{6}}{4 \left(-2 \tilde{\omega}_{0}-4 \tilde{\omega}_{1}\right) \tilde{\omega}_{1}}+\frac{45 \lambda_{6}}{4 \left(-2 \tilde{\omega}_{0}-2 \tilde{\omega}_{1}\right) \tilde{\omega}_{1}}\\-\frac{315 \lambda_{8}}{16 \tilde{\omega}_{0}^{2}\left(\tilde{\omega}_{0}+\tilde{\omega}_{1}\right)}-\frac{315 \lambda_{8}}{16 \tilde{\omega}_{1}{ }^{2}\left(\tilde{\omega}_{0}+\tilde{\omega}_{1}\right)}-\frac{945 \lambda_{8}}{16 \tilde{\omega}_{0} \tilde{\omega}_{1}\left(\tilde{\omega}_{0}+\tilde{\omega}_{1}\right)}+\frac{315 \lambda_{8}}{8 \tilde{\omega}_{0}^{2}\left(2 \tilde{\omega}_{0}+\tilde{\omega}_{1}\right)}+\frac{315 \lambda_{8}}{8 \tilde{\omega}_{0} \tilde{\omega}_{1}\left(2 \tilde{\omega}_{0}+\tilde{\omega}_{1}\right)}-\frac{315 \lambda_{8}}{16 \tilde{\omega}_{0}^{2}\left(3 \tilde{\omega}_{0}+\tilde{\omega}_{1}\right)}\\+\frac{315 \lambda_{8}}{8 \tilde{\omega}_{1}^{2}\left(\tilde{\omega}_{0}+2 \tilde{\omega}_{1}\right)}+\frac{315 \lambda_{8}}{8 \tilde{\omega}_{0} \tilde{\omega}_{1}\left(\tilde{\omega}_{0}+2 \tilde{\omega}_{1}\right)}-\frac{315 \lambda_{8}}{16 \tilde{\omega}_{1}^{2}\left(\tilde{\omega}_{0}+3 \tilde{\omega}_{1}\right)}\bigg]\\$}
	  }\\ \hline 
$\alpha_{4}$ & 
\scalebox{0.70}{
\parbox[t]{20cm}
 {\Large  $\\-2\bigg[\frac{-\lambda_{4}}{8 \tilde{\omega}_{0}}-\frac{5 \lambda_{6}}{32 \tilde{\omega}_{0}^{2}}-\frac{15 \lambda_{6}}{32 \tilde{\omega}_{0} \tilde{\omega}_{1}}-\frac{15 \lambda_{6}}{8 \left(-4 \tilde{\omega}_{0}-2 \tilde{\omega}_{1}\right) \tilde{\omega}_{1}}-\frac{35 \lambda_{8}}{128 \tilde{\omega}_{0}{ }^{3}}-\frac{105 \lambda_{8}}{64 \tilde{\omega}_{0} \tilde{\omega}_{1}^{2}}-\frac{35 \lambda_{8}}{32 \tilde{\omega}_{0}^{2} \tilde{\omega}_{1}}-\frac{105 \lambda_{8}}{64 \tilde{\omega}_{1}^{2}\left(\tilde{\omega}_{0}+\tilde{\omega}_{1}\right)}\\+\frac{105 \lambda_{8}}{16 \tilde{\omega}_{1}{ }^{2}\left(2 \tilde{\omega}_{0}+\tilde{\omega}_{1}\right)}+\frac{105 \lambda_{8}}{16 \tilde{\omega}_{0} \tilde{\omega}_{1}\left(2 \tilde{\omega}_{0}+\tilde{\omega}_{1}\right)}-\frac{105 \lambda_{8}}{16 \tilde{\omega}_{0} \tilde{\omega}_{1}\left(3 \tilde{\omega}_{0}+\tilde{\omega}_{1}\right)}\bigg] \\$}
	  }\\ \hline 
$\alpha_{5}$ & 
\scalebox{0.70}{
\parbox[t]{20cm}
 {\Large $\\-2\bigg[-\frac{\lambda_{4}}{8 \tilde{\omega}_{1}}-\frac{15 \lambda_{6}}{8 \tilde{\omega}_{0}\left(-2 \tilde{\omega}_{0}-4 \tilde{\omega}_{1}\right)}-\frac{5 \lambda_{6}}{32 \tilde{\omega}_{1}{ }^{2}}-\frac{15 \lambda_{6}}{32 \tilde{\omega}_{0} \tilde{\omega}_{1}}-\frac{35 \lambda_{8}}{128 \tilde{\omega}_{1}{ }^{3}}-\frac{35 \lambda_{8}}{32 \tilde{\omega}_{0} \tilde{\omega}_{1}^{2}}-\frac{105 \lambda_{8}}{64 \tilde{\omega}_{0}{ }^{2} \tilde{\omega}_{1}}-\frac{105 \lambda_{8}}{64 \tilde{\omega}_{0}^{2}\left(\tilde{\omega}_{0}+\tilde{\omega}_{1}\right)}+ \frac{105 \lambda_{8}}{16 \tilde{\omega}_{0}{ }^{2}\left(\tilde{\omega}_{0}+2 \tilde{\omega}_{1}\right)}+\frac{105 \lambda_{8}}{16 \tilde{\omega}_{0} \tilde{\omega}_{1}\left(\tilde{\omega}_{0}+2 \tilde{\omega}_{1}\right)}-\frac{105 \lambda_{8}}{16 \tilde{\omega}_{0} \tilde{\omega}_{1}\left(\tilde{\omega}_{0}+3 \tilde{\omega}_{1}\right)}\bigg] \\
$}
	  }\\ \hline 
$\alpha_{6}$ & 
\scalebox{0.70}{
\parbox[t]{20cm}
 {\Large $\\-2\bigg[\frac{15 \lambda_{6}}{4 \left(-4 \tilde{\omega}_{0}-2 \tilde{\omega}_{1}\right)}+\frac{105 \lambda_{8}}{16 \tilde{\omega}_{1}\left(\tilde{\omega}_{0}+\tilde{\omega}_{1}\right)}-\frac{105 \lambda_{8}}{8 \tilde{\omega}_{0}\left(2 \tilde{\omega}_{0}+\tilde{\omega}_{1}\right)}-\frac{105 \lambda_{8}}{8 \tilde{\omega}_{1}\left(2 \tilde{\omega}_{0}+\tilde{\omega}_{1}\right)}+\frac{105 \lambda_{8}}{8 \tilde{\omega}_{0}\left(3 \tilde{\omega}_{0}+\tilde{\omega}_{1}\right)}\bigg]\\$}
	  }\\ \hline 	  
$\alpha_{7}$ & 
\scalebox{0.70}{
\parbox[t]{20cm}
 {\Large $\\-2\bigg[\frac{15 \lambda_{6}}{4 \left(-2 \tilde{\omega}_{0}-4 \tilde{\omega}_{1}\right)}+\frac{105 \lambda_{8}}{16 \tilde{\omega}_{0}\left(\tilde{\omega}_{0}+\tilde{\omega}_{1}\right)}-\frac{105 \lambda_{8}}{8 \tilde{\omega}_{0}\left(\tilde{\omega}_{0}+2 \tilde{\omega}_{1}\right)}-\frac{105 \lambda_{8}}{8 \tilde{\omega}_{1}\left(\tilde{\omega}_{0}+2 \tilde{\omega}_{1}\right)}+\frac{105 \lambda_{8}}{8 \tilde{\omega}_{1}\left(\tilde{\omega}_{0}+3 \tilde{\omega}_{1}\right)}\bigg]\\$}
	  }\\ \hline 	  
$\alpha_{8}$ & 
\scalebox{0.70}{
\parbox[t]{20cm}
 {\Large $\\-2\bigg[\frac{\lambda_{6}}{24 \tilde{\omega}_{0}}-\frac{7 \lambda_{8}}{96 \tilde{\omega}_{0}^{2}}-\frac{7 \lambda_{8}}{24 \tilde{\omega}_{0} \tilde{\omega}_{1}}+\frac{7 \lambda_{8}}{8 \tilde{\omega}_{1}\left(3 \tilde{\omega}_{0}+\tilde{\omega}_{1}\right)}\bigg]\\$}
	  }\\ \hline 	  
$\alpha_{9}$ & 
\scalebox{0.70}{
\parbox[t]{20cm}
 {\Large $ \\-2\bigg[\frac{-\lambda_{6}}{24 \tilde{\omega}_{1}}-\frac{7 \lambda_{8}}{96 \tilde{\omega}_{1}^{2}}-\frac{7 \lambda_{8}}{24 \tilde{\omega}_{0} \tilde{\omega}_{1}}+\frac{7 \lambda_{8}}{8 \tilde{\omega}_{0}\left(\tilde{\omega}_{0}+3 \tilde{\omega}_{1}\right)}\bigg]\\$}
	  }\\ \hline 	
$\alpha_{10}$ & 
\scalebox{0.70}{
\parbox[t]{20cm}
 {\Large $\\ \frac{7 \lambda_{8}}{2 \left(\tilde{\omega}_{0}+3\tilde{\omega}_{1}\right)}\\$}
	  }\\ \hline 
$\alpha_{11}$ & 
\scalebox{0.70}{
\parbox[t]{20cm}
 {\Large $\\ \frac{7 \lambda_{8}}{2 \left(3 \tilde{\omega}_{0}+\tilde{\omega}_{1}\right)} \\ $}
	  }\\ \hline 
$\alpha_{12}$ & 
\scalebox{0.70}{
\parbox[t]{20cm}
{\Large $\\ \frac{35\lambda_{8}}{8 \left( \tilde{\omega}_{0}+\tilde{\omega}_{1}\right)} \\ $}
	  }\\ \hline 
$\alpha_{13}$ & 
\scalebox{0.70}{
\parbox[t]{20cm}
 {\Large $\\ \frac{\lambda_{8}}{32 \tilde{\omega}_{0}} \\$}
	  }\\ \hline 
$\alpha_{14}$ & 
\scalebox{0.70}{
\parbox[t]{20cm}
 {\Large $ \\ \frac{\lambda_{8}}{32 \tilde{\omega}_{1}}\\$}
	  }\\ \hline\hline \hline
	  \end{longtable}

\textcolor{Sepia}{\subsection{\sffamily  Circuit Complexity}\label{sec:3.1}}
We will describe complexity in terms of a quantum circuit model. So 
 to calculate the circuit complexity for the two-oscillator system with even interactions up to $\phi^8$, we need to fix our reference state, target state, and a set of elementary gates. We will construct the unitary transformation using these gates. This unitary transformation will take the system from the reference state ($\ket{\psi}_R$) to the target state($\ket{\psi}_T$), i.e. $\ket{\psi}_T = U\ket{\psi}_R$. The minimum number of gates needed to construct such a unitary transformation is the complexity of the target state. Since our wave functions are nearly Gaussian, we can consider our space of states as the space of positive quadratic forms. This space can be parameterized as a function of a smooth parameter  \lq $s$\rq~ as follows

\begin{equation}
\psi^{s}(\bar{x}_{0}, \bar{x}_{1})=\mathcal{N}^{s} \exp \Big[-\frac{1}{2}\Big(v_{a}  A(s)_{a b} \ v_{b}\Big)\Big]
\label{Eq_3.10}
\end{equation}

    Here, $\mathcal{N}^{s}$ is the normalization constant, and the parameter \lq s\rq~ runs from 0 to 1. If $s=1$, the circuit represents the target state Eq. (\ref{Eq_3.9}) with $\mathcal{N}^{s=1}=\frac{(\bar{\omega}_{0} \bar{\omega}_{1})^{1 / 4}}{\sqrt{\pi}} \exp [\alpha_{0}]$, and at $s=0$ the circuit is in the reference state. The continuous unitary transformation, specified by the \lq s\rq~ parameter, gives us the target state from the reference state. Writing the states in the form of Eq. (\ref{Eq_3.10}) helps us formulate the matrix version of our problem. Now we want to represent the exponent of the wavefunction, which is a polynomial in the matrix form $A(s)$.

\begin{equation}
\psi^{s=0}(x_{1}, x_{2})=\mathcal{N}^{s=0} \exp
\Big[-\frac{\omega_{r e f}}{2}(x_{1}^{2}+x_{2}^{2}+\lambda_{0}^4(x_{1}^{4}+x_{2}^{4})+\lambda_{0}^6(x_{1}^{6}+x_{2}^{6})+\lambda_{0}^8(x_{1}^{8}+x_{2}^{8}))\Big]
\label{Eq_3.11}
\end{equation}

Here $\lambda_0^{4}$, $\lambda_0^{6}$, $\lambda_0^{8}$ are the initial coupling constants for  $\phi^4$, $\phi^6$ and $\phi^8$ respectively. Transforming to the normal coordinates, we get:

\begin{align}\nonumber
\psi^{s=0}(\bar{x}_{0}, \bar{x}_{1})= &\mathcal{N}^{s=0}\exp \Big[-\frac{\tilde{\omega}_{r e f}}{2}(\bar{x}_{0}^{2}+\bar{x}_{1}^{2}+\frac{\lambda_{4}}{2}(\bar{x}_{0}^{4}+\bar{x}_{1}^{4}+6 \bar{x}_{0}^{2} \bar{x}_{1}^{2})+\frac{\lambda_{6}}{4}(\bar{x}_{0}^{6}+\bar{x}_{1}^{6}
+15 \bar{x}_{0}^{4} \bar{x}_{1}^{2}\\ 
&+15 \bar{x}_{1}^{4} \bar{x}_{0}^{2})
+\frac{\lambda_{8}}{8}(\bar{x}_{0}^{8}+\bar{x}_{1}^{8}+28 \bar{x}_{0}^{6} \bar{x}_{1}^{2}+28 \bar{x}_{0}^{2} \bar{x}_{1}^{6}+28 \bar{x}_{0}{ }^{4} \bar{x}_{1}{ }^{4}))\Big]
\label{Eq_3.12}
\end{align}

We represent the exponent of the reference state shown above in a block-diagonal matrix form as
\begin{equation}\label{Eq_3.14}
A(s=0)=\left(\begin{array}{ccccc}
A_{1}^0 & 0  & 0 & 0 \\
0 & A_{2}^0  & 0 & 0 \\
0 & 0 & A_{3}^0 & 0 \\
0 & 0 & 0 & A_{4}^0
\end{array}\right)_{14\times 14}
\end{equation}

The basis chosen for this representation is 
\begin{equation}
  \vec{v}= \left\{\bar{x}_{0}, \bar{x}_{1}, \bar{x}_{0} \bar{x}_{1}, \bar{x}_{0}^{2}, \bar{x}_{1}^{2},\bar{x}_{0}^2 \bar{x}_{1},\bar{x}_{0} \bar{x}_{1}^2, \bar{x}_{0}^3, \bar{x}_{1}^3 ,\bar{x}_{0}\bar{x}_{1}^3, \bar{x}_{0}^3 \bar{x}_{1}, \bar{x}_{0}^2 \bar{x}_{1},\bar{x}_{0}^4, \bar{x}_{1}^4\right\}
  \label{Eq_3.13}
\end{equation}

We need to ensure that the determinants of $A(s=0)$ and $A(s=1)$ matrices are positive so that that wavefunction remains square integrable everywhere. It should be noted that the matrix elements of $A$, i.e. $A_{1}^0-A_{4}^0$, are matrices themselves, as shown below
where \\
\begin{equation}
A_{1}^0= \left( \begin{array}{cc}
\tilde{\omega}_{ref} & 0 \\
0 & \tilde{\omega}_{ref}
\end{array} \right)
\hspace{2cm}
A_{2}^0=\lambda_{0}^4 \tilde{\omega}_{ref}\left( \begin{array}{ccc}
b & 0 &0\\
0 & \frac{1}{2}&\frac{1}{2}(3-b)\\
0&\frac{1}{2}(3-b)&\frac{1}{2}
\end{array} \right) \nonumber
\end{equation}

\begin{equation}
A_{3}^0= \tilde{\omega}_{ref}\lambda_0^6\left( \begin{array}{cccc}
\frac{p}{2} &0&0&\frac{1}{8}(15-2k) \\
0 & k&\frac{1}{8}(15-2p)&0\\
0&\frac{1}{8}(15-2p)&\frac{1}{4}&0\\
\frac{1}{8}(15-2k)&0&0&\frac{1}{4}
\end{array} \right) \nonumber
\end{equation}

\begin{equation}
A_{4}^0=\tilde{\omega}_{ref}\lambda_0^8 \left( \begin{array}{ccccc}
\frac{1}{8} & \frac{1}{4}(\frac{35}{4}-e)&0&0&0 \\
\frac{1}{4}(\frac{35}{4}-e) & \frac{1}{8}&0&0&0\\
0&0&e&\frac{1}{16}(1-c)&\frac{1}{16}(1-d)\\
0&0&\frac{1}{16}(1-c)&\frac{7}{2}&\frac{1}{4}(\frac{35}{4}-e)\\
0&0&\frac{1}{16}(1-d)&\frac{1}{4}(\frac{35}{4}-e)&\frac{7}{2}
\end{array} \right) \nonumber
\end{equation}
We have introduced a few parameters $b,p,k,c,d,e$ to ensure that the determinant of each block diagonal matrix is positive definite.
Because we are considering higher even interactions, it is needed to consider various quadratic and other higher-order terms.
To get the positive determinant of $A_2^0$ block, the value of $b$ must be in the range $2<b<4$. To eliminate the off-diagonal components, we set $b=3$, as it would give the minimum line element. In $A_3^0$ block, we fix $k=\frac{15}{2}$ and the determinant becomes

\begin{equation*}
    \text{Det}(A_0^3)=-\frac{1}{512}p\left(221+4\left(-15+p\right)p \ \omega_{ref}^4\lambda_6^4\right)
\end{equation*}

We set $p$ as $15/2$, in the range $\frac{13}{2}<p<\frac{17}{2}$, to satisfy the condition $\text{Det}(A_3^0)>0$. Similarly, to ensure that the determinant of $A_0^4$ block is positive and the line element is minimum, we set $c=d=1$ and $e=35/4$.\par

Using the same basis as mentioned in \ref{Eq_3.13}, the target state matrix $A(s=1)$ can be written as another $14\times14$ matrix:

\begin{equation}\label{Eq_3.15}
A(s=1)=\left(\begin{array}{ccccc}
A_{1}^1 & 0  & 0 & 0 \\
0 & A_{2}^1  & 0 & 0 \\
0 & 0 & A_{3}^1 & 0 \\
0 & 0 & 0 & A_{4}^1
\end{array}\right)_{14\times 14}
\end{equation}

where we have the following block diagonal entries:
\begin{equation}
A_{1}^1= \left( \begin{array}{cc}
\alpha_1 & 0 \\
0 & \alpha_2
\end{array} \right)
\hspace{2cm}
A_{2}^1=\left( \begin{array}{ccc}
\tilde{b}\alpha_5 & 0 &0\\
0 & \alpha_3&\frac{1}{2}(1-\tilde{b})\alpha_5\\
0&\frac{1}{2}(1-\tilde{b})\alpha_5&\alpha_4
\end{array} \right) \nonumber
\end{equation}

\begin{equation}
A_{3}^1= \left( \begin{array}{cccc}
\tilde{p}\alpha_6 & 0&0&\frac{1}{2}(1-\tilde{k})\alpha_7 \\
0 & \tilde{k}\alpha_7&\frac{1}{2}(1-\tilde{p})\alpha_6\\
0&\frac{1}{2}(1-\tilde{p})\alpha_6&\alpha_8&0\\
\frac{1}{2}(1-\tilde{k})\alpha_7&0&0&\alpha_9
\end{array} \right) \nonumber
\end{equation}
\begin{equation}
A_{4}^1=\left( \begin{array}{ccccc}
\tilde{d}\alpha_{10} & \frac{1}{4}(1-\tilde{e})\alpha_{12} &0&0&0\\
\frac{1}{4}(1-\tilde{e})\alpha_{12} & \tilde{c}\alpha_{11}&0&0&0\\
0&0&\tilde{e}\alpha_{12}&\frac{1}{2}(1-\tilde{c})\alpha_{11}&\frac{1}{2}(1-\tilde{d})\alpha_{10}\\
0&0&\frac{1}{2}(1-\tilde{c})\alpha_{11}&\alpha_{13}&\frac{1}{4}(1-\tilde{e})\alpha_{12}\\
0&0&\frac{1}{2}(1-\tilde{d})\alpha_{10}&\frac{1}{4}(1-\tilde{e})\alpha_{12}&\alpha_{14}
\end{array} \right) \nonumber
\end{equation}\\
Here as well we fix $\tilde{k}$, $\tilde{c}$ and $\tilde{d}$ to be $1$ to make the off-diagonal terms zero and keep $\tilde{b}$, $\tilde{p}$ and $\tilde{e}$ for the positivity of all the block matrices.

As we are considering a closed quantum system, the reference state evolves into the target state via a certain unitary operator. Now, we represent this as: 
\begin{equation}\label{Eq_3.16}
\psi^{s=1}\left(\bar{x}_{0}, \bar{x}_{1}\right)=\mathbb{U}(s=1) \psi^{s=0}\left(\bar{x}_{0}, \bar{x}_{1}\right)
\end{equation}

We represent the unitary matrix in the following form:
\begin{equation}\label{Eq_3.17}
\mathbb{U}=\overleftarrow{\mathcal{P}} \exp \left[\int_{0}^{s} d s Y^{I}(s) \mathcal{O}_{I}\right]
\end{equation}

We have to act the operators $\mathcal{O}_I$'s in a particular order. The $Y_I'$s depend on the specific order in which $\mathcal{O}_I$ are acting on the reference state. To get the minimum complexity, we try to have a geometric understanding of this unitary evolution process. Then we can write the expression in Eq. (\ref{Eq_3.17}) as follows

\begin{equation}\label{Eq_3.18}
\mathbb{U}=\overleftarrow{\mathcal{P}} \exp \left[\int_{0}^{s} Y^{I}(s) M_{I} d s\right]
\end{equation}

where, $\left(M_{I}\right)_{j k}^{\prime} s$ are $\text{GL}(14, \mathbb{R})$ generators satisfying,
\begin{equation}\label{Eq_3.19}
\operatorname{Tr}\left[M_{I} M_{J}^T\right]=\delta_{I J}
\end{equation}
$I, J$ runs from 1 to 196. As mentioned above, $A(s=0)$ is the reference state which undergoes a unitary transformation to get the target state $A(s=1)$. It enables us to calculate the boundary conditions that lead us to calculate the complexity functional. So we have
\begin{equation}\label{Eq_3.20}
A(s=1)=\mathbb{U}(s=1) A(s=0) \mathbb{U}^{T}(s=1) 
\end{equation}
This leads to the expression,
\begin{equation}\label{Eq_3.21}
Y^{I} M_{I}=\partial_{s} \mathbb{U}(s) \mathbb{U}(s)^{-1}
\end{equation}
Hence,
\begin{equation}\label{Eq_3.22}
Y^{I}=\frac{1}{\operatorname{Tr}\left[M^{I}\left(M^{I}\right)^{T}\right]} \operatorname{Tr}\left[\partial_{s} \mathbb{U}(s) \mathbb{U}^{-1}\left(M^{I}\right)^{T}\right]
\end{equation}
Now the line element can be defined in terms of $Y^{I}$\rq s as, 

\begin{align}\label{Eq_3.23}
d s^{2}&=\textstyle G_{I J} d Y^{I} d Y^{J} \\ \nonumber
&=\textstyle G_{I J}\bigg[\frac{1}{\operatorname{Tr}[M^{I}\left(M^{I}\right)^{T}]} \operatorname{Tr}\left[d_{s} \mathbb{U}(s) \mathbb{U}^{-1}\left(M^{I}\right)^{T}\right]\bigg]\left[\frac{1}{\operatorname{Tr}\left[M^{J}\left(M^{J}\right)^{T}\right]} \operatorname{Tr}\left[d_{s} \mathbb{U}(s) \mathbb{U}^{-1}\left(M^{J}\right)^{T}\right]\right] 
\end{align}

Here, we should mention that $dY^I$ does not denote the total differential for $Y^I$. Observing the structure of the matrix $A$, we find that $\mathbb{U}(s)$ can be considered as an element of $\text{GL}(14, \mathbb{R})$ with a positive determinant. Now we will express the $\mathbb{U}$ matrix with a similar structure as it is in the target state matrix and the unitary matrix contains four block-diagonal matrices.

\begin{equation}\label{Eq_3.24}
\mathbb{U}=\left(\begin{array}{ccccc}
\mathbb{U}_{1} & 0  & 0 & 0 \\
0 & \mathbb{U}_{2}  & 0 & 0 \\
0 & 0 & \mathbb{U}_{3} & 0 \\
0 & 0 & 0 & \mathbb{U}_{4}
\end{array}\right)_{14\times 14}
\end{equation}
where, 

\begin{equation}
\mathbb{U}_{1}= \left( \begin{array}{cc}
x_0-x_1 & x_3-x_2 \\
x_3+x_2 & x_0+x_1
\end{array} \right)
\hspace{2cm}
\mathbb{U}_{2}=\left( \begin{array}{ccc}
\tilde{x}_4 & 0 &0\\
0 & \tilde{x}_5-\tilde{x}_6&\tilde{x}_8-\tilde{x}_7\\
0&\tilde{x}_8+\tilde{x}_7&\tilde{x}_5+\tilde{x}_6
\end{array} \right) \nonumber
\end{equation}

\begin{equation}
\mathbb{U}_{3}= \left( \begin{array}{cccc}
\tilde{x}_9 & 0&0&0 \\
0 & \tilde{x}_{10}-\tilde{x}_{11}&\tilde{x}_{13}-\tilde{x}_{12}&0\\
0&\tilde{x}_{13}+\tilde{x}_{12}&\tilde{x}_{10}+\tilde{x}_{11}&0\\
0&0&0&\tilde{x}_{14} 
\end{array} \right)
\hspace{0.3cm}
\mathbb{U}_{4}= \left( \begin{array}{ccccc}
\tilde{x}_{15}-\tilde{x}_{16} & \tilde{x}_{18}-\tilde{x}_{17} &0&0&0\\
x_{18}+x_{17} & x_{15}+x_{16}&0&0&0\\
0&0&\tilde{x}_{19}&0&0\\
0&0&0&\tilde{x}_{20}-\tilde{x}_{21}&\tilde{x}_{23}-\tilde{x}_{22}\\
0&0&0&\tilde{x}_{23}+\tilde{x}_{22}&\tilde{x}_{20}+\tilde{x}_{21}
\end{array} \right) \nonumber
\end{equation}
We have decomposed $\mathbb{U}(s)$ in terms of four block-diagonal matrices. First, we note that the quadratic part or the first block is always diagonal which induces a flat space, so we take $x_3=x_2=0$. In the unitary operator $\mathbb{U}$, we don't allow the off-diagonal terms as in the final state, only block diagonal form remains. So if we allow off-diagonal terms we will be having an increased line element which we don't want. Now $\text{GL}(2, \mathbb{R})$ can be expressed as $\mathbb{R}\times \text{SL}(2, \mathbb{R})$, so we observe that our $\mathbb{U}$ has $\mathbb{R}^{10} \times \text{SL}(2, \mathbb{R})^4$ group structure.  We will parameterize each $2\times2$ block matrix in $\mathbb{U}$ as it has been done in \cite{Bhattacharyya:2018bbv} i.e, we will parameterize as $\text{AdS}_3$ space.

\vspace{0.5cm}
\begin{equation}
\begin{aligned}  \label{Eq_3.25}
    &x_0 = \exp[y_1]\cosh(\rho_1) &  &x_1 = \exp[y_1]\sinh(\rho_1) & \\ 
    &\tilde{x}_4 = \exp[y_2] & &{x}_5 = \exp[y_3]\cos(\tau_3)\cosh(\rho_3) & \\  
    &\tilde{x}_6=\exp[y_3]\sin(\theta_3)\cosh(\rho_3) & &\tilde{x}_7 = \exp[y_3]\sin(\tau_3)\cosh(\rho_3) &\\   &\tilde{x}_8=\exp[y_3]\cos(\theta_3)\sinh(\rho_3) & &\tilde{x}_9 =\exp[y_4] \\   &\tilde{x}_{10}=\exp[y_5]\cos(\tau_5)\cosh(\rho_5) & &\tilde{x}_{11}=\exp[y_5]\sin(\theta_5)\sinh(\rho_5) \\   &\tilde{x}_{12}=\exp[y_5]\sin(\tau_5)\cosh(\rho_5) & &\tilde{x}_{13}=\exp[y_5]\cos(\theta_5)\sinh(\rho_5) &\\   
    &\tilde{x}_{14}=\exp[y_6] & &\tilde{x}_{15}=\exp[y_7]\cos(\tau_7)\cosh(\rho_7) & \\   
    &\tilde{x}_{16}=\exp[y_7]\sin(\theta_7)\sinh(\rho_7) & &\tilde{x}_{17}=\exp[y_7]\sin(\tau_7)\cosh(\rho_7) & \\   
    &\tilde{x}_{18}=\exp[y_7]\cos(\theta_7)\sinh(\rho_7) &  &\tilde{x}_{19}=\exp[y_8] \\    &\tilde{x}_{20}=\exp[y_9]\cos(\tau_9)\cosh(\rho_9)   & &\tilde{x}_{21}=\exp[y_7]\sin(\theta_9)\sinh(\rho_9) \\  
    &\tilde{x}_{22}=\exp[y_9]\sin(\tau_9)\cosh(\rho_9)   & &\tilde{x}_{23}=\exp[y_9]\cos(\theta_9)\sinh(\rho_9) 
\end{aligned}   
\end{equation}

\vspace{0.5cm}

Using these parameters for $\mathbb{U}$ we can then calculate the infinitesimal line element in Eq. (\ref{Eq_3.23}); which now becomes:

\begin{equation}
\begin{aligned} \label{Eq_3.26}
ds^{2} &= \bigg[2  y_{1}^{2}+ y_{2}^{2}+2  y_{3}^{2}+ y_{4}^{2}+2 y_{5}^{2}+ y_{6}^{2}+2  y_{7}^{2}+ y_{8}^{2}+2 y_{9}^{2}+ 2\bigg( \rho_{1}^{2}+ \rho_{3}^{2}&\\ 
& +  \rho_{5}^{2} +  \rho_{7}^{2} +  \rho_{9}^{2}+\cosh(2 \rho_{3})\Big\{\cosh^2(\rho_{3})  \tau_{3}^{2}+\sinh^2 (\rho_{3})  \theta_{3}^{2} \Big\} -\sinh^2(2\rho_{3}) \theta_{3}  \tau_{3}&\\  
& + \cosh(2 \rho_{5})\Big\{\cosh^2(\rho_{5})  \tau_{5}^{2}+\sinh^2(\rho_{5}) \theta_{5}^{2}\Big\} -\sinh^2(2 \rho_{5})  \theta_{5}  \tau_{5}&\\   
& +\cosh(2 \rho_{7})\Big\{\cosh^2(\rho_{7})  \tau_{7}^{2}+ \sinh^2(\rho_{7}) \theta_{7}^{2}\Big\}-\sinh^2(2 \rho_{7})  \theta_{7}  \tau_{7}&\\  
& +\cosh(2\rho_{9})\Big\{\cosh^2(\rho_{9})  \tau_{9}^{2}+\sinh^2(\rho_{9}) \theta_{9}^{2}\Big\}- \sinh^2(2 \rho_{9}) \theta_{9}  \tau_{9}\bigg)\bigg] 
\end{aligned} 
\end{equation}

 We need to find the shortest path between the reference and the target state in this geometry described by metric expressed in Eq. (\ref{Eq_3.26}). This shortest path will be the circuit complexity for our problem. For that purpose, we also need to calculate the proper boundary conditions denoting the reference and target states.

\textcolor{Sepia}{\subsection{\sffamily Boundary Conditions for the geodesic}}

As we mentioned before the minimal geodesic will be equivalent to finding the geodesic in $GL(14, R)$ group manifold. The geodesic can be found by minimizing the following equation on the distance functional.

\begin{equation}\label{Eq_3.27}
    \mathcal{D}(U)=\int_{0}^{1} \sqrt{g_{ij}\dot{x}^i\dot{x}^j} ds
\end{equation}


The boundary conditions from Eq. (\ref{Eq_3.20}) are

\begin{equation}\label{Eq_3.28}
y_i(0)=\rho_j(0)=0
\end{equation}
where, $i=1,2,...,9$ and $j=1,3,5,7,9$
and

For solving the geodesic equations we have to find conserved charges using the results of \cite{Jefferson:2017sdb} as our metric is $\mathbb{R}^{10} \times \text{SL}(2, \mathbb{R})^4$. Using Eq. (\ref{Eq_3.28}) and Eq. ($\ref{Eq_3.29}$) we get 
\begin{equation}\label{Eq_3.30}
    y_i(s)=y_i(1)s \hspace{1cm} \rho_j(s)=\rho_j(1)s
\end{equation}

\vspace{0.6cm}

where, $i=1,2,...,9$ and $j=1,3,5,7,9$
\vspace{0.5cm}
\begin{equation}
\begin{aligned}\label{Eq_3.29}
&2\big(y_1(1)-\rho_1(1))=\ln\left[\frac{\alpha_1}{\tilde{\omega}_{ref}}\right]& 2\big(y_1(1)&+\rho_1(1))=\ln\Big[\frac{\alpha_2}{\tilde{\omega}_{ref}}\Big]\\ 
&2y_2(1)=\ln\left[\frac{\tilde{b}\alpha_5}{3\tilde{\omega}_{ref}\lambda_4}\right] & 2y_3(1)&= \ln\left[\frac{\sqrt{4\alpha_3\alpha_4 - (1-\tilde{b})^2\alpha_5^2}}{\tilde{\omega}_{ref}\lambda_4}\right]\\ 
&2\rho_3(1)=\cosh^{-1}\left[\frac{\alpha_3+\alpha_4}{\sqrt{4\alpha_3\alpha_4 - (1-\tilde{b})^2\alpha_5^2}}\right]&2y_4(1)&=\ln\Big[\frac{4\tilde{p}\alpha_6}{15 \omega_{ref}\lambda_6}\Big]\\ 
&2y_5(1)=\ln\left[\frac{\sqrt{16\alpha_7\alpha_8-4(1-\tilde{p})^2\alpha_6^2}}{\tilde{\omega}_{ref}\lambda_6}\right]
&2y_6(1)&=\ln\left[\frac{4\alpha_9}{\tilde{\omega}_{ref}\lambda_6}\right]\\ 
&2\rho_5(1)=\cosh^{-1}\left[\frac{2(\alpha_7+\alpha_8)}{\sqrt{16\alpha_7\alpha_8-4(1-\tilde{p})^2\alpha_6}}\right]& 2y_7(1)&=\ln\left[\frac{\sqrt{64\alpha_{10}\alpha_{11} - 4(1-\tilde{e})^2\alpha_{12}^2}}{\tilde{\omega}_{ref}\lambda_8}\right]\\ 
&2\rho_7(1)=\cosh^{-1}\left[\frac{\alpha_{10}+\alpha_{11}}{\sqrt{64\alpha_{10}\alpha_{11} - 4(1-\tilde{e})^2\alpha_{12}^2}}\right]& 2y_8(1)&=\ln\left[\frac{4\tilde{e}\alpha_{12}}{35\tilde{\omega}_{ref}\lambda_6}\right]\\ 
&2\rho_9(1)=\cosh^{-1}\left[\frac{\alpha_{13}+\alpha_{14}}{\sqrt{4\alpha_{13}\alpha_{14}-((1-\tilde{e})^2/4)\alpha_{12}^2}}\right]& 2y_9(1)&=\ln\left[\frac{\sqrt{4\alpha_{13}\alpha_{14}-((1-\tilde{e})^2/4)\alpha_{12}^2}}{7\tilde{\omega}_{ref}\lambda_8}\right]
\end{aligned}
\end{equation}

\vspace{0.5cm}

With the same arguments in \cite{Jefferson:2017sdb}, we set

\begin{equation}\label{Eq_3.31}
    \tau_j(s)=0 \hspace{1cm} \theta_j(s)=\theta_{c_j}
\end{equation}

\vspace{0.5cm}
Where $j=3,5,7,9$ and $\theta_{c_j}$ are constants which do not depend on $s$. So, here we have the freedom to choose any constant value of $\theta_{c_j}$ which tells that it would leave the origin in any direction. (Note: When we are calculating $\rho_5$, any arbitrary constant value will not provide us an analytical expression, so we choose $\theta_5$ to be 0 to get the simple analytical expression in Eq. (\ref{Eq_3.29})).
\newpage
Taking all of these terms and conditions we get the complexity functional as: 

\begin{equation}
\begin{aligned} \label{complexity_functional}
\mathcal{D}(U)=&\sqrt{2\left[\sum_{i=1,odd}^9\left[y_i(1)\right]^2 + \frac{1}{2}\sum_{i=2,even}^8\left[y_i(1)\right]^2 + \sum_{j=1, odd}^9\left[\rho_i(1)\right]^2\right]}  \\ 
=&\frac{1}{\sqrt{2}}\Bigg(2\Bigg[\cosh^{-1}\Bigg(\frac{\alpha_3+\alpha_4}{\sqrt{4\alpha_3\alpha_4-\alpha_5^2(-1+\tilde{b})^2}}\Bigg)\Bigg]^2+2\left[\cosh^{-1}\left(\frac{\alpha_{10}+\alpha_{11}}{2\sqrt{16\alpha_{10}\alpha_{11} +(1-\tilde{e})^2\alpha_{12}^2}}\right)\right]^2\\
& +2\left[\cosh^{-1}\left(\frac{\alpha_{13}+\alpha_{14}}{\sqrt{4\alpha_{13}\alpha_{14}-((1-\tilde{e})^2/4)\alpha_{12}}}\right)\right]^2+2\left[\cosh^{-1}\left(\frac{2(\alpha_7+\alpha_8)}{\sqrt{-\alpha_6^2+4\alpha_7\alpha_8+\alpha_6^2\tilde{p}}}\right)\right]^2\\ 
& +\frac{1}{2}\left[\ln\frac{\alpha_2}{\alpha_1}\right]^2+\frac{1}{2}\left[\ln\left(\frac{\alpha_1\alpha_2}{\tilde{\omega}_{ref}^2}\right)\right]^2+\left[\ln\left(\frac{4\alpha_9}{\lambda_6\tilde{\omega}_{ref}}\right)\right]^2+2\left[\ln\left(\frac{\sqrt{4\alpha_3\alpha_4 - (1-\tilde{b})^2\alpha_5^2}}{\tilde{\omega}_{ref}\lambda_4}\right)\right]^2\\ 
& +2\left[\ln\left(\frac{\tilde{b}\alpha_5}{3\lambda_4\tilde{\omega}_{ref}}\right)\right]^2+2\left[\ln\left(\frac{\sqrt{64\alpha_{10}\alpha_{11} - 4(-1+\tilde{e})^2\alpha_{12}^2}}{\tilde{\omega}_{ref}\lambda_8}\right)\right]^2+ \left[\ln\left(\frac{4\alpha_{12}\tilde{e}}{35\lambda_8\tilde{\omega}_{ref}}\right)\right]^2\\ 
& +2 \left[\ln\left(\frac{\sqrt{4\alpha_{13}\alpha_{14}-((-1+\tilde{e})^2/16)\alpha_{12}^2}}{7\tilde{\omega}_{ref}\lambda_8}\right)\right]^2+2\left[\ln\left(\frac{2\sqrt{-\alpha_6^2+4\alpha_7\alpha_8+\alpha_6\tilde{p}}}{\tilde{\omega}_{ref}\lambda_6}\right)\right]^2\\ 
& + \left[\ln\left(\frac{4\alpha_6\tilde{p}}{15\lambda_6\tilde{\omega}_{ref}}\right)\right]^2 \Bigg)^\frac{1}{2}
\end{aligned}
\end{equation}

which is a straight line as there is no off-diagonal term for we set $\tau_i(s)$ to be $0$ and $\theta_j(s)$ to be independent of $s$ according to the Eq. (\ref{Eq_3.30}).

	For the particular choice of a cost function that we took i.e. $\mathcal{F}_2$, the complexity functional is 
	\begin{equation}
		\mathcal{C}_2 = \int_{s=0}^{1} ds \mathcal{F}_2 
	\end{equation}
	
	As it was shown in Eq. (\ref{complexity_functional}) the complexity functional can be written in terms of some boundary values only. It can also be proven that this functional can just involve the eigenvalues of the reference and target matrix. 
	
	\begin{equation}\label{eq4.13}
		\mathcal{C}_2 = \frac{1}{2} \sqrt{\sum_{i=1}^{14}\log\left[\frac{(\lambda_T)_i}{(\lambda_R)_i}\right]^2}
	\end{equation}
	The proof of this expression is explicitly constructed in Appendix \ref{sec:appendixD}. This result is very crucial and we exploit this relation to generalize the complexity to $N$ oscillators.

\textcolor{Sepia}{\section{\sffamily Analysis for $N$ oscillators}\label{sec:5}}
To this point, our discussion in this paper was concerned with two coupled harmonic oscillators involving higher-order interactions. To extend our analysis to effective field theories, we first need to generalize our results to $N$ coupled harmonic oscillators with $\left(\phi^4 + \phi^6 + \phi^8\right)$ interaction terms. Then, we will gradually move toward the continuum limit for this problem. With that in mind, we consider the following Hamiltonian, \\
\begin{equation}\label{Eq_4.1}
    H = \frac{1}{2} \sum_{a= 0}^{N-1}
    \big[p_a^2+\omega^2 x_a^2+\Omega^2(x_a-x_{a+1})^2+2\lambda_4 x_a^4 + 2\lambda_6 x_a^6 + 2\lambda_8 x_a^8  \big] 
\end{equation}

Now, we will assume the periodic boundary condition is valid on this lattice of $N$ oscillators such that $x_{a+N} = x_a$ (we do so as it allows us to impose translational symmetry and use Fourier transform to express in terms of normal mode coordinates). Then, we perform discrete Fourier transform for this lattice using, 
\begin{equation} \label{Eq_4.2}
    x_a = \frac{1}{\sqrt{N}} \sum_{k=0}^{N-1} \exp{\left[i \frac{ 2\pi a }{N} k\right]} \tilde{x}_k 
\end{equation}
\begin{equation} \label{Eq_4.3}
    p_a = \frac{1}{\sqrt{N}} \sum_{k=0}^{N-1} \exp{\left[i \frac{ 2\pi a }{N} k\right]} \tilde{p}_k
\end{equation}

Using the above Eq. (\ref{Eq_4.2}), (\ref{Eq_4.3}), we can transform the spatial coordinates into normal mode coordinates. The resultant Hamiltonian is then,
\begin{equation}\label{Eq_4.4}
\begin{split}
    H & = \frac{1}{2} \sum_{a= 0}^{N-1}
    \big[p_a^2+\omega^2 x_a^2+\Omega^2(x_a-x_{a+1})^2+2\lambda_4 x_a^4 + 2\lambda_6 x_a^6 + 2\lambda_8 x_a^8  \big] \\
    & = \frac{1}{2} \sum_{k=0}^{N-1} \Big [|\tilde{p}_k|^2 + \Big(\omega^2 + 4 \Omega^2 \sin^2{\Big(\frac{\pi k}{N}\Big)}\Big) |\tilde{x}_k|^2\Big] + H'_{\phi^4}+ H'_{\phi^6}+ H'_{\phi^8}
\end{split}   
\end{equation}
where $H'_{\phi^4}$ ,$H'_{\phi^6}$, $H'_{\phi^8}$ are the contributions from $\phi^4$, $\phi^6$, $\phi^8$ interaction terms respectively. Now, 
\begin{equation}\label{Eq_4.5}
   H'_{\phi^4} = \frac{\lambda_4} {N} \sum_{k_1,k_2,k_3 = 0}^{N-1} \tilde{x}_{\alpha} \tilde{x}_{k_1} \tilde{x}_{k_2} \tilde{x}_{k_3} \ ; \ \alpha = N-k_1-k_2-k_3\ \text{mod}\ N 
\end{equation}
\begin{equation}\label{Eq_4.6}
     H'_{\phi^6}  = \frac{\lambda_6} {N^2} \sum_{k_1,k_2,k_3,k_4,k_5 = 0}^{N-1} \tilde{x}_{\alpha} \tilde{x}_{k_1} \tilde{x}_{k_2} \tilde{x}_{k_3} \tilde{x}_{k_4} \tilde{x}_{k_5}\ ; \ \alpha =\left( N-\sum_{i=1}^{5}k_i\right)\ \text{mod}\ N
\end{equation}
\begin{equation}\label{Eq_4.7}
     H'_{\phi^8}  = \frac{\lambda_8}{N^3} \sum_{k_1,k_2,k_3,k_4,k_5,k_6,k_7 = 0}^{N-1} \tilde{x}_{\alpha} \tilde{x}_{k_1} \tilde{x}_{k_2} \tilde{x}_{k_3} \tilde{x}_{k_4} \tilde{x}_{k_5}\tilde{x}_{k_6}\tilde{x}_{k_7}\   ; \ \alpha = \left(N-\sum_{i=1}^{7}k_i\right)\ \text{mod}\ N
\end{equation}

The proof of transformation of interaction Hamiltonian in Fourier space is given in the appendix (\ref{sec:appendixC}).

The target state wavefunction is given by:
\begin{equation}\label{Eq_4.8}
\psi_{0,0, \cdots 0}\left(\bar{x}_{0}, \cdots \tilde{x}_{N-1}\right)=\left(\frac{\tilde{\omega}_{0} \tilde{\omega}_{1} \ldots \tilde{\omega}_{N-1}}{\pi^{N}}\right)^{\frac{1}{4}} \exp \left[-\frac{1}{2} \sum_{k=0}^{N-1} \tilde{\omega}_{k} \tilde{x}_{k}^{2}+\lambda_4 \psi_4^{1}+\lambda_6 \psi_6^{1}+\lambda_8 \psi_8^{1}\right]
\end{equation}

Where, total perturbation wavefunction $\psi^1$ is :
\begin{equation}\label{Eq_4.9}
\psi^{1}= \lambda_4 \psi_4^{1}+\lambda_6 \psi_6^{1}+\lambda_8 \psi_8^{1}
\end{equation}

where $\lambda_4 \psi_4^{1},$ $ \lambda_6 \psi_6^{1},$ $ \lambda_8 \psi_8^{1}$ are first order perturbation corrections for respective $\phi^4,$ $\phi^6,$ $\phi^8$ self interaction terms.

The expression of $\psi_4^1$ along with $B$ terms have been taken from \cite{Bhattacharyya:2018bbv}.
\vspace{10mm}
\textcolor{Sepia}{\subsection*{\sffamily Expression for $\psi_4^1$ is :}\label{sec:N_Oscillators}}

\begin{equation}
\begin{aligned}
\psi_{4}^{1} =
&~~~\sum_{\substack{a=0\\ 4a \bmod N\equiv0}}^{N-1} B_{1}(a) ~~+ \sum_{\substack{a,b=0\\(2 a+2 b) \bmod N \equiv 0\\a \neq b} }^{N-1} \frac{B_{2}(a, b)}{2}+\sum_{\substack{a,b=0\\(3 b+a) \bmod N\equiv 0 \\a \neq b } }^{N-1} B_{3}(a, b)&~~~~~~~~~~~~~~~~~~~~\\ 
+ &\sum_{\substack{a,b,c=0\\(a+2 b+c) \bmod N\equiv0\\  a \neq b \neq c} }^{N-1} \frac{B_{4}(a, b, c)}{2} ~~+ 
\sum_{\substack{a, b, c, d=0\\(a+b+k+d) \bmod N\equiv 0\\a \neq b \neq c \neq d} }^{N-1} \frac{B_{5}(a, b, c, d)}{24}&\\
\end{aligned}
\end{equation}

\newpage
\vspace{10mm}
\textcolor{Sepia}{\subsection*{\sffamily Expression for $\psi_6^1$ is :}\label{sec:N_Oscillators}}

\begin{equation}
\begin{aligned}
\psi_{6}^{1}&=\frac{1}{N^{2}}\Bigg[\sum_{\substack{a=0 \\ ~~~~~~6 a \bmod N \equiv 0}}^{N-1} C_{1}(a) ~~~~~~~~~~~~~~~~~~+& &\sum_{\substack{a, b=0 \\(a+5b) \bmod N\equiv 0 \\a \neq b}}^{N-1} C_{2}(a, b)  &\\ 
+ &\sum_{\substack{a, b=0 \\(3b+3a) \bmod N \equiv 0  \\a \neq b}}^{N-1} \frac{1}{2} C_{3}(a, b)~~~~~~~~~~~~~~+& &\sum_{\substack{a, b=0 \\(2a+4b) \bmod N\equiv 0 \\a \neq b}}^{N-1} C_{4}(a, b)&\\
+ &\sum_{\substack{a,b, c=0 \\(a+b+4c) \bmod N\equiv 0 \\a\neq  b \neq c}}^{N-1} \frac{1}{2}C_{5}(a,b, c)~~~~~~~~~~+& &\sum_{\substack{a,b, c=0 \\(2a+b+3c) \bmod N\equiv 0 \\a\neq b \neq c}}^{N-1} C_{6}(a,b, c) &\\
+ &\sum_{\substack{a,b,c=0 \\(2a+2b+2c) \bmod N\equiv 0 \\ a\neq b\neq c}}^{N-1} \frac{1}{6}C_{7}(a,b,c)~~~~~~~~+& &\sum_{\substack{a,b,c,d=0 \\(a+b+c+3d) \bmod N\equiv 0 \\ a\neq b \neq c \neq d}}^{N-1} \frac{1}{6}C_{8}(a,b,c,d) &\\
+ &\sum_{\substack{a,b,c,d=0 \\(a+b+2c+2d) \bmod N\equiv 0 \\a\neq b \neq c \neq d}}^{N-1} \frac{1}{4}C_{9}(a,b,c,d)~~~~+& &\sum_{\substack{a,b,c,d,e=0\\(a+b+c+d+2e) \bmod N\equiv0 \\a\neq b \neq c \neq d \neq e}}^{N-1} \frac{1}{4!}C_{10}(a,b,c,d,e)~~~~~~~~~~~~&\\
+ &\sum_{\substack{a,b,c,d,e,f=0\\(a+b+c+d+e+f) \bmod N\equiv0 \\ a\neq b \neq c \neq d \neq e \neq f}}^{N-1} \frac{1}{6!}C_{11}(a,b,c,d,e,f)\Bigg]\\ 
\end{aligned}
\end{equation}

\vspace{1cm}
where, the terms $C_1$, $C_2$, ..., $C_{11}$ are given by
\vspace{1cm}
\begin{longtable}{|c|c|}

		\hline
		\rowcolor{Gray}
		  & Expression for $C_i$ Coefficients \\
\hline
$C_1$ & 
\Large\scalebox{0.70}{
\parbox[t]{20cm}
 {$\\ \bigg[\frac{55}{32 \tilde{\omega}_{a}^{4}}-\frac{15 \tilde{x}_{a}^{2}}{8 \tilde{\omega}_{a}^{3}}-\frac{5 \tilde{x}_{a}^{4}}{8 \tilde{\omega}_{a}^{2}}-\frac{\tilde{x}_{a}^{6}}{6 \tilde{\omega}_{a}}\bigg]\\$}
	  }\\ \hline
$C_2$ & 
\Large\scalebox{0.70}{
\parbox[t]{20cm}
 {$\\\bigg[\frac{-180 \tilde{x}_{a} \tilde{x}_{b}}{(\tilde{\omega}_{a}+\tilde{\omega}_{b})(\tilde{\omega}_{a}+3 \tilde{\omega}_{b})(\tilde{\omega}_{a}+5 \tilde{\omega}_{b})}-\frac{60 \tilde{x}_{a} \tilde{x}_{b}^{3}}{(\tilde{\omega}_{a}+3 \tilde{\omega}_{b})(\tilde{\omega}_{a}+5 \tilde{\omega}_{b})}-\frac{6 \tilde{x}_{a} \tilde{x}_{b}^{5}}{\tilde{\omega}_{a}+5 \tilde{\omega}_{b}}\bigg]\\$}
	  }\\ \hline
$C_3$ & 
\Large\scalebox{0.70}{
\parbox[t]{20cm}
 {$\\\bigg[\frac{-12 0 \tilde{x}_{a} \tilde{x}_{b}}{(\tilde{\omega}_{a}+\tilde{\omega}_{b})(3 \tilde{\omega}_{a}+\tilde{\omega}_{b})(\tilde{\omega}_{a}+3 \tilde{\omega}_{b})}-\frac{10 \tilde{x}_{a}^{3} \tilde{x}_{b}}{(\tilde{\omega}_{a}+\tilde{\omega}_{b})(3 \tilde{\omega}_{a}+\tilde{\omega}_{b})}
-\frac{10 \tilde{x}_{a} \tilde{x}_{b}^{3}}{(\tilde{\omega}_{a}+\tilde{\omega}_{b})(\tilde{\omega}_{a}+3 \tilde{\omega}_{b})}-\frac{10 \tilde{x}_{a}^{3} \tilde{x}_{b}^{3}}{3(\tilde{\omega}_{a}+\tilde{\omega}_{b})}\bigg]\\$}
	  }\\ \hline
$C_4$ & 
\Large\scalebox{0.70}{
\parbox[t]{20cm}
 {$\\\bigg[\frac{135}{32 \tilde{\omega}_{a} \tilde{\omega}_{b}^{3}}+\frac{45}{8 \tilde{\omega}_{a}^{2}(\tilde{\omega}_{a}+\tilde{\omega}_{b})(\tilde{\omega}_{a}+2 \tilde{\omega}_{b})}-\frac{45 \tilde{x}_{a}^{2}}{4 \tilde{\omega}_{a}(\tilde{\omega}_{a}+\tilde{\omega}_{b})(\tilde{\omega}_{a}+2 \tilde{\omega}_{b})}
-\frac{45(\tilde{\omega}_{a}+3 \tilde{\omega}_{b}) \tilde{x}_{b}^{2}}{8 \tilde{\omega}_{b}^{2}(\tilde{\omega}_{a}+\tilde{\omega}_{b})(\tilde{\omega}_{a}+2 \tilde{\omega}_{b})}-\frac{45 \tilde{x}_{a}^{2} \tilde{x}_{b}^{2}}{2(\tilde{\omega}_{a}+\tilde{\omega}_{b})(\tilde{\omega}_{a}+2 \tilde{\omega}_{b})}\\-\frac{15 \tilde{x}_{b}^{4}}{8 \tilde{\omega}_{a} \tilde{\omega}_{b}+16 \tilde{\omega}_{b}^{2}}
-\frac{15 \tilde{x}_{a}^{2} \tilde{x}_{b}^{4}}{2 \tilde{\omega}_{a}+4 \tilde{\omega}_{b}}\bigg]\\$}
	  }\\ \hline
	  
$C_5$ & 
\Large\scalebox{0.70}{
\parbox[t]{20cm}
 {$\\\bigg[-\frac{180 \tilde{x}_{a} \tilde{x}_{b}}{(\tilde{\omega}_{a}+\tilde{\omega}_{b})(\tilde{\omega}_{a}+\tilde{\omega}_{b}+2 \tilde{\omega}_{c})(\tilde{\omega}_{a}+\tilde{\omega}_{b}+4 \tilde{\omega}_{c})}
-\frac{180 \tilde{x}_{a} \tilde{x}_{b} \tilde{x}_{c}^{2}}{(\tilde{\omega}_{a}+\tilde{\omega}_{b}+2 \tilde{\omega}_{c})(\tilde{\omega}_{a}+\tilde{\omega}_{b}+4 \tilde{\omega}_{c})}-\frac{30 \tilde{x}_{a} \tilde{x}_{b} \tilde{x}_{c}^{4}}{\tilde{\omega}_{a}+\tilde{\omega}_{b}+4 \tilde{\omega}_{c}}\bigg]\\$}
	  }\\ \hline
$C_6$ & 
\Large\scalebox{0.70}{
\parbox[t]{20cm}
 {$\\\bigg[\frac{-360(\tilde{\omega}_{a}+\tilde{\omega}_{b}+2 \tilde{\omega}_{c}) \tilde{x}_{b} \tilde{x}_{c}}{(\tilde{\omega}_{b}+\tilde{\omega}_{c})(2 \tilde{\omega}_{a}+\tilde{\omega}_{b}+\tilde{\omega}_{c})(\tilde{\omega}_{b}+3 \tilde{\omega}_{c})(2 \tilde{\omega}_{a}+\tilde{\omega}_{b}+3 \tilde{\omega}_{c})}
-\frac{180 \tilde{x}_{a}^{2} \tilde{x}_{b} \tilde{x}_{c}}{(2 \tilde{\omega}_{a}+\tilde{\omega}_{b}+\tilde{\omega}_{c}) (2 \tilde{\omega}_{a}+\tilde{\omega}_{b}+3 \tilde{\omega}_{c})}-\frac{60 \tilde{x}_{b} \tilde{x}_{c}^{3}}{(\tilde{\omega}_{b}+3 \tilde{\omega}_{c})(2 \tilde{\omega}_{a}+\tilde{\omega}_{b}+3 \tilde{\omega}_{c})}
\\-\frac{60 \tilde{x}_{a}^{2} \tilde{x}_{b} \tilde{x}_{c}^{3}}{2 \tilde{\omega}_{a}+\tilde{\omega}_{b}+3 \tilde{\omega}_{c}}\bigg]\\$}
	  }\\ \hline
$C_7$ & 
\Large\scalebox{0.70}{
\parbox[t]{20cm}
 {$\\\bigg[\frac{45}{8 \tilde{\omega}_{a} \tilde{\omega}_{b} \tilde{\omega}_{c}^{2}}+\frac{45}{8 \tilde{\omega}_{a} \tilde{\omega}_{b}^{2} \tilde{\omega}_{c}}+\frac{45}{8 \tilde{\omega}_{a}^{2} \tilde{\omega}_{b} \tilde{\omega}_{c}}-\frac{45}{8 \tilde{\omega}_{a} \tilde{\omega}_{b}(\tilde{\omega}_{a}+\tilde{\omega}_{b}) \tilde{\omega}_{c}}
-\frac{45}{8 \tilde{\omega}_{a} \tilde{\omega}_{b} \tilde{\omega}_{c}(\tilde{\omega}_{a}+\tilde{\omega}_{c})}-\frac{45}{8 \tilde{\omega}_{a} \tilde{\omega}_{b} \tilde{\omega}_{c}(\tilde{\omega}_{b}+\tilde{\omega}_{c})}\\+\frac{45}{8 \tilde{\omega}_{a} \tilde{\omega}_{b} \tilde{\omega}_{c}(\tilde{\omega}_{a}+\tilde{\omega}_{b}+\tilde{\omega}_{c})}
-\frac{45 (2 \tilde{\omega}_{a}+\tilde{\omega}_{b}+\tilde{\omega}_{c}) \tilde{x}_{a}^{2}}{4 \tilde{\omega}_{a}(\tilde{\omega}_{a}+\tilde{\omega}_{b})(\tilde{\omega}_{a}+\tilde{\omega}_{c})(\tilde{\omega}_{a}+\tilde{\omega}_{b}+\tilde{\omega}_{c})}-\frac{45 \tilde{x}_{a}^{2} \tilde{x}_{a}^{2} \tilde{x}_{c}^{2}}{\tilde{\omega}_{a}+\tilde{\omega}_{b}+\tilde{\omega}_{c}}
-\frac{45 \tilde{x}_{a}^{2} \tilde{x}_{b}^{2}}{2(\tilde{\omega}_{a}+\tilde{\omega}_{b})(\tilde{\omega}_{a}+\tilde{\omega}_{b}+\tilde{\omega}_{c})}\\-\frac{45(\tilde{\omega}_{a}+\tilde{\omega}_{b}+2 \tilde{\omega}_{c}) \tilde{x}_{c}^{2}}{4 \tilde{\omega}_{c}(\tilde{\omega}_{a}+\tilde{\omega}_{c})(\tilde{\omega}_{b}+\tilde{\omega}_{c})(\tilde{\omega}_{a}+\tilde{\omega}_{b}+\tilde{\omega}_{c})}
-\frac{45 \tilde{x}_{a}^{2} \tilde{x}_{c}^{2}}{2(\tilde{\omega}_{a}+\tilde{\omega}_{c})(\tilde{\omega}_{a}+\tilde{\omega}_{b}+\tilde{\omega}_{c})}-\frac{45 \tilde{x}_{a}^{2} \tilde{x}_{c}^{2}}{2(\tilde{\omega}_{b}+\tilde{\omega}_{c})(\tilde{\omega}_{a}+\tilde{\omega}_{b}+\tilde{\omega}_{c})} \\
-\frac{45(\tilde{\omega}_{a}+2 \tilde{\omega}_{b}+\tilde{\omega}_{c}) \tilde{x}_{a}^{2}}{4 \tilde{\omega}_{b}(\tilde{\omega}_{a}+\tilde{\omega}_{b})(\tilde{\omega}_{b}+\tilde{\omega}_{c})(\tilde{\omega}_{a}+\tilde{\omega}_{b}+\tilde{\omega}_{c})}\bigg]\\$}
	  }\\ \hline
$C_8$ & 
\Large\scalebox{0.70}{
\parbox[t]{20cm}
 {$\\\bigg[\frac{-360 \tilde{x}_{a} \tilde{x}_{b} \tilde{x}_{c} \tilde{x}_{d}}{(\tilde{\omega}_{a}+\tilde{\omega}_{b}+\tilde{\omega}_{c}+\tilde{\omega}_{d})(\tilde{\omega}_{a}+\tilde{\omega}_{b}+\tilde{\omega}_{c}+3 \tilde{\omega}_{d})}-\frac{12 0 \tilde{x}_{a} \tilde{x}_{b} \tilde{x}_{c} \tilde{x}_{d}^{3}}{\tilde{\omega}_{a}+\tilde{\omega}_{b}+\tilde{\omega}_{c}+3 \tilde{\omega}_{d}}\bigg]\\$}
	  }\\ \hline
$C_9$ & 
\Large\scalebox{0.70}{
\parbox[t]{20cm}
 {$\\\bigg[\frac{-360(\tilde{\omega}_{a}+\tilde{\omega}_{b}+\tilde{\omega}_{c}+\tilde{\omega}_{d}) \tilde{x}_{a} \tilde{x}_{b}}{(\tilde{\omega}_{a}+\tilde{\omega}_{b})(\tilde{\omega}_{a}+\tilde{\omega}_{b}+2 \tilde{\omega}_{c})(\tilde{\omega}_{a}+\tilde{\omega}_{b}+2 \tilde{\omega}_{d})(\tilde{\omega}_{a}+\tilde{\omega}_{b}+2(\tilde{\omega}_{c}+\tilde{\omega}_{d}))}
-\frac{18 0 \tilde{x}_{a} \tilde{x}_{b}\left((\tilde{\omega}_{a}+\tilde{\omega}_{b}+2 \tilde{\omega}_{d}) \tilde{x}_{x}^{2}+(\tilde{\omega}_{a}+\tilde{\omega}_{b}+2 \tilde{\omega}_{c}) \tilde{x}_{d}^{2}\right)}{(\tilde{\omega}_{a}+\tilde{\omega}_{b}+2 \tilde{\omega}_{c})(\tilde{\omega}_{a}+\tilde{\omega}_{b}+2 \tilde{\omega}_{d})(\tilde{\omega}_{a}+\tilde{\omega}_{b}+2(\tilde{\omega}_{c}+\tilde{\omega}_{d}))}\\
-\frac{180 \tilde{x}_{a} \tilde{x}_{b} \tilde{x}_{x}^{2} \tilde{x}_{d}^{2}}{\tilde{\omega}_{a}+\tilde{\omega}_{b}+2(\tilde{\omega}_{c}+\tilde{\omega}_{d})}\bigg]\\$}
	  }\\ \hline
$C_{10}$ & 
\Large\scalebox{0.70}{
\parbox[t]{20cm}
 {$\\\bigg[\frac{ -360\tilde{x}_{a} \tilde{x}_{b} \tilde{x}_{c} \tilde{x}_{d}}{(\tilde{\omega}_{a}+\tilde{\omega}_{b}+\tilde{\omega}_{c}+\tilde{\omega}_{d})(\tilde{\omega}_{a}+\tilde{\omega}_{b}+\tilde{\omega}_{c}+\tilde{\omega}_{d}+2 \tilde{\omega}_{e})}
-\frac{360\tilde{x}_{a} \tilde{x}_{b} \tilde{x}_{c} \tilde{x}_{d} \tilde{x}_{e}^{2}}{(\tilde{\omega}_{a}+\tilde{\omega}_{b}+\tilde{\omega}_{c}+\tilde{\omega}_{d}+2 \tilde{\omega}_{e})}\bigg]\\$}
	  }\\ \hline
$C_{11}$ & 
\Large\scalebox{0.70}{
\parbox[t]{20cm}
 {$\\\bigg[\frac{-720\tilde{x}_{a} \tilde{x}_{b} \tilde{x}_{c} \tilde{x}_{d} \tilde{x}_{e} \tilde{x}_{f}}{(\tilde{\omega}_{a}+\tilde{\omega}_{b}+\tilde{\omega}_{c}+\tilde{\omega}_{d}+\tilde{\omega}_{e}+\tilde{\omega}_{f})}\bigg]\\$}
	  }\\ \hline
\end{longtable}

\vspace{10cm}

\textcolor{Sepia}{\subsection*{\sffamily Expression for $\psi_8^1$ is :}\label{sec:N_Oscillators}}
\vspace{5mm}
\begin{equation}
\begin{aligned}
\psi_{8}^{1}&= \\
&\frac{1}{N^{3}}\Bigg[~~~~~~~~~\sum_{\substack{a=0 \\8a\bmod N\equiv0}}^{N-1} D_{1}(a)~~~~~~~~~~~~~~~~~~~~+&      &\sum_{\substack{a,b=0\\(6a+2b)\bmod N\equiv0\\a\neq b}}^{N-1} D_{2}(a,b) &\\
+&~~~~~~~\sum_{\substack{a,b=0\\(5a+3b) \bmod N\equiv0\\a\neq b}}^{N-1} D_{3}(a,b)~~~~~~~~~~~~~~~+&  &\sum_{\substack{a,b=0 \\(4a+4b) \bmod N\equiv0\\a\neq b}}^{N-1} \frac{1}{2}~D_{4}(a,b) &\\
+&~~~~~~~~\sum_{\substack{a,b=0\\(a+7b)\bmod N\equiv0\\a\neq b}}^{N-1} D_{5}(a,b)~~~~~~~~~~~~~~~+& &\sum_{\substack{a,b,c=0\\(a+b+6c) \bmod N\equiv0}}^{N-1} \frac{1}{2}~D_{6}(a,b,c) &\\
+&~~~~~~\sum_{\substack{a,b,c=0\\(a+2b+5 c) \bmod N\equiv0\\a\neq b \neq c}}^{N-1} D_{7}(a,b,c)~~~~~~~~~~~+& &\sum_{\substack{a,b,c=0\\(a+4b+3c) \bmod N\equiv0\\a\neq b \neq c}}^{N-1} D_{8}(a,b,c)&\\
+&~~~~~\sum_{\substack{a,b,c=0\\(2a+2b+4c)\bmod N\equiv0\\a\neq b \neq c}}^{N-1} \frac{D_{9}(a,b,c)}{2}~~~~~~~~~~~~~+& &\sum_{\substack{a,b,c=0\\(3a+2b+3c)\bmod N\equiv0\\a\neq b \neq c}}^{N-1} \frac{D_{10}(a,b,c)}{2}&\\
+&~~~~\sum_{\substack{a,b,c,d=0\\(a+b+2c+4d)\bmod N\equiv0\\a\neq b \neq c\neq d}}^{N-1} \frac{D_{11}(a,b,c,d)}{2}~~~~~~~~~~~~+& &\sum_{\substack{a,b,c,d=0\\2(a+b+c+d)\bmod N\equiv0 \\a\neq b \neq c\neq d}}^{N-1} \frac{~D_{12}(a,b,c,d)}{24} ~~~&\\
+&~~~\sum_{\substack{a,b,c,d=0\\~(a+2b+2c+3d) \bmod N\equiv0\\a\neq b \neq c\neq d}}^{N-1} \frac{D_{13}(a,b,c,d)}{2}~~~~~~~~~~+& &\sum_{\substack{a,b,c,d=0\\(a+b+c+5d) \bmod N\equiv0\\a\neq b \neq c\neq d}}^{N-1} \frac{D_{14}(a,b,c,d)}{6}&\\
+&~~\sum_{\substack{a,b,c,d=0\\~(a+b+3c+3d) \bmod N\equiv0 \\a\neq b \neq c\neq d}}^{N-1} \frac{D_{15}(a,b,c,d)}{4}~~~~~~~~~~~+& & \sum_{\substack{a,b,c,d,e=0 \\a+b+2(c+d+e)\bmod N\equiv0\\a\neq b \neq c\neq d \neq e}}^{N-1} \frac{D_{16}(a,b,c,d,e)}{12}&\\
+&~\sum_{\substack{a,b,c,d,e=0\\~(a+b+c+2d+3e)\bmod N\equiv0\\a\neq b \neq c\neq d \neq e}}^{N-1} \frac{D_{17}(a,b,c,d,e)}{6}
~~~~~~~+& &\sum_{\substack{a,b,c,d,e=0\\(a+b+c+d+4e) \bmod N\equiv0\\a\neq b \neq c\neq d \neq e}}^{N-1} \frac{D_{18}(a,b,c,d,e)}{24} &\\
+&~~~~~\footnotesize{\sum_{\substack{a,b,c,d,e,f=0\\(a+b+c+d+e\\~~~~~+3f)\bmod N\equiv0\\a\neq b \neq c\neq d \neq e \neq f }}^{N-1} \frac{D_{19}(a,b,c,d,e,f)}{5!}} 
~~~~~~~~~~~+& &~~~\footnotesize{\sum_{\substack{a,b,c,d,e,f=0\\(a+b+c+d+2e\\~~~~~+2f) \bmod N\equiv0 \\ a\neq b \neq c\neq d \neq e \neq f}}^{N-1} \frac{D_{20}(a,b,c,d,e,f)}{48}} &\\
+ &~~~~~\footnotesize{\sum_{\substack{a,b,c,d,e,f,g=0\\(a+b+c+d+e+f\\~~~+2g)\bmod N\equiv0\\a\neq b \neq c\neq d \neq e \neq f\neq g}}^{N-1} \frac{D_{21}(a,b,c,d,e,f,g)}{6!}}~~~~~~~~~+& 
&~~~\footnotesize{\sum_{\substack{a,b,c,d,e,~~~\\f,g,h=0\\
    (a+b+c+d+e+f\\+g+h)\bmod N\equiv0 \\ a\neq b \neq c\neq d \neq e \\ \neq f \neq g \neq h}}^{N-1} \frac{D_{22}(a,b,c,d,e,f,g,h)}{8!}}\Bigg] &\\
\end{aligned}
\end{equation}
\newpage
\vspace{10mm}
The terms $D_{1}, D_{2}, D_{3}\dots D_{22}$ are given in the table below
\vspace{5mm}
\begin{longtable}{|c|c|}

		\hline
		\rowcolor{Gray}
		  & Expression for $D_i$ Coefficients \\
\hline
$D_1$ & 
\Large\scalebox{0.70}{
\parbox[t]{20cm}
 {$\\ \Bigg[\frac{875}{128 \tilde{\omega}_{a}^{5}}-\frac{105  x_{b}^{2}}{16 \tilde{\omega}_{a}^{4}}-\frac{35  x_{a}^{4}}{16 \tilde{\omega}_{a}^{3}}-\frac{7  x_{a}^{6}}{12 \tilde{\omega}_{a}^{2}}-\frac{ x_{a}^{8}}{8 \tilde{\omega}_{a}}\Bigg]\\$}
	  }\\ \hline 	
$D_2$ & 
\Large\scalebox{0.70}{
\parbox[t]{20cm}
 {$\\\frac{8!}{2!6!}\bigg[
\frac{5 \left(36 \tilde{\omega}_{a}^{4}+66 \tilde{\omega}_{a}^{3} \tilde{\omega}_{b}+121 \tilde{\omega}_{a}^{2} \tilde{\omega}_{b}^{2}+66 \tilde{\omega}_{a} \tilde{\omega}_{b}^{3}+11 \tilde{\omega}_{b}^{4}\right)}{64 \tilde{\omega}_{a}^{4} \tilde{\omega}_{b}^{2}\left(\tilde{\omega}_{a}+\tilde{\omega}_{b}\right)\left(2 \tilde{\omega}_{a}+\tilde{\omega}_{b}\right) \left(3 \tilde{\omega}_{a}+\tilde{\omega}_{b}\right)}-\frac{15 \left(11 \tilde{\omega}_{a}^{2}+6 \tilde{\omega}_{a} \tilde{\omega}_{b}+\tilde{\omega}_{b}^{2}\right) x_{a}^{2}}{16 \tilde{\omega}_{a}^{3}\left(\tilde{\omega}_{a}+\tilde{\omega}_{b}\right)\left(2 \tilde{\omega}_{a}+\tilde{\omega}_{b}\right) \left(3 \tilde{\omega}_{a}+\tilde{\omega}_{b}\right)}\\-\frac{45  x_{b}^{2}}{8 \tilde{\omega}_{b}\left(\tilde{\omega}_{a}+\tilde{\omega}_{b}\right)\left(2 \tilde{\omega}_{a}+\tilde{\omega}_{b}\right) \left(3 \tilde{\omega}_{a}+\tilde{\omega}_{b}\right)}-\frac{5 \left(5 \tilde{\omega}_{a}+\tilde{\omega}_{b}\right) x_{a}^{4}}{16 \tilde{\omega}_{a}^{2}\left(2 \tilde{\omega}_{a}+\tilde{\omega}_{b}\right) \left(3 \tilde{\omega}_{a}+\tilde{\omega}_{b}\right)}-\frac{45  x_{a}^{2} x_{b}^{2}}{4\left(\tilde{\omega}_{a}+\tilde{\omega}_{b}\right)\left(2 \tilde{\omega}_{a}+\tilde{\omega}_{b}\right) \left(3 \tilde{\omega}_{a}+\tilde{\omega}_{b}\right)}\\
-\frac{ x_{b}^{6}}{12 \tilde{\omega}_{a}\left(3 \tilde{\omega}_{a}+\tilde{\omega}_{b}\right)}-\frac{15 x_{b}^{4} x_{b}^{2}}{4 \left(2 \tilde{\omega}_{a}+\tilde{\omega}_{b}\right) \left(3 \tilde{\omega}_{a}+\tilde{\omega}_{b}\right)}-\frac{ x_{a}^{6} x_{b}^{2}}{2 \left(3 \tilde{\omega}_{b}+\tilde{\omega}_{a}\right)}\bigg]\\$}
	  }\\ \hline 
$D_3$ & 
\Large\scalebox{0.70}{
\parbox[t]{20cm}
 {$\\\frac{8!}{3!5!}\bigg[\frac{-~30 \left(23 \tilde{\omega}_{a}+13 \tilde{\omega}_{b}\right) x_{a} x_{b}}{\left(\tilde{\omega}_{a}+\tilde{\omega}_{b}\right)\left(3 \tilde{\omega}_{a}+\tilde{\omega}_{b}\right) \left(5 \tilde{\omega}_{a}+\tilde{\omega}_{b}\right)\left(\tilde{\omega}_{a}+3 \tilde{\omega}_{b}\right)\left(5 \tilde{\omega}_{a}+3 \tilde{\omega}_{b}\right)}-\frac{10  x_{a} x_{b}^{3}}{\left(\tilde{\omega}_{a}+\tilde{\omega}_{b}\right)\left(\tilde{\omega}_{a}+3 \tilde{\omega}_{b}\right)\left(5 \tilde{\omega}_{a}+3 \tilde{\omega}_{b}\right)}\\
 \hspace{1cm}-\frac{40 \left(2 \tilde{\omega}_{a}+\tilde{\omega}_{b}\right) x_{a}^{3} x_{b}}{\left(\tilde{\omega}_{a}+\tilde{\omega}_{b}\right)\left(3 \tilde{\omega}_{a}+\tilde{\omega}_{b}\right) \left(5 \tilde{\omega}_{a}+\tilde{\omega}_{b}\right) \left(5 \tilde{\omega}_{a}+3 \tilde{\omega}_{b}\right)}-\frac{10x_{a}^{3} x_{b}^{3}}{3\left(\tilde{\omega}_{a}+\tilde{\omega}_{b}\right)\left(5 \tilde{\omega}_{a}+3 \tilde{\omega}_{b}\right)}-\frac{3 x_{a}^{5} x_{b}}{\left(5 \tilde{\omega}_{a}+\tilde{\omega}_{b}\right) \left(5 \tilde{\omega}_{a}+3 \tilde{\omega}_{b}\right)}-\frac{x_{a}^{5} x_{b}^{3}}{5 \tilde{\omega}_{a}+3 \tilde{\omega}_{b}}\bigg]\\$}
	  }\\ \hline 	
$D_4$ & 
\Large\scalebox{0.70}{
\parbox[t]{20cm}
 {\Large$\\\frac{8!}{4!4!} \bigg[\frac{27\left(2 \tilde{\omega}_{a}^{4}+7 \tilde{\omega}_{a}^{3} \tilde{\omega}_{b}+7 \tilde{\omega}_{a}^{2} \tilde{\omega}_{b}^{2}+7 \tilde{\omega}_{a} \tilde{\omega}_{b}^{3}+2 \tilde{\omega}_{b}^{4}\right)}{64 \tilde{\omega}_{a}^{3} \tilde{\omega}_{b}^{3}\left(\tilde{\omega}_{a}+\tilde{\omega}_{b}\right)\left(2 \tilde{\omega}_{a}+\tilde{\omega}_{b}\right)\left(\tilde{\omega}_{a}+2 \tilde{\omega}_{b}\right)}-\frac{9 \left(7 \tilde{\omega}_{a}+2 \tilde{\omega}_{b}\right) x_{a}^{2}}{16 \tilde{\omega}_{a}^{2}\left(\tilde{\omega}_{a}+\tilde{\omega}_{b}\right)\left(2 \tilde{\omega}_{a}+\tilde{\omega}_{b}\right)\left(\tilde{\omega}_{a}+2 \tilde{\omega}_{b}\right)}\\
 -\frac{9 \left(2 \tilde{\omega}_{a}+7 \tilde{\omega}_{b}\right) x_{b}^{2}}{16 \tilde{\omega}_{b}^{2}\left(\tilde{\omega}_{a}+\tilde{\omega}_{b}\right)\left(2 \tilde{\omega}_{a}+\tilde{\omega}_{b}\right)\left(\tilde{\omega}_{a}+2 \tilde{\omega}_{b}\right)}-\frac{3  x_{b}^{4}}{16 \tilde{\omega}_{b}\left(\tilde{\omega}_{a}+\tilde{\omega}_{b}\right)\left(\tilde{\omega}_{a}+2 \tilde{\omega}_{b}\right)}-\frac{3  x_{b}^{4}}{16 \tilde{\omega}_{b}\left(\tilde{\omega}_{b}+\tilde{\omega}_{b}\right)\left(\tilde{\omega}_{a}+2 \tilde{\omega}_{b}\right)
 }\\-\frac{27 x_{a}^{2} x_{b}^{2}}{4\left(\tilde{\omega}_{a}+\tilde{\omega}_{b}\right)\left(2 \tilde{\omega}_{a}+\tilde{\omega}_{b}\right)\left(\tilde{\omega}_{a}+2 \tilde{\omega}_{b}\right)}-\frac{3  x_{a}^{2} x_{b}^{4}}{4\left(\tilde{\omega}_{a}+\tilde{\omega}_{b}\right)\left(\tilde{\omega}_{a}+2 \tilde{\omega}_{b}\right)}-\frac{3 x_{a}^{4} x_{b}^{2}}{4\left(\tilde{\omega}_{a}+\tilde{\omega}_{b}\right)\left(2 \tilde{\omega}_{a}+\tilde{\omega}_{b}\right)}-\frac{ x_{b}^{4} x_{b}^{4}}{4\left(\tilde{\omega}_{a}+\tilde{\omega}_{b}\right)}\bigg]\\$}
	  }\\ \hline 	  
$D_5$ & 
\Large\scalebox{0.70}{
\parbox[t]{20cm}
 {$\\\frac{8!}{7!}\bigg[\frac{-~630  x_{a} x_{b}}{\left(\tilde{\omega}_{a}+\tilde{\omega}_{b}\right)\left(\tilde{\omega}_{a}+3 \tilde{\omega}_{b}\right)\left(\tilde{\omega}_{a}+5 \tilde{\omega}_{b}\right)\left(\tilde{\omega}_{a}+7 \tilde{\omega}_{b}\right)}-\frac{210  x_{a} x_{b}^{3}}{\left(\tilde{\omega}_{a}+3 \tilde{\omega}_{b}\right)\left(\tilde{\omega}_{a}+5 \tilde{\omega}_{b}\right)\left(\tilde{\omega}_{a}+7 \tilde{\omega}_{b}\right)}-\frac{21  x_{a} x_{b}^{5}}{\left(\tilde{\omega}_{a}+5 \tilde{\omega}_{b}\right)\left(\tilde{\omega}_{a}+7 \tilde{\omega}_{b}\right)}\\-\frac{ x_{a} x_{b}^{7}}{\tilde{\omega}_{a}+7 \tilde{\omega}_{b}}
\bigg]\\$}
	  }\\ \hline 	  
$D_6$ & 

{\scalebox{.70}{
\parbox[t]{20cm}
{\Large $\\\frac{8!}{6!}\bigg[\frac{-~90 x_{a} x_{b}}{\left(\tilde{\omega}_{a}+\tilde{\omega}_{b}\right)\left(\tilde{\omega}_{a}+\tilde{\omega}_{b}+2 \tilde{\omega}_{c}\right)\left(\tilde{\omega}_{a}+\tilde{\omega}_{b}+4 \tilde{\omega}_{c}\right)\left(\tilde{\omega}_{a}+\tilde{\omega}_{b}+6 \tilde{\omega}_{c}\right)}-\frac{90  x_{a} x_{b} x_{c}^{2}}{\left(\tilde{\omega}_{c}+\tilde{\omega}_{b}+2 \tilde{\omega}_{c}\right)\left(\tilde{\omega}_{a}+\tilde{\omega}_{b}+4 \tilde{\omega}_{c}\right)\left(\tilde{\omega}_{a}+\tilde{\omega}_{b}+6 \tilde{\omega}_{c}\right)}\\
-\frac{15 x_{a} x_{b} x_{c}^{4}}{\left(\tilde{\omega}_{a}+\tilde{\omega}_{b}+4 \tilde{\omega}_{c}\right)\left(\tilde{\omega}_{a}+\tilde{\omega}_{b}+6 \tilde{\omega}_{c}\right)}-\frac{ x_{a} x_{b} x_{c}^{6}}{\tilde{\omega}_{a}+\tilde{\omega}_{b}+6 \tilde{\omega}_{c}}\bigg]\\$}}
	  }\\ \hline
$D_7$ & 
\Large\scalebox{0.70}{
\parbox[t]{20cm}
 {\Large $\\\frac{8!}{2! 5!}\bigg[ \frac{-~20 x_{a} x_{c}^{3}\left(\tilde{\omega}_{a}+\tilde{\omega}_{b}+4 \tilde{\omega}_{c}\right)}{\left(\tilde{\omega}_{a}+3 \tilde{\omega}_{c}\right)\left(\tilde{\omega}_{a}+2 \tilde{\omega}_{b}+3 \tilde{\omega}_{c}\right)\left(\tilde{\omega}_{a}+5 \tilde{\omega}_{c}\right)\left(\tilde{\omega}_{a}+2 \tilde{\omega}_{b}+5 \tilde{\omega}_{c}\right)}-\frac{ x_{a} x_{b}^{2} x_{c}}{\left(\tilde{\omega}_{a}+2 \tilde{\omega}_{b}+\tilde{\omega}_{c}\right)\left(\tilde{\omega}_{a}+2 \tilde{\omega}_{b}+3 \tilde{\omega}_{c}\right)\left(\tilde{\omega}_{a}+2 \tilde{\omega}_{b}+5 \tilde{\omega}_{c}\right)}\\
 -\frac{ x_{a} x_{c}^{5}}{\left(\tilde{\omega}_{a}+5 \tilde{\omega}_{c}\right)\left(\tilde{\omega}_{a}+2 \tilde{\omega}_{b}+5 \tilde{\omega}_{c}\right)}-\frac{10 x_{a} x_{b}^{3} x_{c}^{3}}{\left(\tilde{\omega}_{a}+2 \tilde{\omega}_{b}+3 \tilde{\omega}_{c}\right)\left(\tilde{\omega}_{a}+2  \tilde{\omega}_{b}+5 \tilde{\omega}_{c}\right)}-\frac{ x_{a} x_{b}^{2} x_{c}^{5}}{\tilde{\omega}_{a}+2 \tilde{\omega}_{b}+5 \tilde{\omega}_{c}}\\
 -\frac{30 x_{a} x_{c}\left(3 \tilde{\omega}_{a}^{2}+6 \tilde{\omega}_{a} \tilde{\omega}_{b}+4 \tilde{\omega}_{b}^{2}+18 \tilde{\omega}_{a} \tilde{\omega}_{c}+18 \tilde{\omega}_{b} \tilde{\omega}_{c}+23 \tilde{\omega}_{c}^{2}\right)}{\left(\tilde{\omega}_{a}+\tilde{\omega}_{c}\right)\left(\tilde{\omega}_{a}+2 \tilde{\omega}_{b}+\tilde{\omega}_{c}\right)\left(\tilde{\omega}_{a}+3 \tilde{\omega}_{c}\right)\left(\tilde{\omega}_{a}+2 \tilde{\omega}_{b}+3 \tilde{\omega}_{c}\right)\left(\tilde{\omega}_{a}+5 \tilde{\omega}_{c}\right)\left(\tilde{\omega}_{a}+2 \tilde{\omega}_{b}+5 \tilde{\omega}_{c}\right)}
\bigg]\\$}
	  }\\\hline
$D_8$ & 
\Large\scalebox{0.70}{
\parbox[t]{20cm}
{$\\\frac{8!}{3!4!}\bigg[\frac{-~6  x_{a} x_{b}^{3}}{\left(\tilde{\omega}_{a}+3 \tilde{\omega}_{b}\right)\left(\tilde{\omega}_{a}+3 \tilde{\omega}_{b}+2 \tilde{\omega}_{c}\right)\left(\tilde{\omega}_{a}+3 \tilde{\omega}_{b}+4 \tilde{\omega}_{c}\right)}-\frac{36  x_{a} x_{b} x_{c}^{2}\left(\tilde{\omega}_{a}+2 \tilde{\omega}_{b}+3 \tilde{\omega}_{c}\right)}{\left(\tilde{\omega}_{a}+\tilde{\omega}_{b}+2 \tilde{\omega}_{c}\right)\left(\tilde{\omega}_{a}+3 \tilde{\omega}_{b}+2 \tilde{\omega}_{c}\right)\left(\tilde{\omega}_{a}+\tilde{\omega}_{b}+4 \tilde{\omega}_{c}\right)\left(\tilde{\omega}_{a}+3 \tilde{\omega}_{b}+4 \tilde{\omega}_{c}\right)}\\-\frac{3  x_{a} x_{b} x_{c}^{4}}{\left(\tilde{\omega}_{a}+\tilde{\omega}_{b}+4 \tilde{\omega}_{c}\right)\left(\tilde{\omega}_{a}+3 \tilde{\omega}_{b}+4 \tilde{\omega}_{c}\right)}-\frac{6  x_{a} x_{b}^{3} x_{c}^{2}}{\left(\tilde{\omega}_{a}+3 \tilde{\omega}_{b}+2 \tilde{\omega}_{c}\right)\left(\tilde{\omega}_{a}+3 \tilde{\omega}_{b}+4 \tilde{\omega}_{c}\right)}-\frac{ x_{a} x_{b}^{3} x_{c}^{4}}{\tilde{\omega}_{a}+3 \tilde{\omega}_{b}+4 \tilde{\omega}_{c}}  
\\-\frac{18  x_{a} x_{b}\left(3 \tilde{\omega}_{a}^{2}+13 \tilde{\omega}_{b}^{2}+24 \tilde{\omega}_{b} \tilde{\omega}_{c}+8 \tilde{\omega}_{c}^{2}+12 \tilde{\omega}_{a}\left(\tilde{\omega}_{b}+\tilde{\omega}_{c}\right)\right)}{\left(\tilde{\omega}_{a}+\tilde{\omega}_{b}\right)\left(\tilde{\omega}_{a}+3 \tilde{\omega}_{b}\right)\left(\tilde{\omega}_{a}+\tilde{\omega}_{b}+2 \tilde{\omega}_{c}\right)\left(\tilde{\omega}_{a}+3 \tilde{\omega}_{b}+2 \tilde{\omega}_{c}\right)\left(\tilde{\omega}_{a}+\tilde{\omega}_{b}+4 \tilde{\omega}_{c}\right)\left(\tilde{\omega}_{a}+3 \tilde{\omega}_{b}+4 \tilde{\omega}_{c}\right)}\bigg]\\$}
	  }\\ \hline
$D_9$ & 
\Large\scalebox{0.70}{
\parbox[t]{20cm}
 {$\\\frac{8!}{2!2!4!}\bigg[ \frac{9 }{64 \tilde{\omega}_{a} \tilde{\omega}_{b} \tilde{\omega}_{c}^{3}}+\frac{3 }{32 \tilde{\omega}_{a} \tilde{\omega}_{b}^{2} \tilde{\omega}_{c}^{2}}+\frac{3 }{32 \tilde{\omega}_{a}^{2} \tilde{\omega}_{b} \tilde{\omega}_{c}^{2}}-\frac{3 }{32 \tilde{\omega}_{a} \tilde{\omega}_{b}\left(\tilde{\omega}_{a}+\tilde{\omega}_{b}\right) \tilde{\omega}_{c}^{2}}-\frac{3 }{16 \tilde{\omega}_{a} \tilde{\omega}_{b} \tilde{\omega}_{c}^{2}\left(\tilde{\omega}_{a}+\tilde{\omega}_{c}\right)}\\
 -\frac{3 }{16 \tilde{\omega}_{a} \tilde{\omega}_{b} \tilde{\omega}_{c}^{2}\left(\tilde{\omega}_{b}+\tilde{\omega}_{c}\right)}+ \frac{3 }{16 \tilde{\omega}_{a} \tilde{\omega}_{b} \tilde{\omega}_{c}^{2}\left(\tilde{\omega}_{a}+\tilde{\omega}_{b}+\tilde{\omega}_{c}\right)}+\frac{3 }{32 \tilde{\omega}_{a} \tilde{\omega}_{b} \tilde{\omega}_{c}^{2}\left(\tilde{\omega}_{a}+2 \tilde{\omega}_{c}\right)}+\frac{3 }{32 \tilde{\omega}_{a} \tilde{\omega}_{b} \tilde{\omega}_{c}^{2}\left(\tilde{\omega}_{b}+2 \tilde{\omega}_{c}\right)}\\
 -\frac{3 }{32 \tilde{\omega}_{a} \tilde{\omega}_{b} \tilde{\omega}_{c}^{2}\left(\tilde{\omega}_{a}+\tilde{\omega}_{b}+2 \tilde{\omega}_{c}\right)}-\frac{3  x_{a}^{2}}{16 \tilde{\omega}_{a} \tilde{\omega}_{b} \tilde{\omega}_{c}^{2}}-\frac{3  x_{a}^{2}\left(3 \tilde{\omega}_{a}^{2}+\tilde{\omega}_{b}^{2}+3 \tilde{\omega}_{b} \tilde{\omega}_{c}+2 \tilde{\omega}_{c}^{2}+3 \tilde{\omega}_{a}\left(\tilde{\omega}_{b}+2 \tilde{\omega}_{c}\right)\right)}{8 \tilde{\omega}_{a}\left(\tilde{\omega}_{a}+\tilde{\omega}_{b}\right)\left(\tilde{\omega}_{a}+\tilde{\omega}_{c}\right)\left(\tilde{\omega}_{a}+\tilde{\omega}_{b}+\tilde{\omega}_{c}\right)\left(\tilde{\omega}_{a}+2 \tilde{\omega}_{c}\right)\left(\tilde{\omega}_{a}+\tilde{\omega}_{b}+2 \tilde{\omega}_{c}\right)}\\
 -\frac{3  x_{b}^{2}\left(\tilde{\omega}_{a}^{2}+3 \tilde{\omega}_{b}^{2}+6 \tilde{\omega}_{b} \tilde{\omega}_{c}+2 \tilde{\omega}_{c}^{2}+3 \tilde{\omega}_{a}\left(\tilde{\omega}_{b}+\tilde{\omega}_{c}\right)\right)}{8 \tilde{\omega}_{b}\left(\tilde{\omega}_{a}+\tilde{\omega}_{b}\right)\left(\tilde{\omega}_{b}+\tilde{\omega}_{c}\right)\left(\tilde{\omega}_{a}+\tilde{\omega}_{b}+\tilde{\omega}_{c}\right)\left(\tilde{\omega}_{b}+2 \tilde{\omega}_{c}\right)\left(\tilde{\omega}_{a}+\tilde{\omega}_{b}+2 \tilde{\omega}_{c}\right)}-\frac{3  x_{c}^{2}}{16 \tilde{\omega}_{a} \tilde{\omega}_{b} \tilde{\omega}_{c}^{2}}+\frac{3  x_{c}^{2}}{8 \tilde{\omega}_{a} \tilde{\omega}_{b}\left(\tilde{\omega}_{b}+\tilde{\omega}_{c}\right)\left(\tilde{\omega}_{b}+2 \tilde{\omega}_{c}\right)}\\
 +\frac{3  x_{c}^{2}}{8 \tilde{\omega}_{a} \tilde{\omega}_{b}\left(\tilde{\omega}_{a}+\tilde{\omega}_{c}\right)\left(\tilde{\omega}_{a}+2 \tilde{\omega}_{c}\right)}-\frac{ x_{c}^{4}\left(\tilde{\omega}_{a}+\tilde{\omega}_{b}+4 \tilde{\omega}_{c}\right)}{16 \tilde{\omega}_{c}\left(\tilde{\omega}_{a}+2 \tilde{\omega}_{c}\right)\left(\tilde{\omega}_{b}+2 \tilde{\omega}_{c}\right)\left(\tilde{\omega}_{a}+\tilde{\omega}_{b}+2 \tilde{\omega}_{c}\right)}-\frac{3  x_{a}^{2} x_{b}^{2}}{4\left(\tilde{\omega}_{a}+\tilde{\omega}_{b}\right)\left(\tilde{\omega}_{a}+\tilde{\omega}_{b}+\tilde{\omega}_{c}\right)\left(\tilde{\omega}_{a}+\tilde{\omega}_{b}+2 \tilde{\omega}_{c}\right)}\\
 -\frac{ x_{a}^{2} x_{c}^{2}\left(2 \tilde{\omega}_{a}+\tilde{\omega}_{b}+3 \tilde{\omega}_{c}\right)}{4\left(\tilde{\omega}_{a}+\tilde{\omega}_{c}\right)\left(\tilde{\omega}_{a}+\tilde{\omega}_{b}+\tilde{\omega}_{c}\right)\left(\tilde{\omega}_{a}+2 \tilde{\omega}_{c}\right)\left(\tilde{\omega}_{a}+\tilde{\omega}_{b}+2 \tilde{\omega}_{c}\right)}-\frac{3  x_{b}^{2} x_{c}^{2}\left(\tilde{\omega}_{a}+2 \tilde{\omega}_{b}+3 \tilde{\omega}_{c}\right)}{4\left(\tilde{\omega}_{b}+\tilde{\omega}_{c}\right)\left(\tilde{\omega}_{a}+\tilde{\omega}_{b}+\tilde{\omega}_{c}\right)\left(\tilde{\omega}_{b}+2 \tilde{\omega}_{c}\right)\left(\tilde{\omega}_{a}+\tilde{\omega}_{b}+2 \tilde{\omega}_{c}\right)}\\-\frac{ x_{a}^{2} x_{c}^{4}}{4\left(\tilde{\omega}_{a}+2 \tilde{\omega}_{c}\right)\left(\tilde{\omega}_{a}+\tilde{\omega}_{b}+2 \tilde{\omega}_{c}\right)}-\frac{ x_{b}^{2} x_{c}^{4}}{4\left(\tilde{\omega}_{b}+2 \tilde{\omega}_{c}\right)\left(\tilde{\omega}_{a}+\tilde{\omega}_{b}+2 \tilde{\omega}_{c}\right)}-\frac{3 x_{a}^{2} x_{b}^{2} x_{c}^{2}}{2\left(\tilde{\omega}_{a}+\tilde{\omega}_{b}+\tilde{\omega}_{c}\right)\left(\tilde{\omega}_{a}+\tilde{\omega}_{b}+2 \tilde{\omega}_{c}\right)}-\frac{ x_{a}^{2} x_{b}^{2} x_{c}^{4}}{2\left(\tilde{\omega}_{a}+\tilde{\omega}_{b}+2 \tilde{\omega}_{c}\right)}\bigg]\\$}
	  }\\ \hline 
$D_{10}$ & 
\Large\scalebox{0.70}{
\parbox[t]{20cm}
{\Large$\\\frac{8!}{2!3!3!}\bigg[ \frac{9 x_{a} x_{c}}{4 \tilde{\omega}_{b} \tilde{\omega}_{c}\left(\tilde{\omega}_{a}+2 \tilde{\omega}_{b}+\tilde{\omega}_{c}\right)\left(3 \tilde{\omega}_{a}+2 \tilde{\omega}_{b}+\tilde{\omega}_{c}\right)}-\frac{6  x_{a} x_{c}}{\tilde{\omega}_{b}\left(\tilde{\omega}_{a}+\tilde{\omega}_{c}\right)\left(3 \tilde{\omega}_{a}+\tilde{\omega}_{c}\right)\left(\tilde{\omega}_{a}+3 \tilde{\omega}_{c}\right)}-\frac{ x_{a}^{3} x_{b}^{2} x_{c}^{3}}{3 \tilde{\omega}_{a}+2 \tilde{\omega}_{b}+3 \tilde{\omega}_{c}}\\
+\frac{9  x_{a} x_{c}}{4 \sqrt{2} \tilde{\omega}_{a} \tilde{\omega}_{b} \tilde{\omega}_{c}\left(\tilde{\omega}_{a}+2 \tilde{\omega}_{b}+3 \tilde{\omega}_{c}\right)}+\frac{9 x_{a} x_{c}}{8 \tilde{\omega}_{a} \tilde{\omega}_{b} \tilde{\omega}_{c}\left(3 \tilde{\omega}_{a}+2 \tilde{\omega}_{b}+3 \tilde{\omega}_{c}\right)}-\frac{ x_{a}^{3} x_{c}}{2 \tilde{\omega}_{b}\left(\tilde{\omega}_{a}+\tilde{\omega}_{c}\right)\left(3 \tilde{\omega}_{c}+\tilde{\omega}_{c}\right)}\\
+\frac{3  x_{a}^{3} x_{c}}{2 \tilde{\omega}_{b}\left(3 \tilde{\omega}_{a}+2 \tilde{\omega}_{b}+\tilde{\omega}_{c}\right) \left(3 \tilde{\omega}_{a}+2 \tilde{\omega}_{b}+3 \tilde{\omega}_{c}\right)}-\frac{9 x_{a} x_{b}^{2} x_{c}\left(\left(3 \tilde{\omega}_{a}+2 \tilde{\omega}_{b}\right)^{2}+10 \tilde{\omega}_{a} \tilde{\omega}_{c}+4 \tilde{\omega}_{b} \tilde{\omega}_{c}+\tilde{\omega}_{c}^{2}\right)}{4 \tilde{\omega}_{a} \tilde{\omega}_{c}\left(\tilde{\omega}_{a}+2 \tilde{\omega}_{b}+\tilde{\omega}_{c}\right)\left(3 \tilde{\omega}_{a}+2 \tilde{\omega}_{b}+\tilde{\omega}_{c}\right) \left(3 \tilde{\omega}_{a}+2 \tilde{\omega}_{b}+3 \tilde{\omega}_{c}\right)}\\
-\frac{3  x_{c}^{3} x_{b}^{2} x_{c}}{\left(3 \tilde{\omega}_{a}+2 \tilde{\omega}_{b}+\tilde{\omega}_{c}\right) \left(3 \tilde{\omega}_{c}+2 \tilde{\omega}_{b}+3 \tilde{\omega}_{c}\right)}-\frac{ x_{a} x_{c}^{3}}{2 \tilde{\omega}_{b}\left(\tilde{\omega}_{a}+\tilde{\omega}_{c}\right)\left(\tilde{\omega}_{a}+3 \tilde{\omega}_{c}\right)}+\frac{3  x_{a} x_{b}^{2} x_{c}^{3}}{2 \tilde{\omega}_{a}\left(3 \tilde{\omega}_{a}+2 \tilde{\omega}_{b}+3 \tilde{\omega}_{c}\right)}\\
-\frac{3  x_{a} x_{c}^{3}\left(\tilde{\omega}_{a}+3 \sqrt{2} \tilde{\omega}_{a}+2 \tilde{\omega}_{b}+2 \sqrt{2} \tilde{\omega}_{b}+3 \tilde{\omega}_{c}+3 \sqrt{2} \tilde{\omega}_{c}\right)}{4 \tilde{\omega}_{a} \tilde{\omega}_{b}\left(\tilde{\omega}_{a}+2 \tilde{\omega}_{b}+3 \tilde{\omega}_{c}\right)\left(3 \tilde{\omega}_{a}+2 \tilde{\omega}_{b}+3 \tilde{\omega}_{c}\right)}-\frac{ x_{c}^{3} x_{c}^{3}}{3\left(\tilde{\omega}_{a}+\tilde{\omega}_{c}\right)\left(3 \tilde{\omega}_{a}+2 \tilde{\omega}_{b}+3 \tilde{\omega}_{c}\right)}\bigg]\\$}
	  }\\ \hline 
$D_{11}$ & 
\Large\scalebox{0.70}{
\parbox[t]{20cm}
 {$\\\frac{8!}{2!4!}\bigg[ \frac{6 \left(-3\left(\tilde{\omega}_{a}+\tilde{\omega}_{b}\right)^{2}-6\left(\tilde{\omega}_{a}+\tilde{\omega}_{b}\right) \tilde{\omega}_{c}-4 \tilde{\omega}_{c}^{2}-12\left(\tilde{\omega}_{a}+\tilde{\omega}_{b}+\tilde{\omega}_{c}\right) \tilde{\omega}_{d}-8 \tilde{\omega}_{d}^{2}\right) x_{a} x_{b}}{\left(\tilde{\omega}_{a}+\tilde{\omega}_{b}\right)\left(\tilde{\omega}_{a}+\tilde{\omega}_{b}+2 \tilde{\omega}_{c}\right)\left(\tilde{\omega}_{a}+\tilde{\omega}_{b}+2 \tilde{\omega}_{d}\right)\left(\tilde{\omega}_{a}+\tilde{\omega}_{b}+4 \tilde{\omega}_{d}\right)\left(\tilde{\omega}_{a}+\tilde{\omega}_{b}+2 \tilde{\omega}_{c}+4 \tilde{\omega}_{d}\right)\left(\tilde{\omega}_{a}+\tilde{\omega}_{b}+2\left(\tilde{\omega}_{c}+\tilde{\omega}_{d}\right)\right)}\\
 -\frac{6 x_{a} x_{b} x_{c}^{2}}{\left(\tilde{\omega}_{a}+\tilde{\omega}_{b}+2 \tilde{\omega}_{c}\right)\left(\tilde{\omega}_{a}+\tilde{\omega}_{b}+2 \tilde{\omega}_{c}+4 \tilde{\omega}_{d}\right)\left(\tilde{\omega}_{a}+\tilde{\omega}_{b}+2\left(\tilde{\omega}_{c}+\tilde{\omega}_{d}\right)\right)}-\frac{x_{a} x_{b} x_{d}^{4}}{\left(\tilde{\omega}_{a}+\tilde{\omega}_{b}+4 \tilde{\omega}_{d}\right)\left(\tilde{\omega}_{a}+\tilde{\omega}_{b}+2 \tilde{\omega}_{c}+4 \tilde{\omega}_{d}\right)}\\
 -\frac{12\left(\tilde{\omega}_{a}+\tilde{\omega}_{b}+\tilde{\omega}_{c}+3 \tilde{\omega}_{d}\right) x_{a} x_{b} x_{d}^{2}}{\left(\tilde{\omega}_{a}+\tilde{\omega}_{b}+2 \tilde{\omega}_{d}\right)\left(\tilde{\omega}_{a}+\tilde{\omega}_{b}+4 \tilde{\omega}_{d}\right)\left(\tilde{\omega}_{a}+\tilde{\omega}_{b}+2 \tilde{\omega}_{c}+4 \tilde{\omega}_{d}\right)\left(\tilde{\omega}_{a}+\tilde{\omega}_{b}+2\left(\tilde{\omega}_{c}+\tilde{\omega}_{d}\right)\right)}-\frac{x_{a} x_{b} x_{c}^{2} x_{d}^{4}}{\tilde{\omega}_{a}+\tilde{\omega}_{b}+2 \tilde{\omega}_{c}+4 \tilde{\omega}_{d}}\\-\frac{6 x_{a} x_{b} x_{c}^{2} x_{d}^{2}}{\left(\tilde{\omega}_{a}+\tilde{\omega}_{b}+2 \tilde{\omega}_{c}+4 \tilde{\omega}_{d}\right)\left(\tilde{\omega}_{a}+\tilde{\omega}_{b}+2\left(\tilde{\omega}_{c}+\tilde{\omega}_{d}\right)\right)}\bigg]\\$}
	  }

\\\hline	  
$D_{12}$ & 
\Large\scalebox{0.70}{
\parbox[t]{20cm}
 {$\\\frac{8!}{2!2!2!2!}\bigg[ 
\frac{1}{32 \tilde{\omega}_{a} \tilde{\omega}_{b} \tilde{\omega}_{c} \tilde{\omega}_{d}{ }^{2}}+\frac{1}{32 \tilde{\omega}_{a} \tilde{\omega}_{b} \tilde{\omega}_{c}^{2} \tilde{\omega}_{d}}+\frac{1}{32 \tilde{\omega}_{a} \tilde{\omega}_{b}^{2} \tilde{\omega}_{c} \tilde{\omega}_{d}}+\frac{1}{32 \tilde{\omega}_{a}^{2} \tilde{\omega}_{b} \tilde{\omega}_{c} \tilde{\omega}_{d}}-\frac{1}{32 \tilde{\omega}_{a} \tilde{\omega}_{b}\left(\tilde{\omega}_{a}+\tilde{\omega}_{b}\right) \tilde{\omega}_{c} \tilde{\omega}_{d}}\\-\frac{1}{32 \tilde{\omega}_{a} \tilde{\omega}_{b} \tilde{\omega}_{c}\left(\tilde{\omega}_{a}+\tilde{\omega}_{c}\right) \tilde{\omega}_{d}}-\frac{1}{32 \tilde{\omega}_{d} \tilde{\omega}_{b} \tilde{\omega}_{c}\left(\tilde{\omega}_{b}+\tilde{\omega}_{c}\right) \tilde{\omega}_{d}}
+\frac{1}{32 \tilde{\omega}_{a} \tilde{\omega}_{b} \tilde{\omega}_{c}\left(\tilde{\omega}_{a}+\tilde{\omega}_{b}+\tilde{\omega}_{c}\right) \tilde{\omega}_{d}}-\frac{1}{32 \tilde{\omega}_{a} \tilde{\omega}_{b} \tilde{\omega}_{c} \tilde{\omega}_{d}\left(\tilde{\omega}_{a}+\tilde{\omega}_{d}\right)}\\-\frac{1}{32 \tilde{\omega}_{a} \tilde{\omega}_{b} \tilde{\omega}_{c} \tilde{\omega}_{d}\left(\tilde{\omega}_{b}+\tilde{\omega}_{d}\right)}+\frac{1}{32 \tilde{\omega}_{a} \tilde{\omega}_{b} \tilde{\omega}_{c} \tilde{\omega}_{d}\left(\tilde{\omega}_{a}+\tilde{\omega}_{b}+\tilde{\omega}_{d}\right)}-\frac{1}{32 \tilde{\omega}_{a} \tilde{\omega}_{b} \tilde{\omega}_{c} \tilde{\omega}_{d}\left(\tilde{\omega}_{c}+\tilde{\omega}_{d}\right)}
+\frac{1}{32 \tilde{\omega}_{a} \tilde{\omega}_{b} \tilde{\omega}_{c} \tilde{\omega}_{d}\left(\tilde{\omega}_{a}+\tilde{\omega}_{c}+\tilde{\omega}_{c}\right)}\\+
\frac{1}{32 \tilde{\omega}_{a} \tilde{\omega}_{b} \tilde{\omega}_{c} \tilde{\omega}_{d}\left(\tilde{\omega}_{b}+\tilde{\omega}_{c}+\tilde{\omega}_{c}\right)}-\frac{1}{32 \tilde{\omega}_{a} \tilde{\omega}_{b} \tilde{\omega}_{c} \tilde{\omega}_{d}\left(\tilde{\omega}_{a}+\tilde{\omega}_{b}+\tilde{\omega}_{c}+\tilde{\omega}_{d}\right)}-\frac{x_{a}^{2}}{16 \tilde{\omega}_{a} \tilde{\omega}_{b} \tilde{\omega}_{c} \tilde{\omega}_{d}}+\frac{x_{a}^{2}}{16 \tilde{\omega}_{b}\left(\tilde{\omega}_{a}+\tilde{\omega}_{b}\right) \tilde{\omega}_{c} \tilde{\omega}_{d}}\\+\frac{x_{a}^{2}}{16 \tilde{\omega}_{b} \tilde{\omega}_{c}\left(\tilde{\omega}_{a}+\tilde{\omega}_{c}\right) \tilde{\omega}_{d}}-\frac{x_{a}^{2}}{16 \tilde{\omega}_{b} \tilde{\omega}_{c}\left(\tilde{\omega}_{a}+\tilde{\omega}_{b}+\tilde{\omega}_{c}\right) \tilde{\omega}_{d}}+\frac{x_{a}^{2}}{16 \tilde{\omega}_{b} \tilde{\omega}_{c} \tilde{\omega}_{d}\left(\tilde{\omega}_{a}+\tilde{\omega}_{d}\right)}-\frac{x_{a}^{2}}{16 \tilde{\omega}_{b} \tilde{\omega}_{c} \tilde{\omega}_{d}\left(\tilde{\omega}_{a}+\tilde{\omega}_{b}+\tilde{\omega}_{d}\right)}\\-\frac{x_{a}^{2}}{16 \tilde{\omega}_{b} \tilde{\omega}_{c} \tilde{\omega}_{d}\left(\tilde{\omega}_{a}+\tilde{\omega}_{c}+\tilde{\omega}_{d}\right)}+\frac{x_{a}^{2}}{16 \tilde{\omega}_{b} \tilde{\omega}_{c} \tilde{\omega}_{d}\left(\tilde{\omega}_{a}+\tilde{\omega}_{b}+\tilde{\omega}_{c}+\tilde{\omega}_{d}\right)}-\frac{x_{b}^{2}}{16 \tilde{\omega}_{a} \tilde{\omega}_{b} \tilde{\omega}_{c} \tilde{\omega}_{d}}+\frac{x_{b}^{2}}{16 \tilde{\omega}_{a}\left(\tilde{\omega}_{a}+\tilde{\omega}_{b}\right) \tilde{\omega}_{c} \tilde{\omega}_{d}}\\+\frac{x_{b}^{2}}{16 \tilde{\omega}_{a} \tilde{\omega}_{c}\left(\tilde{\omega}_{b}+\tilde{\omega}_{c}\right) \tilde{\omega}_{d}}-\frac{x_{b}^{2}}{16 \tilde{\omega}_{a} \tilde{\omega}_{c}\left(\tilde{\omega}_{a}+\tilde{\omega}_{b}+\tilde{\omega}_{c}\right) \tilde{\omega}_{d}}+\frac{x_{b}^{2}}{16 \tilde{\omega}_{a} \tilde{\omega}_{c} \tilde{\omega}_{d}\left(\tilde{\omega}_{b}+\tilde{\omega}_{d}\right)}-\frac{x_{b}^{2}}{16 \tilde{\omega}_{a} \tilde{\omega}_{c} \tilde{\omega}_{d}\left(\tilde{\omega}_{a}+\tilde{\omega}_{b}+\tilde{\omega}_{d}\right)}\\-\frac{x_{b}^{2}}{16 \tilde{\omega}_{a} \tilde{\omega}_{c} \tilde{\omega}_{d}\left(\tilde{\omega}_{b}+\tilde{\omega}_{c}+\tilde{\omega}_{d}\right)}+\frac{x_{b}^{2}}{16 \tilde{\omega}_{a} \tilde{\omega}_{c} \tilde{\omega}_{d}\left(\tilde{\omega}_{a}+\tilde{\omega}_{b}+\tilde{\omega}_{c}+\tilde{\omega}_{d}\right)}-\frac{x_{a}^{2} x_{b}^{2}}{8\left(\tilde{\omega}_{a}+\tilde{\omega}_{b}\right) \tilde{\omega}_{c} \tilde{\omega}_{d}}+\frac{x_{a}^{2} x_{b}^{2}}{8 \tilde{\omega}_{c}\left(\tilde{\omega}_{a}+\tilde{\omega}_{b}+\tilde{\omega}_{c}\right) \tilde{\omega}_{d}}\\+\frac{x_{a}^{2} x_{b}^{2}}{8 \tilde{\omega}_{c} \tilde{\omega}_{d}\left(\tilde{\omega}_{a}+\tilde{\omega}_{b}+\tilde{\omega}_{d}\right)}-\frac{x_{a}^{2} x_{b}^{2}}{8 \tilde{\omega}_{c} \tilde{\omega}_{d}\left(\tilde{\omega}_{a}+\tilde{\omega}_{b}+\tilde{\omega}_{c}+\tilde{\omega}_{d}\right)}-\frac{x_{c}^{2}}{16 \tilde{\omega}_{a} \tilde{\omega}_{b} \tilde{\omega}_{c} \tilde{\omega}_{d}}+\frac{x_{c}^{2}}{16 \tilde{\omega}_{a} \tilde{\omega}_{b}\left(\tilde{\omega}_{a}+\tilde{\omega}_{c}\right) \tilde{\omega}_{d}}\\+\frac{x_{c}^{2}}{16 \tilde{\omega}_{a} \tilde{\omega}_{b}\left(\tilde{\omega}_{b}+\tilde{\omega}_{c}\right) \tilde{\omega}_{d}}-\frac{x_{c}^{2}}{16 \tilde{\omega}_{a} \tilde{\omega}_{b}\left(\tilde{\omega}_{a}+\tilde{\omega}_{b}+\tilde{\omega}_{c}\right) \tilde{\omega}_{d}}+\frac{x_{c}^{2}}{16 \tilde{\omega}_{a} \tilde{\omega}_{b} \tilde{\omega}_{d}\left(\tilde{\omega}_{c}+\tilde{\omega}_{d}\right)}-\frac{x_{c}^{2}}{16 \tilde{\omega}_{a} \tilde{\omega}_{b} \tilde{\omega}_{d}\left(\tilde{\omega}_{a}+\tilde{\omega}_{c}+\tilde{\omega}_{d}\right)}\\-\frac{x_{c}^{2}}{16 \tilde{\omega}_{a} \tilde{\omega}_{b} \tilde{\omega}_{d}\left(\tilde{\omega}_{b}+\tilde{\omega}_{c}+\tilde{\omega}_{d}\right)}+ \frac{x_{c}^{2}}{16 \tilde{\omega}_{a} \tilde{\omega}_{b} \tilde{\omega}_{d}\left(\tilde{\omega}_{a}+\tilde{\omega}_{b}+\tilde{\omega}_{c}+\tilde{\omega}_{d}\right)}-\frac{x_{a}^{2} x_{c}^{2}}{8 \tilde{\omega}_{b}\left(\tilde{\omega}_{a}+\tilde{\omega}_{c}\right) \tilde{\omega}_{d}}+\frac{x_{a}^{2} x_{c}^{2}}{8 \tilde{\omega}_{b}\left(\tilde{\omega}_{a}+\tilde{\omega}_{b}+\tilde{\omega}_{c}\right) \tilde{\omega}_{d}}\\+\frac{x_{a}^{2} x_{c}^{2}}{8 \tilde{\omega}_{b} \tilde{\omega}_{d}\left(\tilde{\omega}_{a}+\tilde{\omega}_{c}+\tilde{\omega}_{d}\right)}-\frac{x_{a}^{2} x_{c}^{2}}{8 \tilde{\omega}_{b} \tilde{\omega}_{d}\left(\tilde{\omega}_{a}+\tilde{\omega}_{b}+\tilde{\omega}_{c}+\tilde{\omega}_{d}\right)}- \frac{x_{b}^{2} x_{c}^{2}}{8 \tilde{\omega}_{a}\left(\tilde{\omega}_{b}+\tilde{\omega}_{c}\right) \tilde{\omega}_{d}}+\frac{x_{b}^{2} x_{c}^{2}}{8 \tilde{\omega}_{a}\left(\tilde{\omega}_{a}+\tilde{\omega}_{b}+\tilde{\omega}_{c}\right) \tilde{\omega}_{d}}\\+\frac{x_{b}^{2} x_{c}^{2}}{8 \tilde{\omega}_{a} \tilde{\omega}_{d}\left(\tilde{\omega}_{b}+\tilde{\omega}_{c}+\tilde{\omega}_{d}\right)}-\frac{x_{b}^{2} x_{c}^{2}}{8 \tilde{\omega}_{a} \tilde{\omega}_{d}\left(\tilde{\omega}_{a}+\tilde{\omega}_{b}+\tilde{\omega}_{c}+\tilde{\omega}_{d}\right)}-\frac{x_{a}^{2} x_{b}^{2} x_{c}^{2}}{4\left(\tilde{\omega}_{a}+\tilde{\omega}_{b}+\tilde{\omega}_{c}\right) \tilde{\omega}_{d}}+\frac{x_{a}^{2} x_{b}^{2} x_{c}^{2}}{4 \tilde{\omega}_{d}\left(\tilde{\omega}_{a}+\tilde{\omega}_{b}+\tilde{\omega}_{c}+\tilde{\omega}_{d}\right)}-\frac{x_{d}^{2}}{16 \tilde{\omega}_{a} \tilde{\omega}_{b} \tilde{\omega}_{c} \tilde{\omega}_{d}}\\+\frac{x_{d}^{2}}{16 \tilde{\omega}_{a} \tilde{\omega}_{b} \tilde{\omega}_{c}\left(\tilde{\omega}_{a}+\tilde{\omega}_{d}\right)}+\frac{x_{d}^{2}}{16 \tilde{\omega}_{a} \tilde{\omega}_{b} \tilde{\omega}_{c}\left(\tilde{\omega}_{b}+\tilde{\omega}_{d}\right)}-\frac{x_{d}^{2}}{16 \tilde{\omega}_{a} \tilde{\omega}_{b} \tilde{\omega}_{c}\left(\tilde{\omega}_{a}+\tilde{\omega}_{b}+\tilde{\omega}_{d}\right)}+ \frac{x_{d}^{2}}{16 \tilde{\omega}_{a} \tilde{\omega}_{b} \tilde{\omega}_{c}\left(\tilde{\omega}_{c}+\tilde{\omega}_{d}\right)}\\-\frac{x_{d}^{2}}{16 \tilde{\omega}_{a} \tilde{\omega}_{b} \tilde{\omega}_{c}\left(\tilde{\omega}_{a}+\tilde{\omega}_{c}+\tilde{\omega}_{d}\right)}-\frac{x_{d}^{2}}{16 \tilde{\omega}_{a} \tilde{\omega}_{b} \tilde{\omega}_{c}\left(\tilde{\omega}_{b}+\tilde{\omega}_{c}+\tilde{\omega}_{d}\right)}+\frac{x_{d}^{2}}{16 \tilde{\omega}_{a} \tilde{\omega}_{b} \tilde{\omega}_{c}\left(\tilde{\omega}_{a}+\tilde{\omega}_{b}+\tilde{\omega}_{c}+\tilde{\omega}_{d}\right)}- \frac{x_{a}^{2} x_{d}^{2}}{8 \tilde{\omega}_{b} \tilde{\omega}_{c}\left(\tilde{\omega}_{a}+\tilde{\omega}_{d}\right)}\\+\frac{x_{a}^{2} x_{d}^{2}}{8 \tilde{\omega}_{b} \tilde{\omega}_{c}\left(\tilde{\omega}_{a}+\tilde{\omega}_{b}+\tilde{\omega}_{d}\right)}+\frac{x_{a}^{2} x_{d}^{2}}{8 \tilde{\omega}_{b} \tilde{\omega}_{c}\left(\tilde{\omega}_{a}+\tilde{\omega}_{c}+\tilde{\omega}_{d}\right)}-\frac{x_{a}^{2} x_{d}^{2}}{8 \tilde{\omega}_{b} \tilde{\omega}_{c}\left(\tilde{\omega}_{a}+\tilde{\omega}_{b}+\tilde{\omega}_{c}+\tilde{\omega}_{d}\right)}-\frac{x_{b}^{2} x_{d}^{2}}{8 \tilde{\omega}_{a} \tilde{\omega}_{c}\left(\tilde{\omega}_{b}+\tilde{\omega}_{d}\right)}+ \frac{x_{b}^{2} x_{d}^{2}}{8 \tilde{\omega}_{a} \tilde{\omega}_{c}\left(\tilde{\omega}_{a}+\tilde{\omega}_{b}+\tilde{\omega}_{d}\right)}\\+\frac{x_{b}^{2} x_{d}^{2}}{8 \tilde{\omega}_{a} \tilde{\omega}_{c}\left(\tilde{\omega}_{b}+\tilde{\omega}_{c}+\tilde{\omega}_{d}\right)}-\frac{x_{b}^{2} x_{d}^{2}}{8 \tilde{\omega}_{a} \tilde{\omega}_{c}\left(\tilde{\omega}_{a}+\tilde{\omega}_{b}+\tilde{\omega}_{c}+\tilde{\omega}_{d}\right)}-\frac{x_{a}^{2} x_{b}^{2} x_{d}^{2}}{4 \tilde{\omega}_{c}\left(\tilde{\omega}_{a}+\tilde{\omega}_{b}+\tilde{\omega}_{d}\right)}+\frac{x_{a}^{2} x_{b}^{2} x_{d}^{2}}{4 \tilde{\omega}_{c}\left(\tilde{\omega}_{a}+\tilde{\omega}_{b}+\tilde{\omega}_{c}+\tilde{\omega}_{d}\right)}\\- \frac{x_{c}^{2} x_{d}^{2}}{8 \tilde{\omega}_{a} \tilde{\omega}_{b}\left(\tilde{\omega}_{c}+\tilde{\omega}_{d}\right)}+\frac{x_{c}^{2} x_{d}^{2}}{8 \tilde{\omega}_{a} \tilde{\omega}_{b}\left(\tilde{\omega}_{a}+\tilde{\omega}_{c}+\tilde{\omega}_{d}\right)}+\frac{x_{c}^{2} x_{d}^{2}}{8 \tilde{\omega}_{a} \tilde{\omega}_{b}\left(\tilde{\omega}_{b}+\tilde{\omega}_{c}+\tilde{\omega}_{d}\right)}-\frac{x_{c}^{2} x_{d}^{2}}{8 \tilde{\omega}_{a} \tilde{\omega}_{b}\left(\tilde{\omega}_{a}+\tilde{\omega}_{b}+\tilde{\omega}_{c}+\tilde{\omega}_{d}\right)}-\frac{x_{a}^{2} x_{c}^{2} x_{d}^{2}}{4 \tilde{\omega}_{b}\left(\tilde{\omega}_{c}+\tilde{\omega}_{c}+\tilde{\omega}_{d}\right)}\\+\frac{x_{a}^{2} x_{c}^{2} x_{d}^{2}}{4 \tilde{\omega}_{b}\left(\tilde{\omega}_{a}+\tilde{\omega}_{b}+\tilde{\omega}_{c}+\tilde{\omega}_{d}\right)}-\frac{x_{b}^{2} x_{c}^{2} x_{d}^{2}}{4 \tilde{\omega}_{a}\left(\tilde{\omega}_{b}+\tilde{\omega}_{c}+\tilde{\omega}_{d}\right)}+\frac{x_{b}^{2} x_{c}^{2} x_{d}^{2}}{4 \tilde{\omega}_{a}\left(\tilde{\omega}_{a}+\tilde{\omega}_{b}+\tilde{\omega}_{c}+\tilde{\omega}_{d}\right)}-\frac{x_{a}^{2} x_{b}^{2} x_{c}^{2} x_{d}^{2}}{2\left(\tilde{\omega}_{a}+\tilde{\omega}_{b}+\tilde{\omega}_{c}+\tilde{\omega}_{d}\right)}
\bigg]\\$}
	  }\\ \hline 
$D_{13}$ & 
\Large\scalebox{0.70}{
\parbox[t]{20cm}
 {$\\\frac{8!}{2!2!3!}\bigg[ \frac{x_{a} x_{d}}{8 \tilde{\omega}_{b} \tilde{\omega}_{c} \tilde{\omega}_{d}\left(\tilde{\omega}_{a}+\tilde{\omega}_{d}\right)}+\frac{3 x_{a} x_{d}}{8 \tilde{\omega}_{b} \tilde{\omega}_{c} \tilde{\omega}_{d}\left(\tilde{\omega}_{a}+2 \tilde{\omega}_{b}+\tilde{\omega}_{d}\right)}+\frac{3 x_{a} x_{d}}{8 \tilde{\omega}_{b} \tilde{\omega}_{c} \tilde{\omega}_{d}\left(\tilde{\omega}_{a}+2 \tilde{\omega}_{c}+\tilde{\omega}_{d}\right)}\\-\frac{3 x_{a} x_{d}}{8 \tilde{\omega}_{b} \tilde{\omega}_{c} \tilde{\omega}_{d}\left(\tilde{\omega}_{a}+2\left(\tilde{\omega}_{b}+\tilde{\omega}_{c}\right)+\tilde{\omega}_{d}\right)}+\frac{3 x_{a} x_{d}}{8 \tilde{\omega}_{b} \tilde{\omega}_{c} \tilde{\omega}_{d}\left(\tilde{\omega}_{a}+3 \tilde{\omega}_{d}\right)}-\frac{3 x_{a} x_{d}}{8 \tilde{\omega}_{b} \tilde{\omega}_{c} \tilde{\omega}_{d}\left(\tilde{\omega}_{a}+2 \tilde{\omega}_{b}+3 \tilde{\omega}_{d}\right)}-\frac{3 x_{a} x_{d}}{8 \tilde{\omega}_{b} \tilde{\omega}_{c} \tilde{\omega}_{d}\left(\tilde{\omega}_{a}+2 \tilde{\omega}_{c}+3 \tilde{\omega}_{d}\right)}\\+\frac{3 x_{a} x_{d}}{8 \tilde{\omega}_{b} \tilde{\omega}_{c} \tilde{\omega}_{d}\left(\tilde{\omega}_{a}+2\left(\tilde{\omega}_{b}+\tilde{\omega}_{c}\right)+3 \tilde{\omega}_{d}\right)}-\frac{6\left(\tilde{\omega}_{a}+2 \tilde{\omega}_{b}+\tilde{\omega}_{c}+2 \tilde{\omega}_{d}\right) _{a} x_{b}^{2} x_{d}}{\left(\tilde{\omega}_{a}+2 \tilde{\omega}_{b}+\tilde{\omega}_{d}\right)\left(\tilde{\omega}_{a}+2\left(\tilde{\omega}_{b}+\tilde{\omega}_{c}\right)+\tilde{\omega}_{d}\right)\left(\tilde{\omega}_{a}+2 \tilde{\omega}_{b}+3 \tilde{\omega}_{d}\right)\left(\tilde{\omega}_{a}+2\left(\tilde{\omega}_{b}+\tilde{\omega}_{c}\right)+3 \tilde{\omega}_{d}\right)}\\- \frac{6\left(\tilde{\omega}_{a}+\tilde{\omega}_{b}+2\left(\tilde{\omega}_{c}+\tilde{\omega}_{d}\right)\right) x_{a} x_{c}^{2} x_{d}}{\left(\tilde{\omega}_{a}+2 \tilde{\omega}_{c}+\tilde{\omega}_{d}\right)\left(\tilde{\omega}_{a}+2\left(\tilde{\omega}_{b}+\tilde{\omega}_{c}\right)+\tilde{\omega}_{d}\right)\left(\tilde{\omega}_{a}+2 \tilde{\omega}_{c}+3 \tilde{\omega}_{d}\right)\left(\tilde{\omega}_{a}+2\left(\tilde{\omega}_{b}+\tilde{\omega}_{c}\right)+3 \tilde{\omega}_{d}\right)}-\frac{x_{a} x_{b}^{2} x_{c}^{2} x_{d}^{3}}{\tilde{\omega}_{a}+2\left(\tilde{\omega}_{b}+\tilde{\omega}_{c}\right)+3 \tilde{\omega}_{d}}\\-\frac{3 x_{a} x_{b}^{2} x_{c}^{2} x_{d}}{\left(\tilde{\omega}_{a}+2\left(\tilde{\omega}_{b}+\tilde{\omega}_{c}\right)+\tilde{\omega}_{d}\right)\left(\tilde{\omega}_{a}+2\left(\tilde{\omega}_{b}+\tilde{\omega}_{c}\right)+3 \tilde{\omega}_{d}\right)}- \frac{2\left(\tilde{\omega}_{d}+\tilde{\omega}_{b}+\tilde{\omega}_{c}+3 \tilde{\omega}_{d}\right) x_{a} x_{d}^{3}}{\left(\tilde{\omega}_{d}+3 \tilde{\omega}_{d}\right)\left(\tilde{\omega}_{a}+2 \tilde{\omega}_{b}+3 \tilde{\omega}_{d}\right)\left(\tilde{\omega}_{a}+2 \tilde{\omega}_{c}+3 \tilde{\omega}_{d}\right)\left(\tilde{\omega}_{a}+2\left(\tilde{\omega}_{b}+\tilde{\omega}_{c}\right)+3 \tilde{\omega}_{d}\right)}\\-\frac{x_{a} x_{b}^{2} x_{d}^{3}}{\left(\tilde{\omega}_{a}+2 \tilde{\omega}_{b}+3 \tilde{\omega}_{d}\right)\left(\tilde{\omega}_{a}+2\left(\tilde{\omega}_{b}+\tilde{\omega}_{c}\right)+3 \tilde{\omega}_{d}\right)}- \frac{x_{a} x_{c}^{2} x_{d}^{3}}{\left(\tilde{\omega}_{a}+2 \tilde{\omega}_{c}+3 \tilde{\omega}_{d}\right)\left(\tilde{\omega}_{a}+2\left(\tilde{\omega}_{b}+\tilde{\omega}_{c}\right)+3 \tilde{\omega}_{d}\right)}\bigg]\\$}
	  }\\ \hline 
$D_{14}$ & 
\Large\scalebox{0.70}{
\parbox[t]{20cm}
 {$\\\frac{8!}{5!}\bigg[\frac{-~30 x_{a} x_{b} x_{c} x_{d}}{\left(\tilde{\omega}_{a}+\tilde{\omega}_{b}+\tilde{\omega}_{c}+\tilde{\omega}_{d}\right)\left(\tilde{\omega}_{a}+\tilde{\omega}_{b}+\tilde{\omega}_{c}+3 \tilde{\omega}_{d}\right)\left(\tilde{\omega}_{a}+\tilde{\omega}_{b}+\tilde{\omega}_{c}+5 \tilde{\omega}_{d}\right)}-\frac{10 x_{a} x_{b} x_{c} x_{d}^{3}}{\left(\tilde{\omega}_{a}+\tilde{\omega}_{b}+\tilde{\omega}_{c}+3 \tilde{\omega}_{d}\right)\left(\tilde{\omega}_{a}+\tilde{\omega}_{b}+\tilde{\omega}_{c}+5 \tilde{\omega}_{d}\right)}\\-\frac{x_{a} x_{b} x_{c} x_{d}^{5}}{\tilde{\omega}_{a}+\tilde{\omega}_{b}+\tilde{\omega}_{c}+5 \tilde{\omega}_{d}}\bigg]\\$}
	  }\\ \hline 
$D_{15}$ & 
\Large\scalebox{0.70}{
\parbox[t]{20cm}
 {$\\\frac{8!}{3!3!}\bigg[\frac{-~18\left(\tilde{\omega}_{a}+\tilde{\omega}_{b}+2\left(\tilde{\omega}_{c}+\tilde{\omega}_{d}\right)\right) x_{a} x_{b} x_{c} x_{d}}{\left(\tilde{\omega}_{a}+\tilde{\omega}_{b}+\tilde{\omega}_{c}+\tilde{\omega}_{d}\right)\left(\tilde{\omega}_{a}+\tilde{\omega}_{b}+3 \tilde{\omega}_{c}+\tilde{\omega}_{d}\right)\left(\tilde{\omega}_{a}+\tilde{\omega}_{b}+\tilde{\omega}_{c}+3 \tilde{\omega}_{d}\right)\left(\tilde{\omega}_{a}+\tilde{\omega}_{b}+3\left(\tilde{\omega}_{c}+\tilde{\omega}_{d}\right)\right)}-\frac{x_{a} x_{b} x_{c}^{3} x_{d}^{3}}{\tilde{\omega}_{a}+\tilde{\omega}_{b}+3\left(\tilde{\omega}_{c}+\tilde{\omega}_{d}\right)}\\-\frac{3 x_{a} x_{b} x_{c}^{3} x_{d}}{\left(\tilde{\omega}_{a}+\tilde{\omega}_{b}+3 \tilde{\omega}_{c}+\tilde{\omega}_{d}\right)\left(\tilde{\omega}_{a}+\tilde{\omega}_{b}+3\left(\tilde{\omega}_{c}+\tilde{\omega}_{d}\right)\right)}-\frac{3 x_{a} x_{b} x_{c} x_{d}^{3}}{\left(\tilde{\omega}_{a}+\tilde{\omega}_{b}+\tilde{\omega}_{c}+3 \tilde{\omega}_{d}\right)\left(\tilde{\omega}_{a}+\tilde{\omega}_{b}+3\left(\tilde{\omega}_{c}+\tilde{\omega}_{d}\right)\right)}\bigg]\\$}
	  }\\ \hline 
$D_{16}$ & 
\Large\scalebox{0.70}{
\parbox[t]{20cm}
 {\Large$\\\frac{8!}{2!2!2!}\bigg[\frac{-~ x_{a} x_{b}}{8 \tilde{\omega}_{c} \tilde{\omega}_{d}\left(\tilde{\omega}_{a}+\tilde{\omega}_{b}+2\left(\tilde{\omega}_{c}+\tilde{\omega}_{d}\right)\right) \tilde{\omega}_{e}}+\frac{ x_{a} x_{b}}{8 \tilde{\omega}_{c} \tilde{\omega}_{d} \tilde{\omega}_{e}\left(\tilde{\omega}_{a}+\tilde{\omega}_{b}+2 \tilde{\omega}_{e}\right)}-\frac{ x_{a} x_{b}}{8 \tilde{\omega}_{c} \tilde{\omega}_{d} \tilde{\omega}_{e}\left(\tilde{\omega}_{a}+\tilde{\omega}_{b}+2\left(\tilde{\omega}_{c}+\tilde{\omega}_{e}\right)\right)}\\-\frac{ x_{a} x_{b}}{8 \tilde{\omega}_{c} \tilde{\omega}_{d} \tilde{\omega}_{e}\left(\tilde{\omega}_{a}+\tilde{\omega}_{b}+2\left(\tilde{\omega}_{d}+\tilde{\omega}_{e}\right)\right)}+\frac{x_{a} x_{b}}{8 \tilde{\omega}_{c} \tilde{\omega}_{d} \tilde{\omega}_{e}\left(\tilde{\omega}_{a}+\tilde{\omega}_{b}+2\left(\tilde{\omega}_{c}+\tilde{\omega}_{d}+\tilde{\omega}_{e}\right)\right)}-\frac{x_{a} x_{b} x_{c}^{2} x_{d}^{2} x_{e}^{2}}{\tilde{\omega}_{a}+\tilde{\omega}_{b}+2\left(\tilde{\omega}_{c}+\tilde{\omega}_{d}+\tilde{\omega}_{e}\right)}\\-\frac{2  x_{a} x_{b} x_{c}^{2} \quad\left(\tilde{\omega}_{a}+\tilde{\omega}_{b}+2 \tilde{\omega}_{c}+\tilde{\omega}_{d}+\tilde{\omega}_{e}\right)}{\left(\tilde{\omega}_{a}+\tilde{\omega}_{b}+2 \tilde{\omega}_{c}\right)\left(\tilde{\omega}_{a}+\tilde{\omega}_{b}+2 \tilde{\omega}_{c}+2 \tilde{\omega}_{d}\right)\left(\tilde{\omega}_{a}+\tilde{\omega}_{b}+2 \tilde{\omega}_{c}+2 \tilde{\omega}_{e}\right)\left(\tilde{\omega}_{a}+\tilde{\omega}_{b}+2 \tilde{\omega}_{c}+2 \tilde{\omega}_{d}+2 \tilde{\omega}_{e}\right)}+\frac{ x_{a} x_{b}}{8 \tilde{\omega}_{c}\left(\tilde{\omega}_{a}+\tilde{\omega}_{b}+2 \tilde{\omega}_{c}\right) \tilde{\omega}_{d} \tilde{\omega}_{e}}\\-\frac{2 x_{a} x_{b} x_{d}^{2}\left(\tilde{\omega}_{a}+\tilde{\omega}_{b}+\tilde{\omega}_{c}+2 \tilde{\omega}_{d}+\tilde{\omega}_{e}\right)}{\left(\tilde{\omega}_{a}+\tilde{\omega}_{b}+2 \tilde{\omega}_{d}\right)\left(\tilde{\omega}_{a}+\tilde{\omega}_{b}+2 \tilde{\omega}_{c}+2 \tilde{\omega}_{d}\right)\left(\tilde{\omega}_{a}+\tilde{\omega}_{b}+2 \tilde{\omega}_{d}+2 \tilde{\omega}_{e}\right)\left(\tilde{\omega}_{a}+\tilde{\omega}_{b}+2 \tilde{\omega}_{c}+2 \tilde{\omega}_{d}+2 \tilde{\omega}_{e}\right)}+\frac{ x_{a} x_{b}}{8 \tilde{\omega}_{c} \tilde{\omega}_{d}\left(\tilde{\omega}_{a}+\tilde{\omega}_{b}+2 \tilde{\omega}_{d}\right) \tilde{\omega}_{e}}\\-\frac{2 x_{a} x_{b} x_{e}^{2}\left(\tilde{\omega}_{a}+\tilde{\omega}_{b}+\tilde{\omega}_{c}+\tilde{\omega}_{d}+2 \tilde{\omega}_{e}\right)}{\left(\tilde{\omega}_{a}+\tilde{\omega}_{b}+2 \tilde{\omega}_{e}\right)\left(\tilde{\omega}_{a}+\tilde{\omega}_{b}+2 \tilde{\omega}_{c}+2 \tilde{\omega}_{e}\right)\left(\tilde{\omega}_{a}+\tilde{\omega}_{b}+2 \tilde{\omega}_{d}+2 \tilde{\omega}_{e}\right)\left(\tilde{\omega}_{a}+\tilde{\omega}_{b}+2 \tilde{\omega}_{c}+2 \tilde{\omega}_{d}+2 \tilde{\omega}_{e}\right)}-\frac{x_{a} x_{b}}{8\left(\tilde{\omega}_{a}+\tilde{\omega}_{b}\right) \tilde{\omega}_{c} \tilde{\omega}_{d} \tilde{\omega}_{e}}\\-\frac{x_{a} x_{b} x_{c}^{2} x_{d}^{2}}{\left(\tilde{\omega}_{a}+\tilde{\omega}_{b}+2 \tilde{\omega}_{c}+2 \tilde{\omega}_{d}\right)\left(\tilde{\omega}_{a}+\tilde{\omega}_{b}+2 \tilde{\omega}_{c}+2 \tilde{\omega}_{d}+2 \tilde{\omega}_{e}\right)}-\frac{ x_{a} x_{b} x_{d}^{2} x_{e}^{2}}{\left(\tilde{\omega}_{a}+\tilde{\omega}_{b}+2 \tilde{\omega}_{d}+2 \tilde{\omega}_{e}\right)\left(\tilde{\omega}_{a}+\tilde{\omega}_{b}+2 \tilde{\omega}_{c}+2 \tilde{\omega}_{d}+2 \tilde{\omega}_{e}\right)}\\-\frac{ x_{a} x_{b} x_{c}^{2} x_{e}^{2}}{\left(\tilde{\omega}_{a}+\tilde{\omega}_{b}+2 \tilde{\omega}_{c}+2 \tilde{\omega}_{e}\right)\left(\tilde{\omega}_{a}+\tilde{\omega}_{b}+2 \tilde{\omega}_{c}+2 \tilde{\omega}_{d}+2 \tilde{\omega}_{e}\right)}\bigg]\\$}
	  }\\ \hline 
$D_{17}$ & 
\Large\scalebox{0.70}{
\parbox[t]{20cm}
 {$\\\frac{8!}{2!3!}\bigg[\frac{-6 x_{a} x_{b} x_{c} x_{e}\left(\tilde{\omega}_{a}+\tilde{\omega}_{b}+\tilde{\omega}_{c}+\tilde{\omega}_{d}+2 \tilde{\omega}_{e}\right)}{\left(\tilde{\omega}_{a}+\tilde{\omega}_{b}+\tilde{\omega}_{c}+\tilde{\omega}_{e}\right)\left(\tilde{\omega}_{a}+\tilde{\omega}_{b}+\tilde{\omega}_{c}+2 \tilde{\omega}_{d}+\tilde{\omega}_{e}\right)\left(\tilde{\omega}_{a}+\tilde{\omega}_{b}+\tilde{\omega}_{c}+3 \tilde{\omega}_{e}\right)\left(\tilde{\omega}_{a}+\tilde{\omega}_{b}+\tilde{\omega}_{c}+2 \tilde{\omega}_{d}+3 \tilde{\omega}_{e}\right)}\\-\frac{3 x_{a} x_{b} x_{c} x_{d}^{2} x_{e}}{\left(\tilde{\omega}_{a}+\tilde{\omega}_{b}+\tilde{\omega}_{c}+2 \tilde{\omega}_{d}+\tilde{\omega}_{e}\right)\left(\tilde{\omega}_{a}+\tilde{\omega}_{b}+\tilde{\omega}_{c}+2 \tilde{\omega}_{d}+3 \tilde{\omega}_{e}\right)}-\frac{ x_{a} x_{b} x_{c} x_{e}^{3}}{\left(\tilde{\omega}_{a}+\tilde{\omega}_{b}+\tilde{\omega}_{c}+3 \tilde{\omega}_{e}\right)\left(\tilde{\omega}_{a}+\tilde{\omega}_{b}+\tilde{\omega}_{c}+2 \tilde{\omega}_{d}+3 \tilde{\omega}_{e}\right)}\\-\frac{ x_{a} x_{b} x_{c} x_{d}^{2} x_{e}^{3}}{\tilde{\omega}_{a}+\tilde{\omega}_{b}+\tilde{\omega}_{c}+2 \tilde{\omega}_{d}+3 \tilde{\omega}_{e}}\bigg]\\$}
	  }\\ \hline 
$D_{18}$ & 
\Large\scalebox{0.70}{
\parbox[t]{20cm}
 {$\\\frac{8!}{4!}\bigg[\frac{-~6 x_{a} x_{b} x_{c} x_{d}}{\left(\tilde{\omega}_{a}+\tilde{\omega}_{b}+\tilde{\omega}_{c}+\tilde{\omega}_{d}\right)\left(\tilde{\omega}_{a}+\tilde{\omega}_{b}+\tilde{\omega}_{c}+\tilde{\omega}_{d}+2 \tilde{\omega}_{e}\right)\left(\tilde{\omega}_{a}+\tilde{\omega}_{b}+\tilde{\omega}_{c}+\tilde{\omega}_{d}+4 \tilde{\omega}_{e}\right)}-\frac{x_{a} x_{b} x_{c} x_{d} x_{e}^{4}}{\tilde{\omega}_{a}+\tilde{\omega}_{b}+\tilde{\omega}_{c}+\tilde{\omega}_{d}+4 \tilde{\omega}_{e}}\\-\frac{6 x_{a} x_{b} x_{c} x_{d} x_{e}^{2}}{\left(\tilde{\omega}_{a}+\tilde{\omega}_{b}+\tilde{\omega}_{c}+\tilde{\omega}_{d}+2 \tilde{\omega}_{e}\right)\left(\tilde{\omega}_{a}+\tilde{\omega}_{b}+\tilde{\omega}_{c}+\tilde{\omega}_{d}+4 \tilde{\omega}_{e}\right)}\bigg]\\$}
	  }\\ \hline 
$D_{19}$ & 
\Large\scalebox{0.70}{
\parbox[t]{20cm}
 {$\\\frac{8!}{3!}\bigg[\frac{-~3 x_{a} x_{b} x_{c} x_{d} x_{e} x_{f}}{\left(\tilde{\omega}_{a}+\tilde{\omega}_{b}+\tilde{\omega}_{c}+\tilde{\omega}_{d}+\tilde{\omega}_{e}+\tilde{\omega}_{f}\right)\left(\tilde{\omega}_{a}+\tilde{\omega}_{b}+\tilde{\omega}_{c}+\tilde{\omega}_{d}+\tilde{\omega}_{e}+3 \tilde{\omega}_{f}\right)}-\frac{x_{a} x_{b} x_{c} x_{d} x_{e} x_{f}^{3}}{\tilde{\omega}_{a}+\tilde{\omega}_{b}+\tilde{\omega}_{c}+\tilde{\omega}_{d}+\tilde{\omega}_{e}+3 \tilde{\omega}_{f}}\bigg]\\$}
	  }\\ \hline 
$D_{20}$ & 
\Large\scalebox{0.70}{
\parbox[t]{20cm}
 {$\\\frac{8!}{2!2!}\bigg[\frac{-~2\left(\tilde{\omega}_{a}+\tilde{\omega}_{b}+\tilde{\omega}_{c}+\tilde{\omega}_{d}+\tilde{\omega}_{e}+\tilde{\omega}_{f}\right) x_{a} x_{b} x_{c} x_{d}}{\left(\tilde{\omega}_{a}+\tilde{\omega}_{b}+\tilde{\omega}_{c}+\tilde{\omega}_{d}\right)\left(\tilde{\omega}_{a}+\tilde{\omega}_{b}+\tilde{\omega}_{c}+\tilde{\omega}_{d}+2 \tilde{\omega}_{e}\right)\left(\tilde{\omega}_{a}+\tilde{\omega}_{b}+\tilde{\omega}_{c}+\tilde{\omega}_{d}+2 \tilde{\omega}_{f}\right)\left(\tilde{\omega}_{a}+\tilde{\omega}_{b}+\tilde{\omega}_{c}+\tilde{\omega}_{d}+2\left(\tilde{\omega}_{e}+\tilde{\omega}_{f}\right)\right)}\\
 -\frac{x_{a} x_{b} x_{c} x_{d} x_{e}^{2}}{\left(\tilde{\omega}_{a}+\tilde{\omega}_{b}+\tilde{\omega}_{c}+\tilde{\omega}_{d}+2 \tilde{\omega}_{e}\right)\left(\tilde{\omega}_{a}+\tilde{\omega}_{b}+\tilde{\omega}_{c}+\tilde{\omega}_{d}+2\left(\tilde{\omega}_{e}+\tilde{\omega}_{f}\right)\right)}-\frac{x_{a} x_{b} x_{c} x_{d} x_{f}^{2}}{\left(\tilde{\omega}_{a}+\tilde{\omega}_{b}+\tilde{\omega}_{c}+\tilde{\omega}_{d}+2 \tilde{\omega}_{f}\right)\left(\tilde{\omega}_{a}+\tilde{\omega}_{b}+\tilde{\omega}_{c}+\tilde{\omega}_{d}+2\left(\tilde{\omega}_{e}+\tilde{\omega}_{f}\right)\right)}\\- \frac{x_{a} x_{b} x_{c} x_{d} x_{e}^{2} x_{f}^{2}}{\left(\tilde{\omega}_{a}+\tilde{\omega}_{b}+\tilde{\omega}_{c}+\tilde{\omega}_{d}+2\left(\tilde{\omega}_{e}+\tilde{\omega}_{f}\right)\right)}\bigg]\\$}
	  }\\ \hline 
$D_{21}$ & 
\Large\scalebox{0.70}{
\parbox[t]{20cm}
 {$\\\frac{8!}{2!}\bigg[\frac{-~x_{a} x_{b} x_{c} x_{d} x_{e} x_{f}}{\left(\tilde{\omega}_{a}+\tilde{\omega}_{b}+\tilde{\omega}_{c}+\tilde{\omega}_{d}+\tilde{\omega}_{e}+\tilde{\omega}_{f}\right)\left(\tilde{\omega}_{a}+\tilde{\omega}_{b}+\tilde{\omega}_{c}+\tilde{\omega}_{d}+\tilde{\omega}_{e}+\tilde{\omega}_{f}+2 \tilde{\omega}_{g}\right)}-\frac{x_{a} x_{b} x_{c} x_{d} x_{e} x_{f} x_{g}^{2}}{\tilde{\omega}_{a}+\tilde{\omega}_{b}+\tilde{\omega}_{c}+\tilde{\omega}_{d}+\tilde{\omega}_{e}+\tilde{\omega}_{f}+2 \tilde{\omega}_{g}}\bigg]\\$}
	  }\\ \hline 
$D_{22}$ & 
\Large\scalebox{0.70}{
\parbox[t]{20cm}
{$\\\frac{-~8!~x_{a} x_{b} x_{c} x_{d} x_{e} x_{f} x_{g} x_{h}}{\tilde{\omega}_{a}+\tilde{\omega}_{b}+\tilde{\omega}_{c}+\tilde{\omega}_{d}+\tilde{\omega}_{e}+\tilde{\omega}_{f}+\tilde{\omega}_{g}+\tilde{\omega}_{h}}\\$}
	  }\\ \hline\hline \hline
	  \end{longtable}

Now for finding the complexity, we represent the $N$-oscillator wavefunction in the following way:
\begin{equation}
\psi_{0,0, \cdots 0}^{s=0}\left(\tilde{x}_{0}, \cdots ,\tilde{x}_{N-1}\right) \approx \exp \left[-\frac{1}{2} v_{a}  A_{a b}^{s=1}  v_{b}\right]
\end{equation}

Once again, we have to choose a particular basis. Now, there are many choices for bases, but we consider the choice of bases in the following way :
\begin{multline}
     \vec{v} = \{\tilde{x}_{0}, \cdots \tilde{x}_{N-1},\tilde{x}_{0}^2,\cdots ,\tilde{x}_{N-1}^2, \cdots ,\tilde{x}_{a}\tilde{x}_{b}, \cdots,\tilde{x}_{0}^3,\cdots ,\tilde{x}_{N-1}^3, \cdots  ,\tilde{x}_{a}\tilde{x}_{b}\tilde{x}_{c}, \cdots,\tilde{x}_{0}^4,\cdots,\tilde{x}_{N-1}^4,
     \cdots ,\\\tilde{x}_{a}\tilde{x}_{b}\tilde{x}_{c}\tilde{x}_{d}, \cdots ,\tilde{x}_{a}^2\tilde{x}_{b}^2 \cdots,\tilde{x}_{0}^5,\cdots ,\tilde{x}_{N-1}^5,\tilde{x}_{0}^6,\cdots ,\tilde{x}_{N-1}^6,\cdots ,\tilde{x}_{a}\tilde{x}_{b}\tilde{x}_{c}\tilde{x}_{d}\tilde{x}_{e}\tilde{x}_{f}, \cdots,
     \tilde{x}_{a}^3\tilde{x}_{b}^3, \cdots,\\\tilde{x}_{a}\tilde{x}_{b}\tilde{x}_{c}\tilde{x}_{d}\tilde{x}_{e}\tilde{x}_{f}\tilde{x}_{g}\tilde{x}_{h}, \cdots, \tilde{x}_a^{1/2}\tilde{x}_b\tilde{x}_c^{1/2}, \cdots \}
\end{multline}
Here, $a,b,c,d,e,f,g,h$ are indices that can have any value in the range $0$ to $N-1$ and must not be equal to each other. In the last term in $\vec{v}$, we mention a term that can be used to kill off-diagonal entries just as we did it for the two-oscillator case. There will be many more terms like this on this basis. Expressing them explicitly isn't necessary for our current work, so we have not mentioned them. 

Now, we will represent the matrix $A(s=1)$ for $N$ oscillators in a block diagonal fashion. In this format, the matrix will look like this, 
\begin{equation}
   A_{ab}^{s=1} = \begin{pmatrix}
    A_1 & 0 \\
    0 &  A_2
    \end{pmatrix}
\end{equation} where $A_1$ and $A_2$ are the so-called \textit{unambiguous} and \textit{ambiguous} blocks. Once we fix the target or reference stats, the coefficients in the \textit{unambiguous} blocks are fixed. However, it is not the case for \textit{the ambiguous} block as it contains numerous parameters which are not fixed beforehand. \\
In the \textit{unambiguous} block $A_1$ we have all of the coefficients of terms like $x_a^2$ and $x_a x_b$ in Eq. (\ref{Eq_4.8}) multiplied by $-2$. On the other hand, the coefficients (multiplied by $-2$) for terms like 
\begin{equation}
    x_a^2 x_b^2,  x_a^2 x_b^2 x_c^2, x_a x_b x_c x_d 
\end{equation} and so on are there on the $A_2$ block.

To compute the complexity, we choose a particular non-entangled reference state for arbitrary $N$ oscillators:
\begin{equation}
    \psi^{s=0}(x_1,x_2,....,x_n) = \mathcal{N}^{s=0} \exp{\Big[-\sum_{i=0}^{N-1}\frac{\tilde{\omega}_{ref}}{2}\big(x_{i}^2+\lambda_4^0 x_{i}^4+\lambda_6^0 x_{i}^6+\lambda_8^0 x_{i}^8\Big)\Big]}
\end{equation}
which can be represented as
S\begin{equation}
    \psi^{s=0}(\tilde{x}_1,\tilde{x}_2,....,\tilde{x}_n) = \mathcal{N}^{s=0} \exp{\Big[-\frac{1}{2}\Big(v_{a} A_{ab}^{s=0}v_b\Big)\Big]}
\end{equation} where the matrix $A_{ab}^{s=0}$ can be written as in the normal mode basis:

\begin{equation}
     A_{ab}^{s=0} = \begin{pmatrix}
    \tilde{\omega}_{ref}\mathbb{I}_{N\times N} & 0 \\
    0 &  A_2^{s=0}
    \end{pmatrix}
\end{equation}

Here, $\mathbb{I}_{N\times N}$ is the $N$ dimensional unit matrix. We are assuming that all the natural frequencies are i.e. for all $x_i$ it's true that $\omega_0 = \tilde{\omega}_{ref}$. However, $A_2^{s=0}$ cannot be represented as easily as the first block because there are many undetermined parameters. Nevertheless, we can choose these parameters in such a way that the $A_2^{s=0}$ block becomes diagonal, just as we did for the $2$ oscillator case.


The complexity functional depends on the particular cost function that we choose. For different cost functions mentioned in Eq. (\ref{cost_function}) we get a different expression for the complexity functional. However, we will work with the following cost function for the rest of the paper:
\begin{equation} \label{4.16}
    \mathcal{F}_{\kappa}(s) = \sum_I p_I |Y^I|^{\kappa} 
 \end{equation}
With respect to this particular choice of the cost function, the complexity functional becomes:
\begin{equation}
    \mathcal{C}_{\kappa} = \int_{s=0}^{1} \mathcal{F}_{\kappa} \ ds
\end{equation}

Here, we set all the $p_I$ to be $1$ to put all directions in the circuit space on equal footing. Now, if we choose the parameters of $ A_2^{s=0}$  such that $A^{s=0}$ is diagonal, then obviously $A^{s=1}$ and $A^{s=0}$ will commute. If this is the case, then all $\mathcal{C}_{\kappa}$ can be written in a single equation as mentioned in \cite{Bhattacharyya:2018bbv}

\begin{equation} \label{comp_N}
\begin{aligned}
    \mathcal{C}_\kappa &= \mathcal{C}_{\kappa}^{(1)} + \mathcal{C}_{\kappa}^{(2)} \\
    &= \frac{1}{2^\kappa} \sum_{i=0}^{N-1}\Big|\log\Big(\frac{\lambda_i^{(1)}}{\tilde{\omega}_{ref}}\Big) \Big|^{\kappa} + \mathcal{C}_{\kappa}^{(2)}
\end{aligned}
\end{equation} 

Here, $\lambda_i^{(1)}$ are the eigenvalues of the unambiguous block of the $A^{s=1}$ matrix and $\mathcal{C}_{\kappa}^{(1)}, \mathcal{C}_{\kappa}^{(2)}$ denote the contribution to the complexity functional for the unambiguous and ambiguous block, respectively. From here on we will use the $\mathcal{C}_1$ complexity functional.

\textcolor{Sepia}{\subsection{\sffamily Comment on $\mathcal{C}_{1}^{(2)}$ and Ambiguous block}}

 Here, we comment on the difficulties and issues with defining ambiguous block $A_{2}$ as it has also been discussed in \cite{Bhattacharyya:2018bbv} for the $\phi^4$ interaction theory. One of the reasons for calling the $A_{2}$ matrix ambiguous is that there is a lot of arbitrariness in defining this block of the matrix, that is, there are many possible choices for defining the coefficients of the $A_{2}$ block, such as some terms can be defined in the diagonal entries as well as in the off-diagonal entries and several higher-order cross terms like ${\tilde{x}_{a}\tilde{x}_{b}\tilde{x}_{c}\tilde{x}_{d}\tilde{x}_{e}\tilde{x}_{f}\tilde{x}_{g}\tilde{x}_{h}}$ which could be defined in several forms. One possible solution to this is to try to define the $A_{2}$ matrix with the most general entries in which the coefficients are placed among all possible places in the $A_{2}$ block so that the determinant of the matrix should be positive definite.
 For ambiguous block, the Complexity $\mathcal{C}_{1}^{(2)}$ could be defined with eigenvalues $\lambda_j^{(2)}$ and the total complexity will be given by Eq. (\ref{comp_N}). However, due to many arbitrariness or ambiguities in defining the $A_2$ block, we cannot easily define the complexity $\mathcal{C}_{1}^{(2)}$. One could think of using the renormalization approach to get the general form of $\mathcal{C}_{1}^{(2)}$ as it has been done in \cite{Bhattacharyya:2018bbv} for the $\phi^4$ interaction, but the theory in our case is non-renormalizable beyond $\phi^4$ term, so it is also not possible to use the standard renormalization procedure for our case.
 
  Here, we calculate the complexity of the unambiguous block, which is easy to analyze. We use this expression to evaluate complexity functional in the next section.

\textcolor{Sepia}{\section{\sffamily Numerical evaluation of the Complexity Functional }\label{sec:continuous}}
Up to this point, we have always set the value of $M=1$ in the $2$ oscillator Hamiltonian and $N$ oscillator Hamiltonian. However, for a generic analysis and also for the continuum limit, we need to put back the $M$ factor in $H$
If we reinstate the factor of $M$ in the Hamiltonian, we get the following expression for the Hamiltonian:
\begin{multline}
H= \frac{1}{M}\sum_{\vec{n}}\Big\{\frac{P(\vec{n})^{2}}{2 }+\frac{1}{2} M^2\Big[\omega^{2} X(\vec{n})^{2}+\Omega^{2} \sum_{i}\left(X(\vec{n})-X\left(\vec{n}-\hat{x}_{i}\right)\right)^{2}+
2\big\{ \lambda_4 X(\vec{n})^{4}
+ \lambda_6 X(\vec{n})^{6}+\lambda_8 X(\vec{n})^{8}\big\}\Big]\Big\}
\label{Eq_3.3}
\end{multline} The overall factor in front of the Hamiltonian doesn't have any effect on the structure of eigenfunctions of this Hamiltonian. However, some of the factors need to be re-scaled in presence of the $M$ factor which is given below: 

\begin{align*}
\omega \rightarrow \frac{\omega}{\delta} \quad
\Omega \rightarrow \frac{\Omega}{\delta} \quad
\lambda_4 \rightarrow \frac{\lambda_4}{\delta^2} \quad
\lambda_6 \rightarrow \frac{\lambda_6}{\delta^2} \quad
\lambda_8 \rightarrow \frac{\lambda_8}{\delta^2} \quad
\tilde{\omega}_{ref} \rightarrow \frac{\tilde{\omega}_{ref}}{\delta} \quad
\lambda_4^0 \rightarrow \frac{\lambda_4^0}{\delta}
\end{align*}

\begin{align*}
\lambda_6^0 \rightarrow \frac{\lambda_6^0}{\delta} \quad
\lambda_8^0 \rightarrow \frac{\lambda_8^0}{\delta}
\end{align*}

Here, we would like to mention again that $M = \frac{1}{\delta}$. Using these rescaled parameters, we assume that the general form of eigenvalues of $A_1$ represent the $N$ oscillator Hamiltonian with first-order perturbative correction:

\begin{equation}
\begin{split}
    \Lambda_{i_k} & = \Lambda_{4_{i_k}}  + \lambda_6 f_{i_k}\left(N,\tilde{\omega}_{i_p}\right)+\lambda_8 g_{i_k}\left(N,\tilde{\omega}_{i_p}\right), \quad N \text{: Even} \\ 
    & = \Lambda_{4_{i_k}} + \lambda_6 f'_{i_k}\left(N,\tilde{\omega}_{i_p}\right)+\lambda_8 g'_{i_k}\left(N,\tilde{\omega}_{i_p}\right), \quad N \text{: Odd}\\
\end{split}
\end{equation}
where $N$ denotes the number of lattice points in each spatial dimension and $i_k$ indices run from $0$ to $N-1$ for each dimension. Then, the $d-1$ dimensional spatial volume becomes $L^{d-1} = (N\delta)^{d-1}$.

Here, $\Lambda_{4_{i_k}}$ is the contribution from $\phi^4$ interaction, and $f,g,f',g'$ denote the additional contribution to the eigenvalues for the presence of $\phi^6$ and $\phi^8$ interaction. The form of $\Lambda_{4_{i_k}}$ as mentioned in \cite{Bhattacharyya:2018bbv}:

\begin{equation}  \label{5.2}
\begin{aligned}
    \Lambda_{4_{i_k}} & = \frac{\tilde{\omega}_{i_k}}{\delta} + \frac{3\lambda_4}{2N}\Big(\frac{2}{\tilde{\omega}_{i_k}(\tilde{\omega}_{i_k}+\tilde{\omega}_{N - i_k})}+\frac{2}{\tilde{\omega}_{i_k}(\tilde{\omega}_{i_k}+\tilde{\omega}_{\frac{N}{2}- i_k})}\Big), \quad N \text{: Even} \\ 
    & = \frac{\tilde{\omega}_{i_k}}{\delta} + \frac{3\lambda_4}{2N}\Big(\frac{2}{\tilde{\omega}_{i_k}(\tilde{\omega}_{i_k}+\tilde{\omega}_{N - i_k})}\Big), \quad N \text{: Odd}
\end{aligned}
\end{equation}

These additional terms $f,g,f',g'$ cannot be calculated analytically; therefore, we resort to numerical methods to calculate these.


The work done in \cite{Bhattacharyya:2018bbv} had a proper analytical expression for eigenvalues, which made it easier to study RG flows. However, when we consider higher-order interactions such as $\phi^6$ and $\phi^8$, such analytic expressions for the RG flows and complexity cannot be found. This makes it difficult to study the RG flows and MERA for us and is beyond the scope of our model. Instead, we will focus only on complexity.  The eigenvalues we obtain are small corrections to the one obtained in \cite{Bhattacharyya:2018bbv}, so the connection they made will not be affected by the addition of higher interacting terms. Now, we will resort to numerical methods in the next section.

\textcolor{Sepia}{\subsection{\sffamily Numerical analysis of the complexity functional}\label{sec:numerical}}

We will calculate the complexity for the unambiguous block first for the increasing number of oscillators. We have already found the wavefunction for the Hamiltonian in Eq. (\ref{Eq_4.1}). As we reinserted the $M$ term, we will just update the complexity using the rescaled parameters mentioned in the previous subsection. We have set the following relevant parameter values:
\begin{align*}
\lambda_4&=0.5 & \lambda_6&=0.2 &  \lambda_8&=0.001 & \omega_0 = m&=4.0\\
\Omega&=0.25 & L&=200 & \tilde{\omega}_{ref}&=1.6
\end{align*}
where $L$ is the length of the periodic chain. We choose $N$ and $\delta$ so that $N\delta = L$ is always satisfied. We will use the $\mathcal{C}_1^{(1)}$ functional for the unambiguous block. \\

\vspace{5mm}
\textbf{Case I: }{Increasing the Interactions}\\
In Fig. \ref{fig_5.1}, we have plotted numerically the behavior of complexity of unambiguous block as a function of $N$, the number of oscillators in $d=2$ dimensions. In Fig. \ref{fig_5.1}(a), we have two complexities, the points in blue represent the complexity of the theory, which has no interaction term, and this complexity is due to the self-interaction 
\begin{figure}[H] 
\includegraphics[width=1\linewidth]{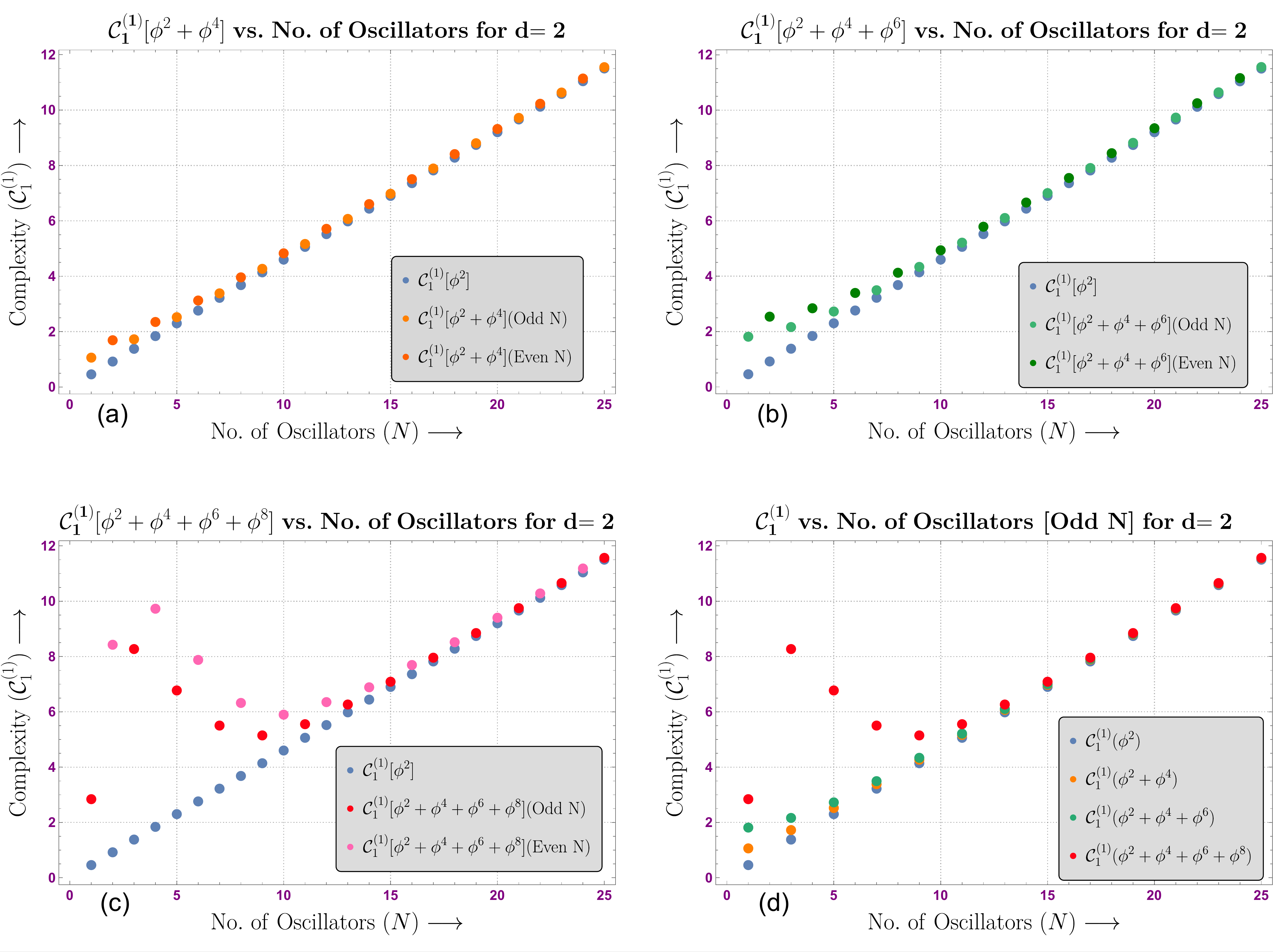}
\caption{Plot: (a), (b), (c) represents the Complexity $\mathcal{C}_{1}^{(1)}$ (from unambiguous block) vs Number of oscillators ($N$) for dimension $d=2$ with different interactions. In plot 4, Complexity $\mathcal{C}_{1}^{(1)}$ vs  Odd No. of oscillators (even resembles the same pattern) from all the interactions are placed together in the same plot, showing the contribution from each of the interaction.}
\label{fig_5.1}
\end{figure}   
between pairs of oscillators. We also see the points in orange and light orange, which is the complexity of the theory with $\lambda_4\phi^4$ interaction.
We notice that there is a bump initially in the graph for small $N$ but in Fig \ref{fig_5.1}(a), \ref{fig_5.1}(b), \ref{fig_5.1}(c), we can observe that the values of complexity with free theory and complexity with interactions become the same as we increase the value of $N$. We see that $\mathcal{C}_1^{(1)}$ grows linearly with increasing $N$ and the contributions to the $\mathcal{C}_1^{(1)}$ due to even interaction terms become negligible and behavior of complexity for the unambiguous block will be same as if we are dealing only with the free theory. In Fig. \ref{fig_5.1}(d), we have plotted $\mathcal{C}_1^{(1)}$ for $N$=odd number of oscillators for even interactions of $\lambda_4\phi^4 +\lambda_6\phi^6+\lambda_8\phi^8$, and we see that the initial values of complexity increase as we include higher-order terms in theory, but when we increase $N$ the contribution from these perturbative terms dies out and graph follows $\phi^2$ linear pattern of $\mathcal{C}_1^{(1)}$.\linebreak
\vspace{2mm}

\textbf{Case II: }{Increasing the Dimension}\\
 In Fig \ref{fig_5.2}, we have shown six different plots. In the first two plots, the complexity for unambiguous block (up to $\phi^4$ interaction) is plotted with respect to the number of oscillators in dimensions $d=3$ and 4 cases. Here, we notice that as we increase the dimension the contribution to $\mathcal{C}_1^{(1)}$ due to the interaction term increases and we see a similar pattern as we include other higher-order even terms, i.e, third and fourth graph have $(\lambda_4\phi^4+\lambda_6\phi^6)$ interactions and fifth and sixth graphs contain $(\lambda_4\phi^4+\lambda_6\phi^6+\lambda_8\phi^8)$ interactions. But in higher dimensions, also the contributions of these interactions to complexity $\mathcal{C}_1^{(1)}$ become negligible when we increase the value of $N$ and the behavior of this complexity becomes similar to the case where we have only the $\phi^2$ term and it grows linearly.\linebreak
\vspace{1mm}
\begin{figure}[H] 
\includegraphics[width=1\linewidth]{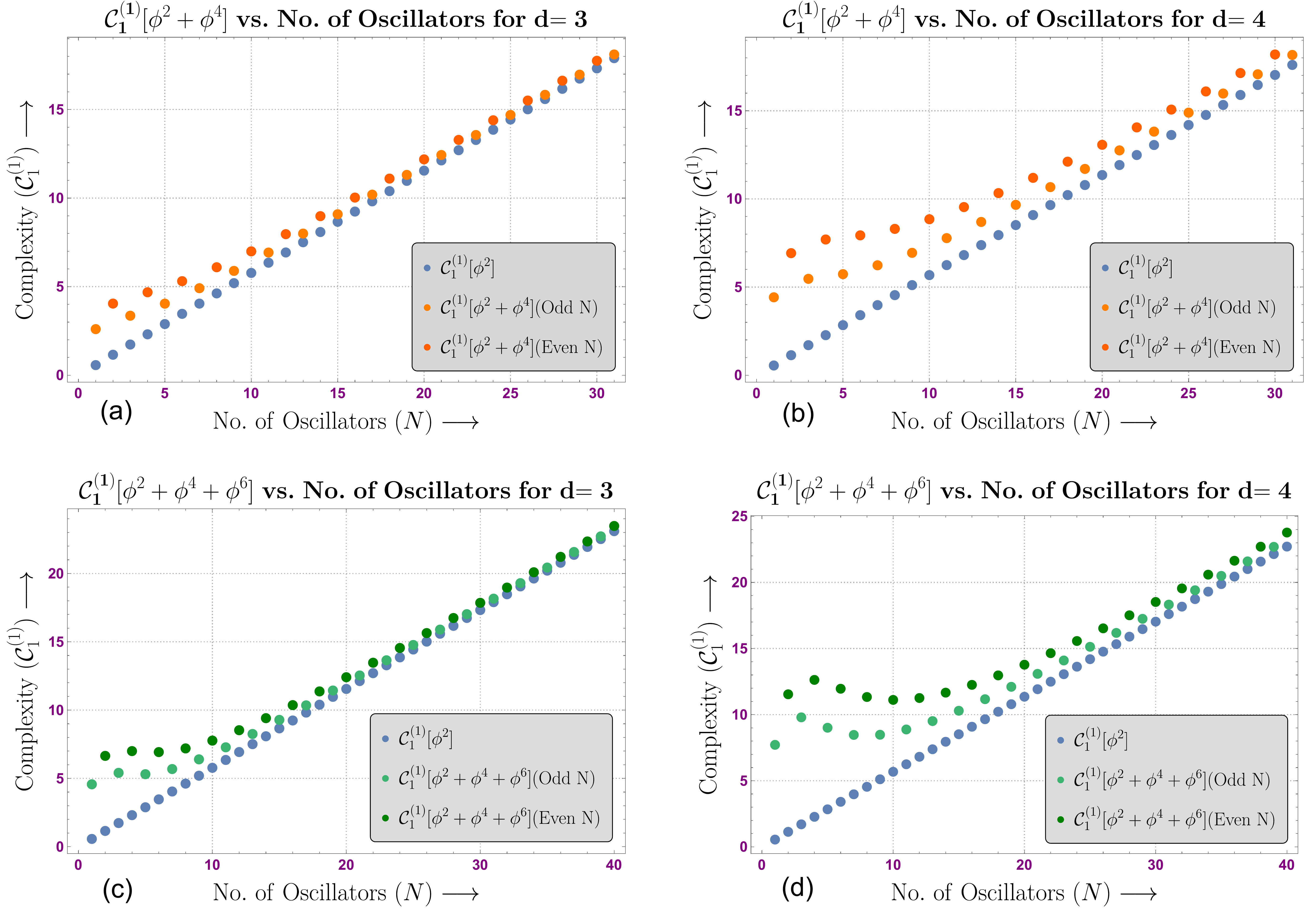}
\label{fig_5.2}
\end{figure}

\begin{figure}[H] 
\includegraphics[width=1\linewidth]{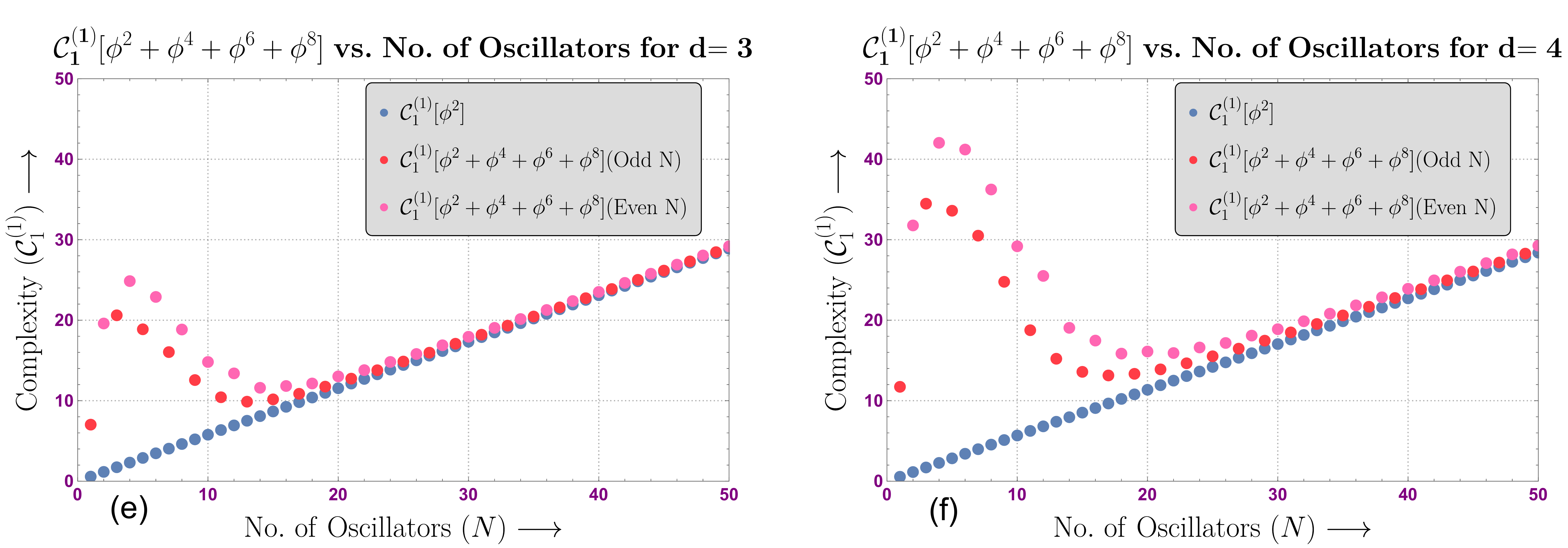}
\caption{Plot of Complexity $\mathcal{C}_{1}^{(1)}$ vs No. of Oscillator in $d=3$ \& $d=4$ respectively for ($\lambda_{2}~\phi^{2}+\lambda_{4}~\phi^{4}+\lambda_{6}~\phi^{6}+\lambda_{8}~\phi^{8}$)}
\label{fig_5.2}
\end{figure}

\textbf{Case III: } $\mathcal{C}_{1}^{(1)}$ vs $\omega_{0}$\\
In Fig. \ref{c-vs-w0} , we have plotted the variation of complexity $\mathcal{C}_{1}^{(1)}$ versus $\omega_{0}$ for a particular value of oscillator, $N= 15$ and we also have shown the variation of the same plot for different dimensions $(d= 2,3,4)$. As we increase the number of
\begin{figure}[H]
\centering
\begin{minipage}{.5\textwidth}
\includegraphics[width=\linewidth]{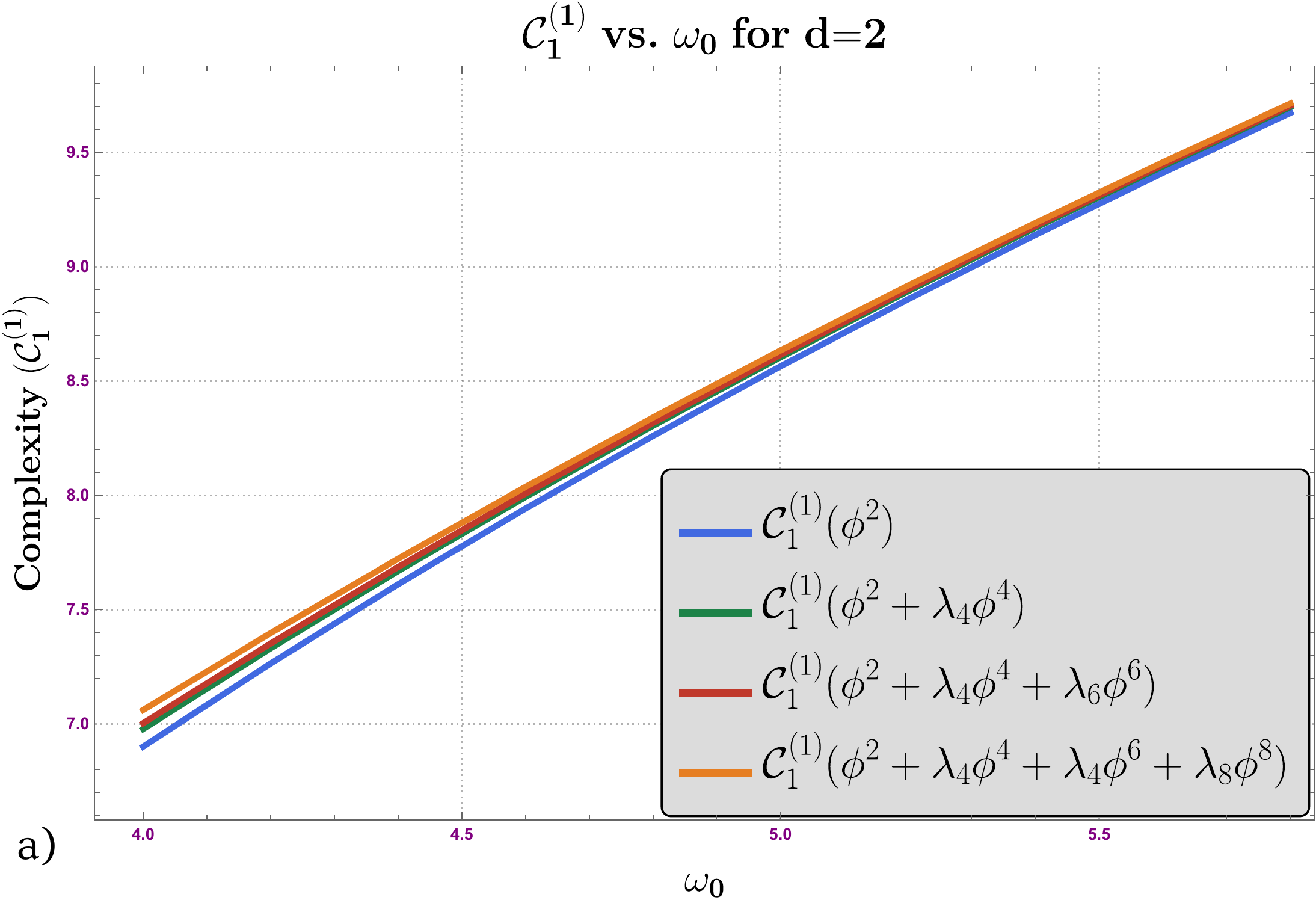}
\end{minipage}\hfill
\begin{minipage}{.5\textwidth}
\includegraphics[width=\linewidth]{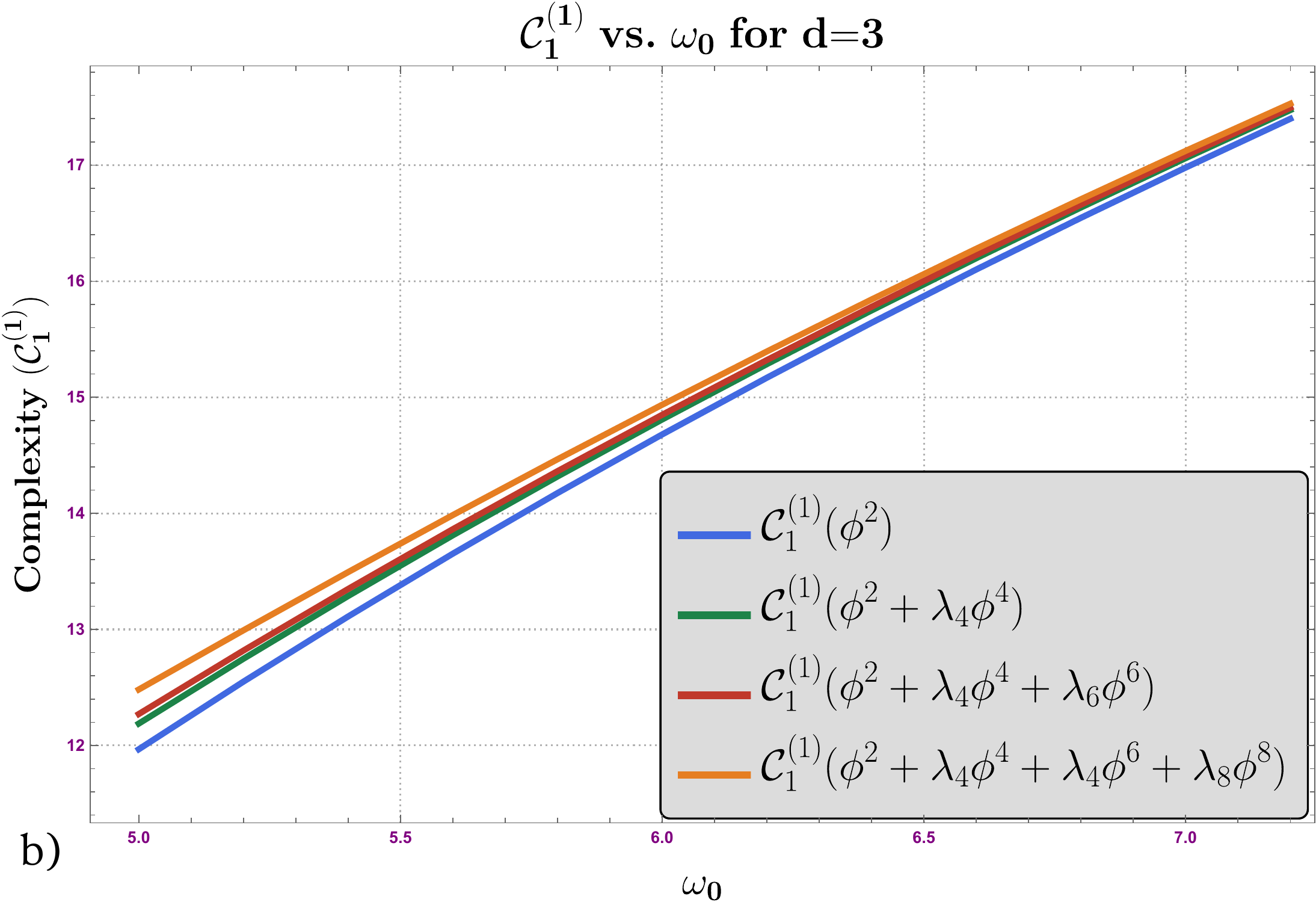}
\end{minipage}\par
\vskip\floatsep
\includegraphics[width=0.5\linewidth]{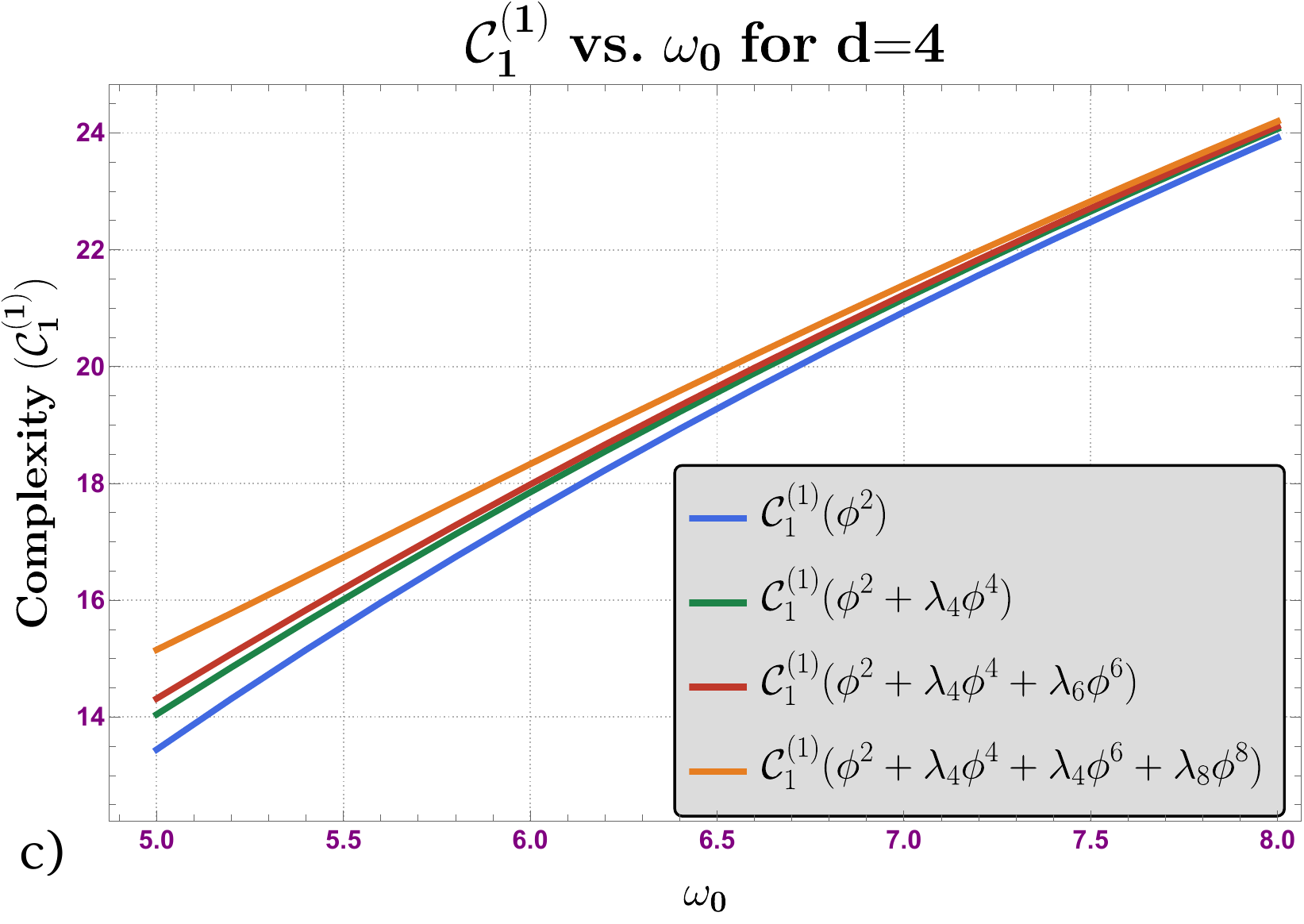}
\caption{Plot of Complexity $\mathcal{C}_{1}^{(1)}$ vs $\omega_0$, Fig: $a)$ if for $d=2$, Fig: $b)$ if for $d=3$ and Fig: $c)$ if for $d=4$ respectively.}
\label{c-vs-w0}
\end{figure}

dimensions the complexity of an unambiguous block $\mathcal{C}_{1}^{(1)}$ increases and in a particular dimension the complexity value increases as we increase the number of interactions, which is noticeable for low values of $\omega_{0}$, but as we increase the value of $\omega_{0}$ the behavior becomes similar to the free scalar theory.

\textbf{Case IV: Fractional change in $\mathcal{C}_{1}^{(1)}$ }
We define the fractional change in complexity $\mathcal{C}_1$ for a particular $N$ as: 
\begin{equation*}
    \frac{\mathcal{C}_1 (N+2)-\mathcal{C}_1 (N)}{\mathcal{C}_1 (N)}
\end{equation*}

Here, we have an increment by $2$ in the definition because odd and even branches of $N$ can possibly show different behaviour as was the case for complexity.
\begin{figure}[ht!]
    \centering
    \includegraphics[width=1\textwidth]{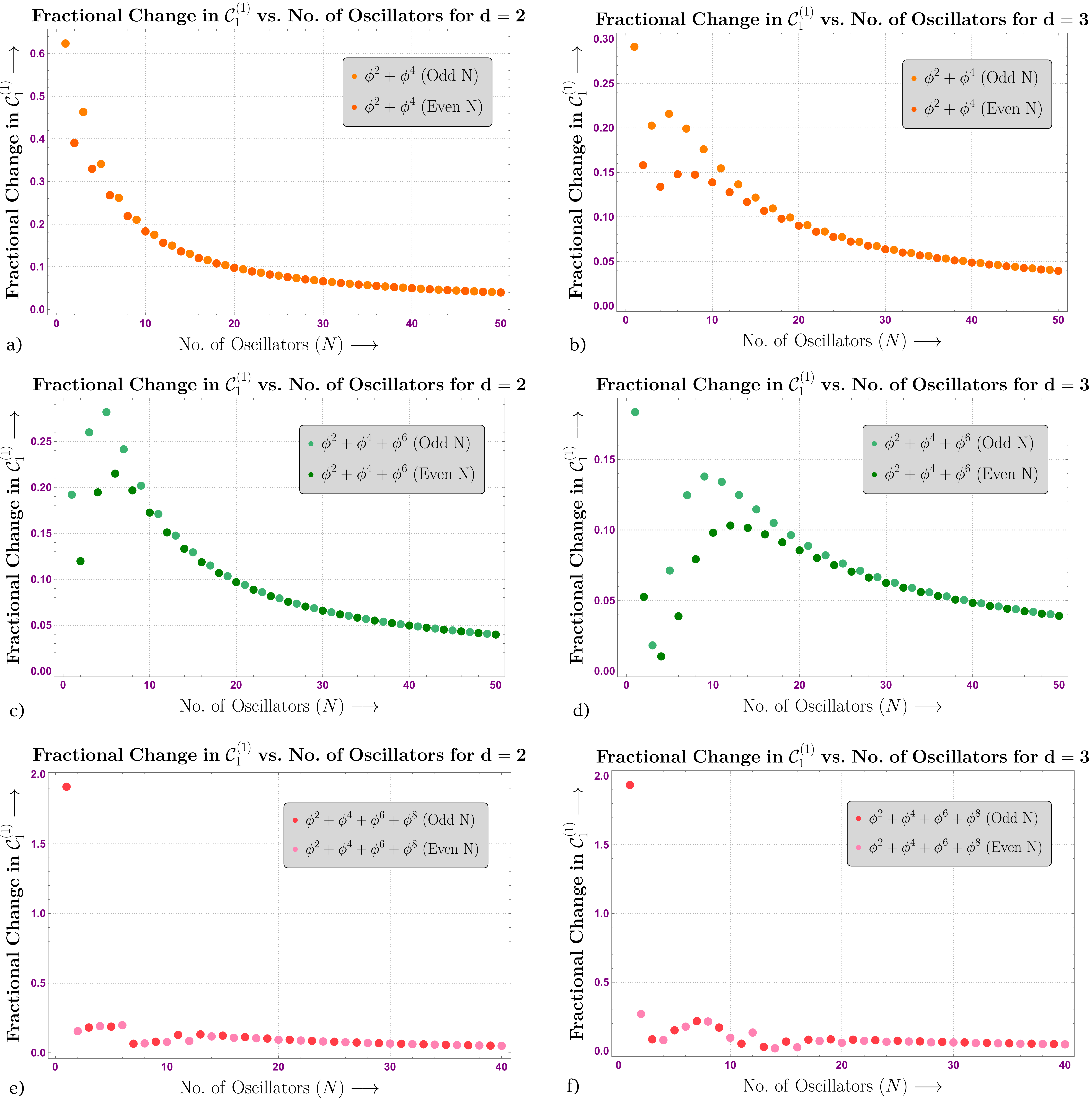}
  \caption{The   Plot of Fractional Change in Complexity vs. No. of Oscillators}
    \label{fig:frac_change}
\end{figure}

For small values of $N$, the even and odd complexities are different from each other. This is directly related to the fact that one can distinguish the system with an even or odd number of oscillators, but as we go for a large number of oscillators or in the continuum limit, the distinction between the even and odd number of oscillators fades away. In Fig. \ref{fig:frac_change} we have plotted the complexity of the unambiguous block and we find that initially, the fractional change in complexity is large for small $N$, but it decreases continuously as we move towards a large number of oscillators.
\textcolor{Sepia}{\section{\sffamily Conclusion and Future Prospects}\label{sec:Conclusion}}

This work has studied the circuit complexity for weakly interacting scalar field theory with $\phi^4$, $\phi^6$, and $\phi^8$ Wilsonian operators, coupled via $\lambda_4$, $\lambda_6$ and $\lambda_8$ to free scalar field theory. The values of the coupling constants have been chosen in the framework of EEFT, such that the perturbation analysis is valid. The reference state is an unentangled, nearly Gaussian state, and the target state is an entangled nearly Gaussian state which has been calculated using first-order perturbation theory. First, we have worked with the case of two oscillators, where the unitary evolution $\mathbb{U}$, which takes us from the reference state to the target state, has been parameterized using the AdS parameters. Using this, we calculated the line element and got the complexity functional by imposing the appropriate boundary conditions. Then we proceeded to the $N$-oscillator case. Now, the circuit complexity depends on the ratio of the eigenvalues of the target and the reference states of the $N$ oscillators. Since we could not observe any analytical expression of the eigenvalues of the target state of $N$ oscillators, we resorted to numerical analysis. The target matrix for $N$ oscillators has a part where the bases can be uniquely determined (unambiguous part) and another part where the bases cannot be (ambiguous part). The contribution to the total complexity comes from the ambiguous as well as the unambiguous parts. In our work, we have focused mainly on the computation of complexity for the unambiguous part, denoted by the $A_2$ matrix. The following are the results that we observed:
\vspace{0.5cm}

\begin{enumerate}

    \item From our numerical analysis, the QCC, with $\kappa = 1$, for the free field theory increases linearly with the number of oscillators. As we include the higher even Wilsonian terms, the growth of complexity (contribution from the unambiguous part) is no longer linear for a small number of oscillators. For the large $N$-limit, the contribution to the complexity from the interacting part vanishes, and the linearity resorts.  
    \item From the graph of complexity vs $\omega_0$, we see that on fixing the dimension and the number of oscillators, the complexity from the unambiguous part increases with increasing value of $\omega_0$.
    
    \item 
    Another pattern inferred from our analysis is that, as the dimension increases, the contribution to $\mathcal{C}_1^{(1)}$ due to the interaction term increases for a fixed number of oscillators. We observed this pattern using degenerate frequencies for higher dimensions. One would expect a similar pattern, even if the frequencies were non-degenerate.

\end{enumerate}

 	In \cite{Bhattacharyya:2018bbv}, the eigenvalues had a proper analytical expression, which makes it easier to study RG flows. On the other hand, after adding higher-order corrections, there is no analytical expression of the eigenvalues. This makes it very challenging to study the RG and MERA connection. The eigenvalues we obtained are small corrections to the one obtained in \cite{Bhattacharyya:2018bbv}, so the connection they made will not be affected by the addition of higher interacting terms. In the upcoming works, we will address this issue.
 
 In our analysis, we have used $\kappa=1$ in our complexity functional $\mathcal{C}_{\kappa}$, but there are other different and useful kinds of measures that one could explore to gain new insights into circuit complexity.\\

Our approach to computing complexity is based on Nielsen's geometric approach, which suffers from ambiguity in choosing the elementary quantum gates and states. Recent works have attempted to develop a new notion of complexity that is independent of these choices. As the future goals we have in mind:
\begin{itemize}
\item We can calculate the circuit complexity for odd Wilsonian terms in the effective theory such as $\phi^3$,$\phi^4$ and $\phi^7$. We can further generalize the study by adding both even and odd interactions term together.
\item We can study the behaviour of circuit complexity in similar theory when there is a quantum quench in the interaction and mass. We have already done this for an $\phi^4$ interacting theory \cite{Choudhury:2022xip}.
\item We can further analyze circuit complexity in fermionic field theories and gauge theories.
\item We can explore this problem in the context of Krylov Complexity \cite{Caputa:2021sib},\cite{Balasubramanian:2022tpr},\cite{Adhikari:2022whf} which is currently a melting pot in this research area.
\item We can compare the Krylov Complexity and Circuit Complexity for such theories to know which is a better measure of information for such cases.
\end{itemize}
\textcolor{Sepia}{\subsection*{\sffamily Acknowledgements}\label{sec:Acknowledge}}
The research fellowship of SC is supported by the J. C. Bose National Fellowship of Sudhakar Panda. SC also would like to thank the School of Physical Sciences, National Institute for Science Education and Research (NISER), Bhubaneswar, for providing a work-friendly environment. SC also thanks all the members of our newly formed virtual international non-profit consortium ``Quantum Structures of the Space-Time \& Matter" (QASTM), for elaborative discussions. SC also would like to thank all the speakers of the QASTM zoominar series from different parts of the world  (For the uploaded YouTube link, look at \textcolor{red}{\url{https://www.youtube.com/playlist?list=PLzW8AJcryManrTsG-4U4z9ip1J1dWoNgd}}) for supporting my research forum by giving outstanding lectures and their valuable time during this COVID pandemic time.   Kiran Adhikari would like to thank TTK, RWTH, and JARA, Institute of Quantum Information for fellowships. Saptarshi Mandal, Nilesh Pandey, Abhishek Roy, Soumya Sarker, Partha Sarker, Sadaat Salman Shariff would like to express their heartiest thanks to Jadavpur University, University of Dhaka, NIT Karnataka, IIT Jodhpur, University of Madras, Delhi Technological University, respectively for imparting knowledge and the enthusiasm for research. Abhishek Roy would like to thank Sujit Damase for discussions related to Group Generators. Partha Sarker would like to thank Dr. Syed Hasibul Hasan Chowdhury for relevant discussions. KA would also like to thank Dr. David Di Vincenzo for his help in understanding quantum information theoretic concepts such as entanglement entropy and complexity. Finally, we would like to acknowledge our debt to the people belonging to the various part of the world for their generous and steady support for research in natural sciences. 
\vspace{1.5cm}

\hrule width 0.5pt

\PRLsep

	\appendix
	\textcolor{Sepia}{\section{\sffamily Interacting part of the Hamiltonian in Fourier basis}\label{sec:appendixC}}
The interacting part in the $N$ oscillator Hamiltonian is 
\begin{equation}
    H' = \sum_{a=0}^{N-1} \lambda_4 x_a^4 + \lambda_6 x_a^6 + \lambda_8 x_a^8 = H'_{\phi^4}+ H'_{\phi^6}+ H'_{\phi^8}
\end{equation} 
Now, we apply the discrete Fourier transform as in Eq. (\ref{Eq_4.2}) we get the $\phi^4$ interaction.
\begin{equation}
    H'_{\phi^4} = \sum_{a=0}^{N-1} \frac{ \lambda_4}{N^2} \sum_{k',k_1,k_2,k_3 = 0}^{N-1} \exp{\left[i\frac{2\pi a}{N}(k'+k_1+k_2+k_3)\right]}\tilde{x}_{k'} \tilde{x}_{k_1} \tilde{x}_{k_2} \tilde{x}_{k_3} 
\end{equation}
Applying the sum over-index $a$ and using the relation 
\begin{equation}
    \sum_{a=0}^{N-1} \exp{\Big[-i\Big(\frac{2\pi a(k-k')}{N}\Big)\Big]} = N \delta_{k,k'}
\end{equation} we get, 
\begin{equation}
    H'_{\phi^4} = \frac{ \lambda_4}{N} \sum_{k',k_1,k_2,k_3 = 0}^{N-1} \delta_{k'+k_1+k_2+k_3,0}\tilde{x}_{k'} \tilde{x}_{k_1} \tilde{x}_{k_2} \tilde{x}_{k_3} 
\end{equation} 
Now, the Kronecker delta will reduce one of the indices, say $k'$ to $-k_1-k_2-k_3$. Now, $k'$ only runs from $[0,N-1]$, whereas $-k_1-k_2-k_3$ has possible values in the range $[-3N,0]$. To get a valid index value for $k'$ we use the relation $\tilde{x}_{k+N}= \tilde{x}_k$ and write $k' = N-k_1-k_2-k_3\ mod N$. This will return a valid index value for $k'$. Then, we have 
\begin{equation}
     H'_{\phi^4} = \frac{\lambda_4} {N} \sum_{k_1,k_2,k_3 = 0}^{N-1} \tilde{x}_{\alpha} \tilde{x}_{k_1} \tilde{x}_{k_2} \tilde{x}_{k_3}
\end{equation}
Using similar arguments, we can get $H'_{\phi^6}$ and $H'_{\phi^8}$.

\textcolor{Sepia}{\section{\sffamily $\mathcal{C}_2$ in terms of the ratio of target and reference matrix eigenvalues}\label{sec:appendixD}}	

We claimed in Eq. (\ref{eq4.13}) that $\mathcal{C}_2$ can be expressed in terms of the ratio of eigenvalues of the target and reference matrix, i.e. $A(s=1)$ and $A(s=0)$. This is due to the nature of the unitary operator $U$ and the diagonal block structure of $A(s=1)$ and $A(s=0)$. 

Now to prove this, let's look at the complexity functional in Eq. (\ref{complexity_functional}). The parameters in $2 \times 2$ blocks on the $U$ matrix have AdS parametrization and they appear in $2[dy_i(1)^2+d\rho_i(1)^2]$ in $\mathcal{C}_2$, where $i = 1,3,5,7,9$. We can get these values of $y_i(1)$ and $\rho_i(1)$ from the boundary conditions we obtained in the equation. (\ref{Eq_3.29}). These values can be represented with the eigenvalues of $A(s=0)$ and $A(s=1)$ in the following way:

\begin{equation}
\begin{split}
y_i &= \frac{1}{4} \log\left[\frac{\lambda_1\lambda_2}{\Omega_1\Omega_2}\right] \\
\rho_i &= \frac{1}{2}\text{cosh}^{-1}\left[\frac{\lambda_1+\lambda_2}{2\sqrt{\lambda_1\lambda_2}}\right]
 \end{split}
\end{equation} 
Here, $\lambda_1$ and $\lambda_2$ are eigenvalues of the $2 \times 2$ block in $A(s=1)$ matrix corresponding to the block in $U$. Whereas $\Omega_1$ and $\Omega_2$ are diagonal elements of the similar block $2 \times 2$ in $A(s=0)$. Using the relation
\begin{equation}
    \text{cosh}^{-1}(x) = \text{ln}(x+\sqrt{x^2-1})
\end{equation}
we can get for $\rho_i$, 
\begin{equation}
    \rho_i = \frac{1}{4} \text{ln}\left[\frac{\lambda_2}{\lambda_1}\right]
\end{equation}

Then, our desired part in $\mathcal{C}_2$ will be
\begin{equation}
    2(y_i(1)^2+\rho_i(1))^2 = 2 \left[ \text{ln}\left[\frac{\lambda_1}{\Omega_1}\right]^2+\text{ln}\left[\frac{\lambda_2}{\Omega_2}\right]^2 \right]
\end{equation}
Now, $i = 2,4,6,8$ we have a different scenario. These are the lone diagonal parameters in the $U$ matrix and have boundary conditions such as: 

\begin{equation}
    y_i = \frac{1}{2} \text{ln}\left[\frac{\lambda_T}{\Omega_R}\right]
\end{equation}
Here, $\lambda_T$ and $\Omega_R$ denote the particular diagonal elements in $A(s=0)$ and $A(s=1)$, respectively, corresponding to the parameter $y_i$ here. With these parameter values in hand, we can get from the complexity functional Eq. (\ref{complexity_functional}) the expression for Eq. (\ref{eq4.13}).
\phantomsection
\addcontentsline{toc}{section}{References}
\bibliographystyle{utphys}
\bibliography{references}

\end{document}